\documentclass{aa}
\usepackage[varg]{txfonts}
\usepackage{graphicx}

\usepackage[normalem]{ulem}
\usepackage{caption}

\usepackage{graphicx}
\usepackage{hyperref}
\usepackage{xcolor}
\hypersetup{
    colorlinks,
    linkcolor={red!50!black},
    citecolor={blue!50!black},
    urlcolor={blue!80!black}
}
\usepackage{color}
\usepackage[square, numbers]{natbib}
\usepackage{multirow}
\usepackage{adjustbox}
\usepackage{natbib}
\usepackage{footnote}
\makesavenoteenv{tabular}
\makesavenoteenv{table}
\usepackage{booktabs}
\usepackage{subcaption}
\usepackage{url}

\bibpunct{(}{)}{;}{a}{}{,} 

\newcommand{\RN}[1]{\uppercase\expandafter{\romannumeral#1}}

\definecolor{RED}{rgb}{1,0,0}\definecolor{BLUE}{rgb}{0,0,1} 

\definecolor{forestgreen}{RGB}{34,139,34}
\definecolor{purple}{RGB}{75,0,130}

\definecolor{darkgreen}{RGB}{0,100,0}

\graphicspath{{./figures/}}

\newcommand{\CII}{CII}
\newcommand{\HII}{HII}

\def\kms{$\mbox{km\,s}^{-1}$}

\usepackage{amstext}

\begin{document}

\title{Observation and calibration strategies for large-scale multi-beam velocity-resolved mapping of the [CII] emission in the Orion molecular cloud\thanks{Final data cube is available at the NASA/IRSA \href{https://irsa.ipac.caltech.edu/applications/sofia/?api=search&spatialConstraints=allsky&planId=04_0066&processingLevel=LEVEL_4&execute=true}{https://irsa.ipac.caltech.edu/} or at the CDS via anonymous ftp to \href{cdsarc.u-strasbg.fr}{cdsarc.u-strasbg.fr} (\href{130.79.128.5}{130.79.128.5}) or via \href{{http://cdsweb.u-strasbg.fr/cgi-bin/qcat?J/A+A/}}{http://cdsweb.u-strasbg.fr/cgi-bin/qcat?J/A+A/}}}

\author{R. Higgins\inst{1},
S. Kabanovic\inst{1},
C. Pabst\inst{2},
D. Teyssier\inst{3},
J. R. Goicoechea\inst{4},
O. Berne\inst{5},
E. Chambers\inst{6},
M. Wolfire\inst{7},
S. Suri\inst{8},
C. Buchbender\inst{1},
Y. Okada\inst{1},
M. Mertens\inst{1},
A. Parikka\inst{6},
R. Aladro\inst{9},
H. Richter\inst{10},
R. G\"{u}sten\inst{9},
J. Stutzki\inst{1},
A.G.G.M. Tielens\inst{2}
}

\institute{
I. Physikalisches Institut der Universit\"at zu K\"oln, Z\"ulpicher Stra\ss e 77, 50937, K\"oln, Germany
\and Leiden Observatory, Leiden University, P.O. Box 9513, 2300 RA Leiden, Netherlands
\and Telespazio Vega UK Ltd for ESA/ESAC, Camino bajo del Castillo, s/n, Urbanizacion Villafranca del Castillo, Villanueva de la Ca\~nada, 28692 Madrid, Spain
\and Instituto de Fisica Fundamental, CSIC, Calle Serrano 121-123, 28006 Madrid, Spain
\and CNRS, IRAP, 9 Av. Colonel Roche, BP 44346, 31028 Toulouse Cedex 4, France
\and SOFIA-USRA, NASA Ames Research Center, MS 232-12, Moffett Field, CA 94035-0001, USA
\and Department of Astronomy, University of Maryland, College Park, MD, USA
\and Max Planck Institute for Astronomy, K\"onigstuhl 
17, 69117 Heidelberg, Germany
\and Max Planck Institut f\"{u}r Radioastronomie, Auf dem H\"{u}gel 69, 53121 Bonn, Germany
\and German Aerospace Center (DLR), Institute of Optical Sensor Systems, Rutherfordstr. 2, 12489 Berlin, Germany
}

\date{\today}
\abstract
{The  [CII]  158$\mu$m  far-infrared fine-structure  line  is  one  of the dominant cooling lines of the star-forming interstellar medium (ISM). Hence 
 [CII]
 emission 
originates in and thus can be used to trace 
a range of ISM processes. Velocity-resolved large-scale mapping of [CII] in star-forming regions provides a unique perspective of the kinematics of these regions and their interactions with the exciting source of radiation.}
{We explore the scientific applications of large-scale mapping of velocity-resolved [CII] observations. With the [CII] observations, we investigate the effect of stellar feedback on the ISM. We present the details of observation, calibration, and data reduction using a heterodyne array receiver mounted on an airborne observatory.}
{A 1.15 square degree velocity-resolved map of the Orion molecular cloud centred on the bar region was observed using the upGREAT heterodyne receiver flying on board the Stratospheric Observatory for Infrared Astronomy (SOFIA). The data were acquired using the 14 pixels of the German REceiver for Astronomy at Terahertz Frequencies (upGREAT) that were observed in an on-the-fly mapping mode. 2.4 million spectra were taken in total. These spectra were gridded into a three-dimensional cube with a spatial resolution of 14.1 arcseconds and a spectral resolution of 0.3 km/s.}
{A square-degree [CII] map with a spectral resolution of 0.3 \kms \ is presented. The scientific potential of this data is summarized with discussion of mechanical and radiative stellar feedback, filament tracing using [CII], [CII] opacity effects, [CII] and carbon recombination lines, and [CII] interaction with the large molecular cloud. The data quality and calibration is discussed in detail, and new techniques are presented to mitigate 
the effects of unavoidable instrument deficiencies (e.g.\ baseline stability) and thus to improve the 
data quality. A comparison with a smaller [CII] map taken with the Herschel/Heterodyne Instrument for the Far-Infrared (HIFI) spectrometer is presented. }
{Large-scale [CII] mapping provides new insight into the kinematics of the ISM. The interaction between massive stars and the ISM is probed through [CII] observations. Spectrally resolving the [CII] emission is necessary to probe the microphysics induced by the feedback of massive stars. 
We show that certain heterodyne instrument data quality issues can be resolved using a  spline-based technique, and better data correction routines allow for more efficient observing strategies.}
{}


\authorrunning{Higgins et al.}
\titlerunning{Large-scale [CII] emission from the OMC}
\maketitle

\section{Introduction}
Massive stars have a profound impact on their environment. They ionize and heat the surrounding gas, creating HII and photodissociation regions (PDRs) \citep{2006agna.book.....O,1999RvMP...71..173H}. 
The resulting 
stellar heating creates overpressurized regions that will expand into their surroundings \citep{1978ppim.book.....S}, creating large-scale ionized gas and photoevaporative flows \citep{1981A&A....98...85B,1997ApJ...476..166W}. This expansion may also be assisted by radiation pressure from the massive stars \citep{2009ApJ...703.1352K,2010ApJ...709..191M}. In addition to this radiative feedback, strong stellar winds of massive stars inject mechanical energy into the interstellar medium, sweeping up dense shells of gas with sizes of some 10 pc \citep{1975ApJ...200L.107C,1977ApJ...218..377W}. Moreover, massive stars will end their life in an explosion that ejects most of their mass at $\sim$10,000 \kms \ into their surroundings. This will result in a supernova remnant filled with hot gas that will expand, further sweeping up surrounding material.
\\
Stellar feedback also affects the interstellar medium on large scales as radiative interaction creates a two-phase medium characterized by ``dense'' clouds subtended in a tenuous intercloud phase \citep{1969ApJ...155L.149F,1995ApJ...443..152W}. The concerted effect of the many supernovae in an OB association will lead to the formation of a collisionally heated, hot intercloud phase \citep{1974ApJ...189L.105C,1977ApJ...218..148M}. These super bubbles may break out of the Galactic plane, venting their hot gas into the lower halo \citep{1987ApJ...317..190M,1988ApJ...324..776M,1989ApJ...345..372N}. This sets up a global circulation of gas over the disk that thoroughly mixes the Interstellar medium (ISM) over large scales.
\\
The radiative and mechanical feedback by massive stars has a strong effect on the star formation 
efficiency 
of the ISM \citep{1997ApJ...476..166W,2013ApJ...776....1K}. On the one hand, this interaction will erode the molecular clouds in which these massive stars were formed, thus limiting the reservoir of molecular gas from which new stars can form. It has been suggested \citep{2017MNRAS.471.4844G,2017MNRAS.466.1903G} that these gas-dispersal processes may be one cause for the observed low star formation efficiency of molecular clouds \citep{1974ApJ...192L.149Z,2008AJ....136.2782L}. On the other hand, the dense swept-up shells can become gravitationally unstable, and thus feedback can trigger new sites of star formation \citep{1977ApJ...214..725E}. On a global scale, galactic outflows limit star formation by removing gas from the disk, and this is a key ingredient in cosmological models of galaxy evolution.
\\
Clearly, feedback by massive stars has a profound effect on the phase structure, physical characteristics, and evolution of the interstellar medium of galaxies. Observationally, this feedback has been studied through X-ray 
emission of the hot gas component in supernova remnants and stellar wind bubbles \citep{2003ApJ...593..874T,2017hsn..book.1981R}. These studies provide a direct measure of the thermal energy of hot gas bubbles involved in the expansion. For young supernova remnants, expansion velocities can be measured through Doppler shifts of UV, optical, and infrared emission lines. However, the kinematics and kinetics of the expanding stellar wind shells are more difficult to trace as velocities involved are quite 
low 
(1$-$20 \kms) and the gas is 
relatively 
cool. The [CII] 1.9 THz $^{2}P_{3/2}$-$^{2}P_{1/2}$ fine-structure transition provides an ideal probe of these shells. The largely evacuated cavities allow the stellar photons to travel unimpeded and create a PDR of warm ($\sim$200 K), largely neutral gas. Except for the densest regions, this PDR gas mainly cools through the [CII] line \citep{1999RvMP...71..173H}. The observed [CII] intensity therefore provides a direct measure of the thermal response of the gas to stellar far-UV photons.
Moreover, the sub-\kms spectral resolution of heterodyne receivers allows a detailed study of the kinematics of wind-blown shells. 
\par
At 414 pc (\citep{2007A&A...474..515M}; Gaia), the Orion molecular cloud core 1 (OMC 1) is the nearest region of massive star formation and has been observed at a multitude of wavelengths in minute detail \citep{2008hsf1.book..459B,2001ARA&A..39...99O,1989ARA&A..27...41G}. The OMC 1 houses the Orion nebula cluster (ONC) of young stellar objects \citep{1997AJ....113.1733H,2016AJ....151....5M}. The O7V star, ${\Theta}^{1}$ Ori C, dominates the ionizing photon flux and luminosity of this cluster \citep{2017ApJ...837..151O}. Interaction of this star with its birth site, Orion molecular core 1, has created the M42 HII region \citep{2009AJ....137..367O} and its associated PDR \citep{1985ApJ...291..747T, 1993Sci...262...86T, 2015ApJ...812...75G}. The stellar wind from ${\Theta}^{1}$ Ori C has excavated a 4 pc diameter cavity, filled with a tenuous, hot plasma emitting at X-ray wavelengths \citep{2008Sci...319..309G}. At optical and UV wavelengths, the region is dominated by HI recombination (e.g., H$\alpha$, H$\beta$) and cooling (e.g., [OIII], [OII], and [NII]) lines from the ionized gas in the M42 HII region. The near-, mid-, and far-infrared regions of the spectrum are dominated by fine-structure lines of abundant species perched on the strong continuum from warm dust and broad emission features due to fluorescence of polycyclic aromatic hydrocarbon molecules \citep{2002A&A...381..571P}. The [CII] 1.9 THz line is one of the brightest transitions in this spectral window \citep{1993ApJ...404..219S,1997ApJ...481..343H,2015ApJ...812...75G} and provides an excellent tracer of the interaction of massive stars and the surrounding swept-up dense shell.
\\
In this paper we present the currently largest  velocity-resolved [CII] map. 
This data set was observed using the upGREAT receiver \citep{2016A&A...595A..34R} mounted on board the flying 
observatory 
SOFIA \citep{2012ApJ...749L..17Y}. The first science results from this project have already been reported by \cite{2019Natur.565..618P}. Here we present the details of the observing strategy and the sophisticated data analysis necessary to provide the highest-quality science data. 
The paper is split into five sections. The observing strategy section details 
the  square-degree mapping strategy with upGREAT. The data reduction section discusses the generation of the final data product and details some of the procedures necessary to mitigate the effects of data artifacts resulting from unavoidable instrument deficiencies. These are partially  
unique to a high-frequency ($>$ 1THz) heterodyne receiver on board an airborne observatory. The data integrity section compares the upGREAT map with the Herschel/HIFI map of the central Orion region. In addition, the repeatability of observations over multiple flights 
is 
investigated in this section. The scientific outlook previews some upcoming scientific work on this unique dataset. A summary section closes the paper with an outlook of upcoming large-scale [CII] mapping projects.
\\
This paper is part of a series of papers studying [CII] in the OMC. This paper focuses on the data acquisition and reduction details. Follow-up papers are in preparation \citep{Kabanovic2020,Suri2020} while some have already been published \citep{Goico20,2019A&A...626A..70S,2019Natur.565..618P,2020A&A...639A...2P}.
\section{Observation overview}
\subsection{upGREAT}
The upGREAT\footnote{upGREAT is a development by the MPI f{\"u}r Radioastronomie and KOSMA/Universit{\"a}t zu K{\"o}ln, in cooperation with the DLR Institut f\"{u}r Optische Sensorsysteme}
Low Frequency Array (LFA) receiver is a heterodyne dual-polarization array with 7 pixels per polarization, that is, 14 pixels in total.
The pixels are placed in a hexagonal format with a pixel at the center of the array; the two polarizations are coaligned on the sky. At 1.9 THz, the hexagon side length is $\sim$32 arcseconds, and each pixel has a beam size of 14.1 arcseconds at 1.9 THz. The pixel spacing is approximately 2.3 times the beam size. This is unavoidable and the necessary minimum spacing for a Gaussian optics instrument. This property leads to gaps in the array pixel distribution on the sky and requires elaborate observing schemes to fully sample a given area.
Figure \ref{fig:array_layout} provides an overview of the pixel layout on the sky during a mapping scan. The upGREAT receiver is coupled to an image rotator that allows the array footprint to be rotated on the sky. Each array of 7 pixels is supplied with the necessary monochromatic signal by an individual local oscillator unit providing frequency coverage from 1.81 to 2.07 THz \citep{2016A&A...595A..34R}.
\par 
Table \ref{tab:main_beam_effeciency} provides an overview of the main-beam efficiency per pixel. All data presented in this paper are in the Rayleigh-Jeans main-beam temperature scale. The overall main-beam efficiency is 0.65 on average because of Gaussian-coupling losses, illumination of the subreflector, and blockage by feed legs (see \citep{2016ITTST...6..199R} for more details). The main-beam efficiency is determined at the start of each flight series with observations of a known calibration source (typically Mars). After four years of LFA operations, the average main-beam efficiencies between flight series  vary within 5\%.
\\
The LFA has an average  single-sideband (SSB) receiver temperature of 2200K. For a nominal 10 micron precipitable water vapor atmospheric burden, this results in a total system temperature of 2600K average over both arrays. The system temperature is defined as the sum of the thermal and receiver noise (for a definition, see \citep{2012A&A...542L...4G}, for a more complete discussion of the receiver temperature breakdown, see \citep{2016ITTST...6..199R}).
Figure \ref{fig:trec_summary} provides a summary of the receiver temperature over the course of the project. Figure \ref{fig:tsys_summary_hifi_upgreat} shows the distribution of the system temperature for both arrays. Figure \ref{fig:trec_example_upgreat} shows an example of an individual receiver temperature spectrum during a single flight. The local oscillator (LO) was tuned so that the OMC line center at a V$_{LSR}$ of 10 \kms \ is at an IF frequency of 1.9 GHz in the lower side band. This setting was chosen during preflight receiver tuning as the best compromise between receiver temperature, receiver stability, and atmospheric transmission. 

\begin{table}[t!]
\centering
     \begin{adjustbox}{width=0.5\textwidth}
        \begin{tabular}{|r|lllllll|}
           \hline
            Pixel         &  0    &   1  & 2    &   3  &  4   &   5  &   6\\\hline
            LFAV          & 0.66  & 0.65 & 0.62 & 0.64 & 0.60 & 0.66 & 0.67\\            
            LFAH          & 0.68  & 0.64 & 0.61 & 0.67 & 0.62 & 0.67 & 0.67\\\hline
        \end{tabular}
     \end{adjustbox}
    \caption{Average main-beam efficiencies of the November 2016 and February 2017 flight series.}
    \label{tab:main_beam_effeciency}
\end{table}

\subsection{Observing strategy}
\label{sect:obs_strat}
\begin{figure}
\centering
\includegraphics[width=1.0\linewidth]{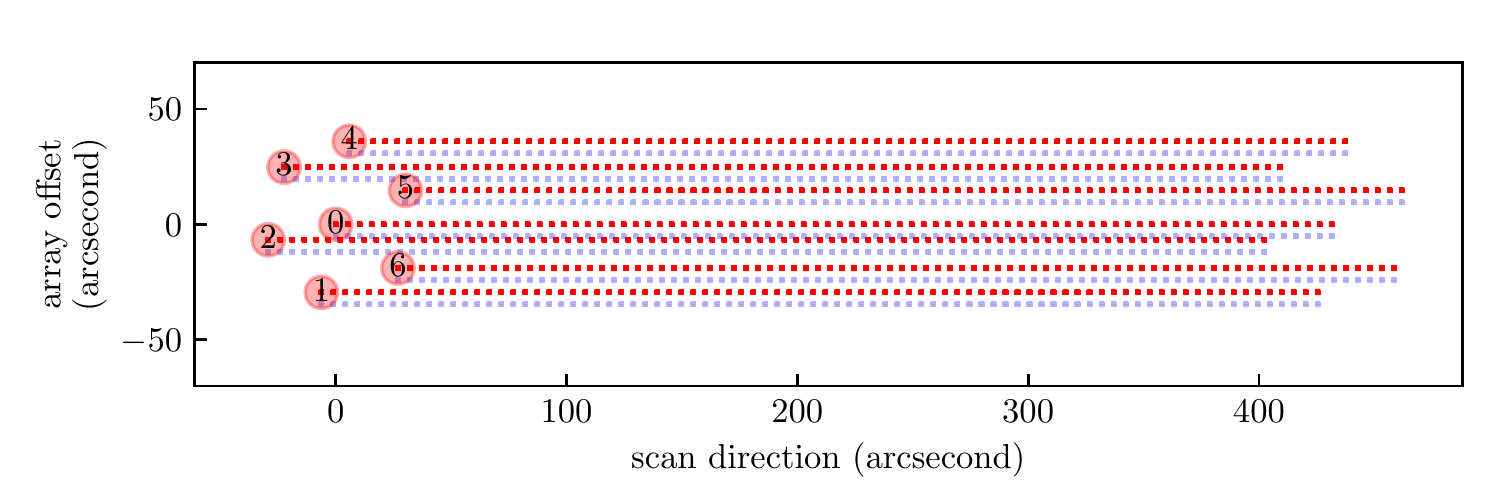}
     \caption{Overview of the upGREAT array pixel layout and an array OTF scan pattern. Red circles show the upGREAT beam size of 14.1 arcseconds rotated to the 19.1$^{\circ}$ array angle, which gives an equidistant spacing orthogonal to the scan direction between the tracks of the individual pixels along the scan direction. The dashed red line shows a typical 84 dump OTF scan, and the dashed blue line shows the subsequent scan offset at 5.2 arcseconds. The pixel positions are actual sky positions from an imperfect alignment, which causes the larger vertical gap between pixel 0 and 5 compared to pixel 0 and 2.}
     \label{fig:array_layout}
\end{figure}

\begin{figure}
\centering
   \includegraphics[width=0.9\linewidth]{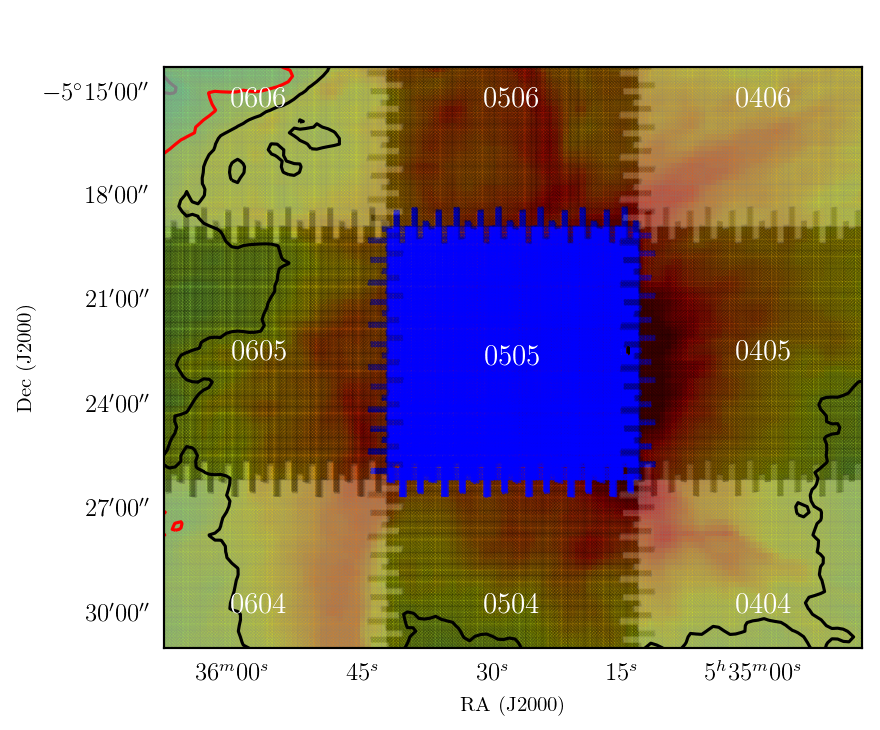}
     \caption{Overview of tile coverage in array OTF mode. Note the interleaving tile edges.}
     \label{fig:tile_overlap}
\end{figure}

\begin{figure}
\centering
   \includegraphics[width=0.9\linewidth]{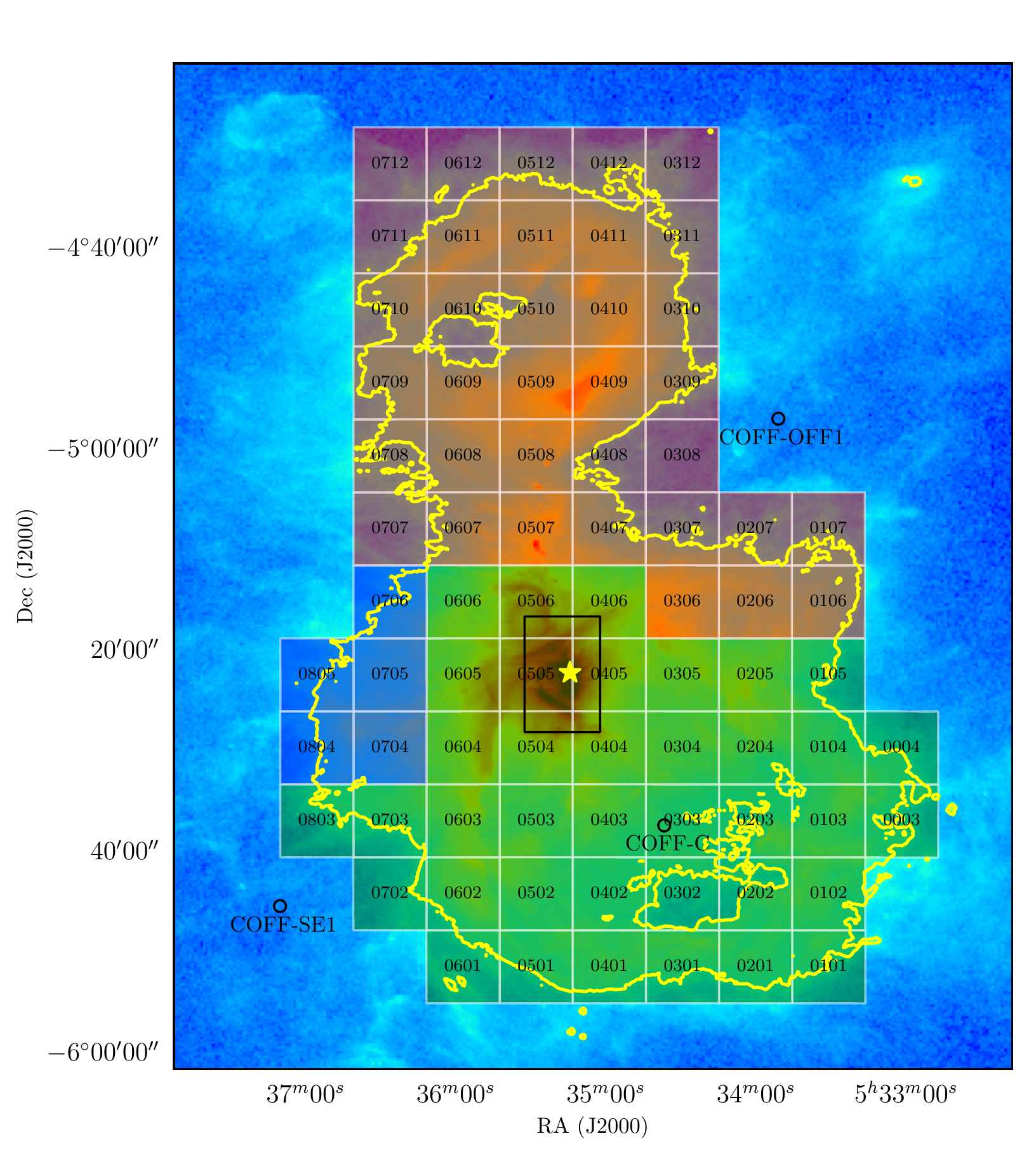}
     \caption{Overview of 78 tiles and their associated OFF positions. Green tiles are associated with the OFF position COFF-C, blue tiles with COFF-SE1, and red tiles with COFF-OFF1. The background image is a 70 micron from Herschel/PACS. The yellow contour denotes regions at a G$_{0}$ of 50 Habing. The HIFI CII map is highlighted with the black box at the center of the map \citep{2015ApJ...812...75G}. The Orion KL outflow is marked with a yellow star.}
     \label{fig:tile_overview}
\end{figure}

The raw data from a typical heterodyne observation are dominated by the instrument response and require calibration observations to remove these effects and determine the instrument response. The simplest heterodyne observation is made up of four phases: an ON phase, an OFF phase, and a HOT and COLD calibration phase. The ON phase is an observation of the astronomical target. The OFF phase is normally a region close to the ON target that ideally is devoid of emission at the target frequency. The calibration phase is taken on an internal 
hot and cold load source 
with known temperature and receiver coupling. In the case of GREAT, the hot load is at the ambient temperature, and the cold load is  at a temperature of about 70 K (cooled by a closed-cycle cryostat Stirling cooler). By combining these four phases, a calibrated spectrum can be created (see \cite{2012A&A...542L...4G} for further details).
\par
The duration of the integration in each observing phase, the frequency of the calibration measurements on the hot and cold load, and the pattern of ON- and OFF-source observations are key parameters for the observing strategy. The duration of the hot- and cold-load measurement has to be sufficiently long so that the noise of the hot-cold observation does not significantly contribute to the noise of the calibrated spectra when applying the gain factor; typically, 10-second integrations on the hot- and cold load are sufficient and also imply that the movement of the receiver optics components to steer the beams to the hot- and cold-load, typically of 1 to 2 seconds duration, does not contribute an excessive dead time. The hot-cold measurement needs to be repeated each time the receiver setup changes, for instance, because of Doppler-tracking of the LO frequency, but also on timescales, where the gain profile of the receivers slowly drifts. Typically, the hot-cold measurements are repeated every few to ten minutes, also by making use of the necessary interruption of the observations by line-of-sight (LOS) rewinds of the inertially tracking three-axis SOFIA telescope. This cadence is sufficient to correct for worst-case gain drifts and Doppler-tracking shifts for all possible source positions.
\par
The regular observation of an off-source position is necessary in order to compensate for total power offset drifts, which are possibly frequency dependent across the receiver reception band, by subtracting the on- from the off-source measurement. By sharing the off-source measurement between several on-source map positions, the dead-time on the off-source position can be minimized because the minimum signal-to-noise ratio of the resulting ON-OFF spectra is achieved by increasing the OFF-source observation $\sqrt{n}$-time, where $n$ is the number of on-source positions in each ON-OFF cycle (see \citep{2007A&A...474..679M} for more information). This leads to a raster mapping scheme with a few ONs per OFF, and ultimately leads to the on-the-fly (OTF) mapping scheme with many (about 10, up to 100) ONs per OFF. The relative fraction of observing time spent as \emph{\textup{dead}} time thus decreases $\propto \frac{1}{\sqrt n}$.
\par
In principle, OTF mapping with fast dumps and many dumps per OFF is therefore most efficient, and only limited by the data rate that the data acquisition hardware can handle. The ON-OFF cycle has to be short enough so that the signal drifts, either due to instrument effects or due to atmospheric changes, do not dominate the radiometric noise. 
The stability of a heterodyne receiver can be determined by Allan variance measurements, which help determine the time between OFF and ON measurements \citep{2009A&A...495..677O}. The spectroscopic stability of upGREAT LFA was determined  to be better than 40 seconds in ideal conditions (see \cite{2016A&A...595A..34R}). For in-flight observation, the stability time is assumed to be 30 seconds given the nonideal thermal stability of an airborne observatory. 
This holds for the frequency resolution that is necessary for the Orion [CII] map observations; it is about 0.3 \kms\  or $1.9\ {\rm MHz}$.
\par
However, for extended sources, the OFF source telescope slews also take time, and they have to be incorporated into the ON-OFF OTF cycle. In practice, this results in OTF dump times of about a few tenths of a second up to a few seconds and some 10 to 50 ON-source dumps per OTF cycle. The overall on-source efficiency is then up to about 80\%.
\par
The stability time, that is, the timescale on which the drifts (independent of frequency channel width $\Delta \nu$) start to dominate the radiometric noise (decreasing with the $\frac{1}{\sqrt{\Delta \nu}}$), decreases for broader frequency resolution, for example, for broad-line observations. When the stability time approaches the order of the dead time of the ON-OFF moves, the observing scheme obviously becomes completely inefficient. For compact sources, chopped observations provide a useful alternative, where the ON-OFF movement is provided by the wobbling secondary mirror and not by the whole telescope assembly. This is correspondingly much faster. However, the ON-OFF throw is limited to small angles on the sky. For SOFIA, the maximum is 10 arcminutes. The ON-source observing efficiency in chopped mode is limited to slightly below 50\% because only single pairs of on- and off-source pointings have to be observed with equal integration time in both, in addition to a short dead time due to the chopper transition time.
\par
Taking the extended nature of the Orion molecular cloud and receiver performance into account, the only choice of an observing mode was the OTF position-switch observation. A further two options of OTF mapping are available for upGREAT: the classical OTF mapping, and array-mapping OTF. Classical mode behaves much like a single-pixel observing mode in that a chosen pixel traces a region in the sky with a fully sampled coverage and all other array pixels follow this pattern. A central core of the map is fully sampled by all pixels, while an edge region is undersampled because of the hexagonal nature of the upGREAT array. In contrast, the array-mapping OTF 
mode, illustrated in Figure \ref{fig:array_layout}, 
takes advantage of the hexagonal nature of the array receiver. With careful selection of the angle between the array longitudinal axis and OTF scan direction, it is possible with two OTF scans separated by 5.2 arcseconds to cover 
a fully sampled 72.6-arcsecond wide strip at 1.9 THz \citep{2016A&A...595A..34R}.
Figure \ref{fig:array_layout} shows a typical array-mapping observation with the array tilted to the appropriate angle.  When a similar double OTF scan in the orthogonal direction is observed, a fully sampled square region of 72.6 arcseconds is generated. This square region forms the base unit of the array OTF mapping scheme. The 
OTF 
array-mapping approach allows mapping of large areas in a shorter time while sacrificing signal-to-noise ratio and pixel redundancy compared to the classical OTF approach. By adding an orthogonal scan to the array OTF mode, the redundancy shortcoming can be compensated for by covering each portion of sky with at least 4 pixels, 2 coaligned H and V pixels in the X direction, and then 2 different H and V pixels in the Y direction. Figure \ref{fig:tile_overlap} shows a typical fully sampled tile with an X and Y scan direction. 
Furthermore, the hexagonal array footprint symmetry can be used by regularly rotating the array by additional multiples of 60 degree in the repetition of observing tiles (see below), thus ensuring that the same area of the sky is observed by different pixels each time. This distributing the pixels equally over the map area and averages out any pixel performance differences over the map. 
\par
The basic tile is six times the base unit tile length of 72.6 arcseconds. Each tile area is covered 
twice, first in X and then in Y 
scan direction. Each OTF scan is 435.6 arcseconds long. An OTF spectrum is taken every 5.2 arcseconds, resulting in 84 OTF spectra per scan. 
This results in a slightly higher than Nyquist sampling, where the Nyquist sampling as derived from the telescope aperture would be 6.4 arcseconds for a 14.1 arcsecond beam size, that is, 2.7 samples per beam versus 2.2. \cite{2007A&A...474..679M} 
recommended a sampling rate of at least twice the Nyquist sampling to reduce elongation issues to less than 1\% in the scanning direction. Elongation is considered in the map-making process, where a convolution kernel 2-3\% larger than the upGREAT beam of 14.1 arcseconds is used to account for this.
\par
In order to stay within the stability time of the system, each OTF dump has an integration time of 0.3 seconds. This results in a total scan 
duration 
of 25.1 seconds. An OFF measurement of 2.8 seconds is then taken every OTF scan to minimize system drift effects where $t_{off} = \sqrt{N_{otf}}*t_{on}$, where $N_{otf}$ is 84 and the scan dump time, $t_{on}$, is 0.3 seconds (see \cite{2007A&A...474..679M} for more details). The final tile layout is shown in Fig.  \ref{fig:tile_overlap}, while the final map layout with all tiles is shown in Fig. \ref{fig:tile_overview}. The boundaries of the map are set by the estimated UV field of G$_{0}$ of 50 Habing. The UV field is determined using the conversion factor from 70 micron to UV field discussed in \cite{2015ApJ...812...75G}. The 50 Habing UV region is shown with a yellow contour in Fig. \ref{fig:tile_overview}. The final map contains 78 square tiles of side length 435.6 arcsecond. 
\par
Figure \ref{fig:tile_overlap} shows the overlap regions between tiles. The hexagonal layout of the upGREAT LFA array leads to jagged edges at the tile edges. Based on our knowledge of previous heterodyne observations (see HIFI map, \cite{2015ApJ...812...75G}), there is the risk that the interface between adjacent tiles will be apparent in the final map. We show in the reduction section that this was not an issue in the final map.

\subsection{OFF selection}
Figure \ref{fig:tile_overview} shows an overview of which tiles are 
linked 
to which OFF positions. Different color tiles are associated with different OFF positions. Three OFF positions were used during the mapping campaign, see table \ref{tab:position_summary} for an overview of the position coordinates. Three OFF positions were chosen to minimize the telescope slew times. During the pilot program, dedicated observations were taken in December 2015 against far-OFF observations to determine the amount of emission from the OFF positions. Clean far-OFF positions were based on previous HIFI observations. It was known in advance that the nearby OFF positions contained some level of emission and that special processing would be required to remove this emission from the final map. An additional complication in the OFF acquisition is the position angle of the array on the sky. Each map tile is made up of an X and Y OTF scan. The position angle of the array between the X and Y scan differs by 30 degrees, resulting in slightly different OFF emission in the X and Y scan directions. This meant that dedicated OFF observation were needed at both the X and Y map 
angles, as is further explained in detail in section 3.5.


\begin{table}
\centering
\begin{tabular}{llll}
\toprule
                 Name &          RA &                                          Dec & Role\\
\midrule
CENTER    & 5\textsuperscript{h}35\textsuperscript{m}27.6\textsuperscript{s} & $-5^\circ22{}^\prime33.7{}^{\prime\prime}$    & Map center\\
COFF-C    & 5\textsuperscript{h}34\textsuperscript{m}36.5\textsuperscript{s} & $-5^\circ37{}^\prime32.7{}^{\prime\prime}$    & Green OFF position \\
COFF-OFF1 & 5\textsuperscript{h}33\textsuperscript{m}51.0\textsuperscript{s} & $-4^\circ57{}^\prime05.2{}^{\prime\prime}$    & Red OFF position \\
COFF-SE1  & 5\textsuperscript{h}37\textsuperscript{m}10.0\textsuperscript{s} & $-5^\circ45{}^\prime33.7{}^{\prime\prime}$    & Blue OFF position \\
FOFF-E    & 5\textsuperscript{h}39\textsuperscript{m}21.6\textsuperscript{s} & $-4^\circ58{}^\prime29.7{}^{\prime\prime}$    & Far OFF east\\
FOFF-W    & 5\textsuperscript{h}31\textsuperscript{m}15.5\textsuperscript{s} & $-5^\circ52{}^\prime27.4{}^{\prime\prime}$    & Far OFF west \\
BAR$\_$PEAK & 5\textsuperscript{h}35\textsuperscript{m}20.9\textsuperscript{s} & $-5^\circ25{}^\prime04.8{}^{\prime\prime}$ & Calibration position \\
\bottomrule
\end{tabular}
\caption{Summary of positions we used during the project. All map tiles were offset from the CENTER position.}
\label{tab:position_summary}
\end{table}

\subsection{Flight summary}
\begin{figure}
\centering
   \includegraphics[width=1.0\linewidth]{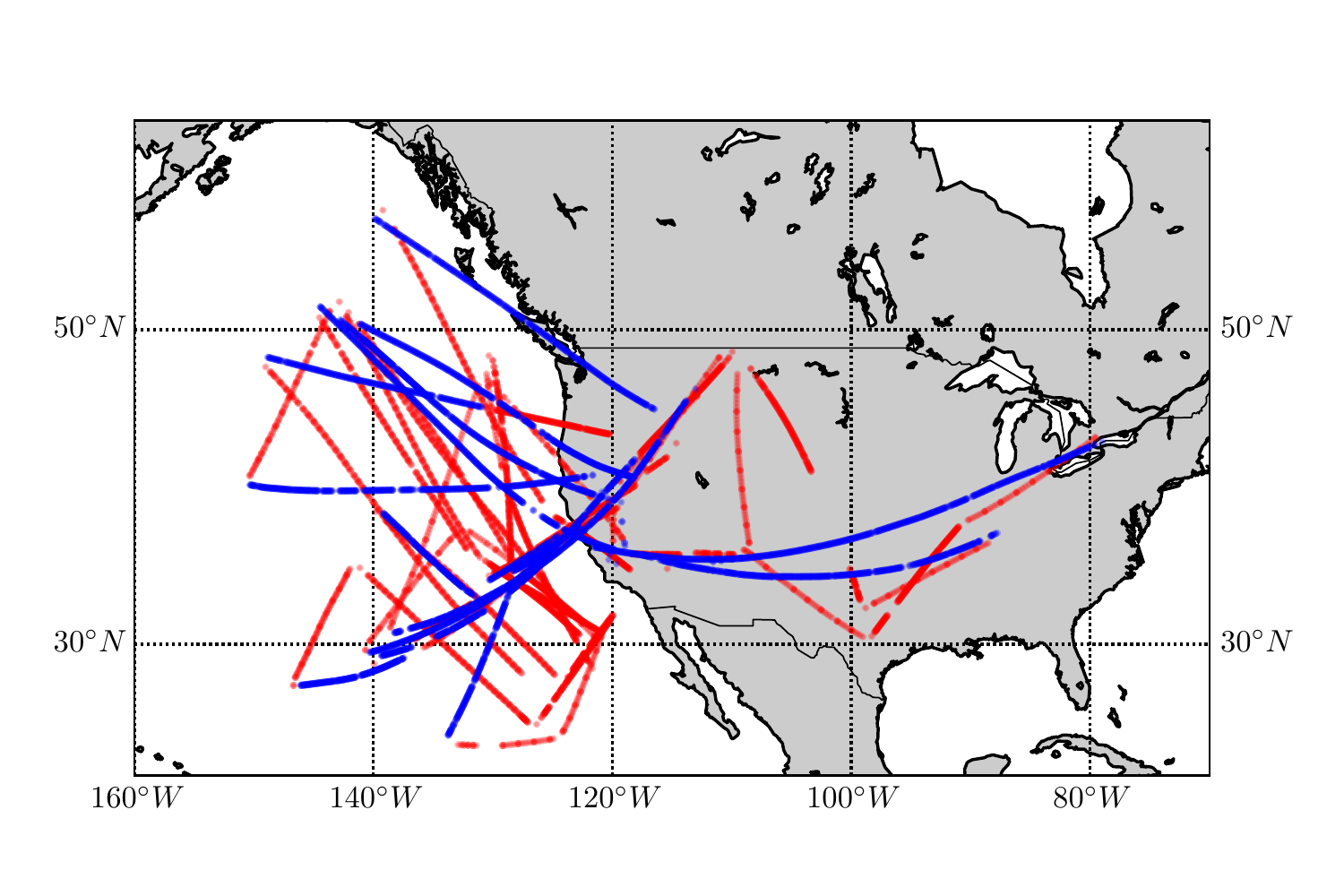}
     \caption{Overview of the flight plans for the 13 Orion project flights. Each point corresponds to the start of an upGREAT observation block. Blue points highlight the Orion legs, while red points show non-Orion flight legs. Each flight returned to Palmdale. The return leg is a not apparent in the map because no observations are taken during the setup and Palmdale return legs.}
     \label{fig:flight_leg_overview}
\end{figure}
The total project was observed over 13 flights (12 flights for map observations, and 1 flight dedicated to calibration and OFF measurements). These flights were divided into two flight series in November 2016 and February 2017. Table \ref{table:flight_summary} provides an overview of the 
tiles 
observed per flight. Eight tiles were observed per flight on average. In total, 2.4 million spectra were recorded over 42 hours of SOFIA flight time. Figure \ref{fig:flight_leg_overview} shows an overview of the flight path, and each Orion flight leg is highlighted in blue.

\section{Data reduction}

\subsection{Overview}
During
the observation, raw spectrometer counts from each observation phase (ON, OFF, HOT, and COLD) are written to a FITS-file \citep{2010A&A...524A..42P}.
The raw data is then converted into antenna temperature using the methods detailed in \cite{2012A&A...542L...4G}. In addition to the  
gain calibration through the HOT-COLD scans, which converts the backend counts into antenna temperature, 
the atmospheric transmission is determined by fitting an atmospheric model to an OFF-HOT spectrum. The atmospheric transmission corrected data 
is then written to a GILDAS (Grenoble Image and Line Data Analysis Software/Continuum and Line Analysis Single-dish Software) class file and is further processed within the GILDAS environment \citep{2005sf2a.conf..721P}. The data calibration process makes use of the dedicated SOFIA user section \citep{2011usersection} and also the associated array infrastructure available with the GILDAS spectrum data format \citep{2015associate}.
\par
The first step in the reduction is the removal of the OFF emission contamination (see section \ref{sec:off_correction}). The next step in the data reduction was the correction of baseline features in the calibrated data. This process was undertaken using a novel scaled spline approach. This is discussed 
in detail 
in section \ref{sec:spline_correction}. The next step in the reduction is the application of the main-beam efficiency for each pixel. The spectra are then down sampled to a suitable velocity resolution of 0.3 \kms\ 
    from the native resolution of 0.04 \kms \ and are then cropped to the velocity 
range of interest, -70 to 85 \kms (in order to cover the [$^{13}$CII] satellite transitions, see table \ref{tab:transition_parameter}). The spectral resolution of 0.3 \kms \ is a compromise between final data size and not compromising on the average line width expected in the OMC. Generating maps at the native resolution would lead to longer processing times without any added benefit to the astronomical interpretation. 
\par
Before we proceeded to generate a map, bad data were filtered out based on a number of criteria, such as high system temperature or large deviations in noise compared to that expected based on the radiometer equation. Further filters were developed to track radio frequency interference (RFI) (discussed section \ref{sec:rfi}) and gain instability (see section \ref{sec:gain_instability}). Finally, some spectra were flagged based on a visual inspection of the final map, in which spectra showed a mismatch with surrounding map 
pixels.
\par
The final step before map generation is to subtract a mean offset from each 
spectrum 
to remove any continuum offsets that are not corrected during the spline 
baseline subtraction 
step. The mean is calculated from a region outside of the main line emission between -10 and 30 \kms. 
This approach is prone to RFI effects outside of the -10 to 30 \kms \ region that skew the mean offset of a 
spectrum. This 
effect can be countered by flagging regions of a spectrum above a given RMS value. This flagging is achieved using the associate array functionality within GILDAS, which allows the user to flag channels that are ignored when the baseline offset is calculated \citep{2015associate}.
The associated arrays are a useful addition to the standard GILDAS \textit{ry} array that contains the astronomical signal. GILDAS allows the user to associate a number of arrays with the \textit{ry} array. Support is built in for bad channels (known as BLANKED array) and line flagging. These arrays are stored as a 2-bit array to minimize the memory footprint. Up to 64 bit doubles are supported, which could be useful for storing atmospheric transmission or system temperature alongside the astronomical signal. This will incur a significant storage penalty, however.
\par
The final filtered set of spectra is then passed to the map maker within the GILDAS software. This takes all map positions and 
interpolates them onto a fixed square grid using a Gaussian kernel based on the beam size of upGREAT at the [CII] transition frequency 
($\sim$14.1 arcseconds). The final map resolution is 18 arcseconds with a pixel size of 3.5 arcseconds. The pixel size is chosen to fully sample the kernel size and allow for smoother contours in the final map. Channels affected by RFI are weighted down to zero weighting 
in the map-generation step by multiplying the associated array weighting 
array by the spectrum intensity. Because each pixel of the map is observed by at least 4 upGREAT pixels and RFI affects just a subset of data, this selective weight approach was possible. The advantage here also is that only the affected channels are discarded and not the entire spectrum. 
\par
The next sections describe the data processing in more detail. We discuss the baseline correction, gain instability, RFI mitigation, and OFF correction.

\subsection{Baseline correction using a scaled spline approach}
\label{sec:spline_correction}

\begin{figure*}

\includegraphics[width=0.9\linewidth]{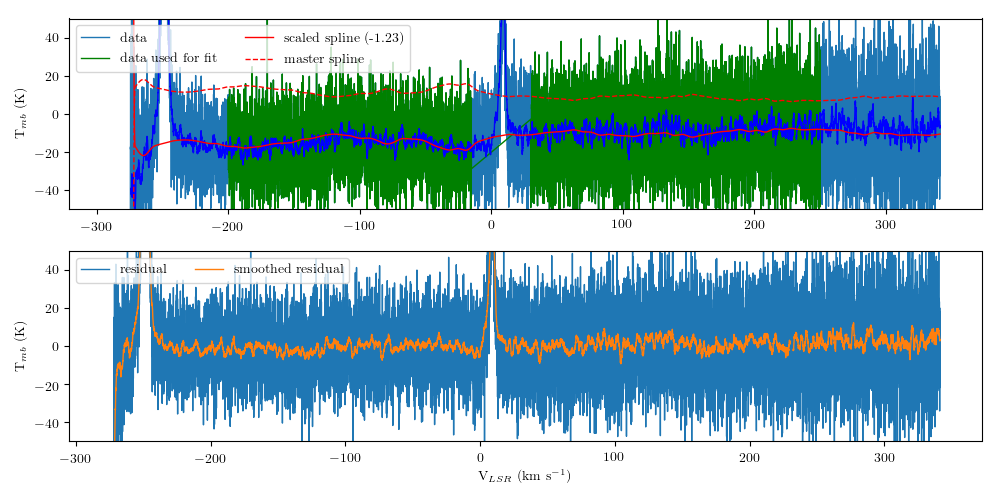}
\caption{Example of the spline baseline correction process. The dashed red line shows the master spline generated from the residual of two OFF sky measurements. The solid red line shows a scaled version of the master spline that matches the smoothed sky data best (dark blue). The green region shows the channel range over which the spline is fitted. The lower panel shows the residual of the spline correction. Light blue shows the original data resolution, and orange shows a smoothed version of the corrected data.}
\label{fig:spline_correction_process}
\end{figure*}

\begin{figure*}
\includegraphics[width=0.9\linewidth]{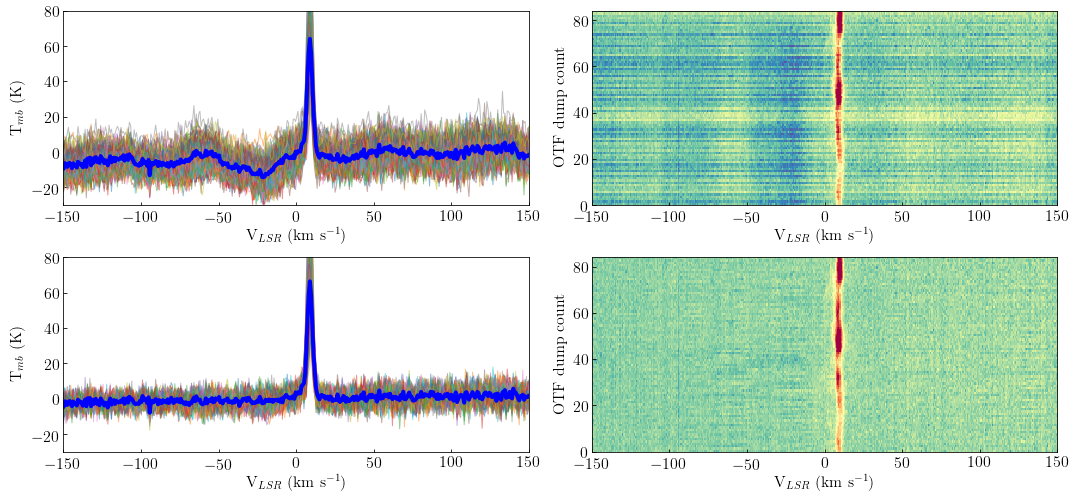}
\caption{Example spline baseline correction for a typical OTF scan of 84 dumps length. \textit{Top left panel:} Baseline after standard polynomial order 3 baseline correction. The average over 84 dumps is shown in dark blue. \textit{Top right panel:} Waterfall plot of 84 spectra taken during an OTF scan. The dip at -25\kms \ is common to all spectra. \textit{Bottom left panel:} Baseline after spline correction. The corresponding waterfall plot is shown on the right side. \textit{Bottom right panel:} Waterfall map of 84 spectra after spline correction.}
\label{fig:spline_otf_scan}
\end{figure*}

\begin{figure*}
\includegraphics[width=0.32\linewidth]{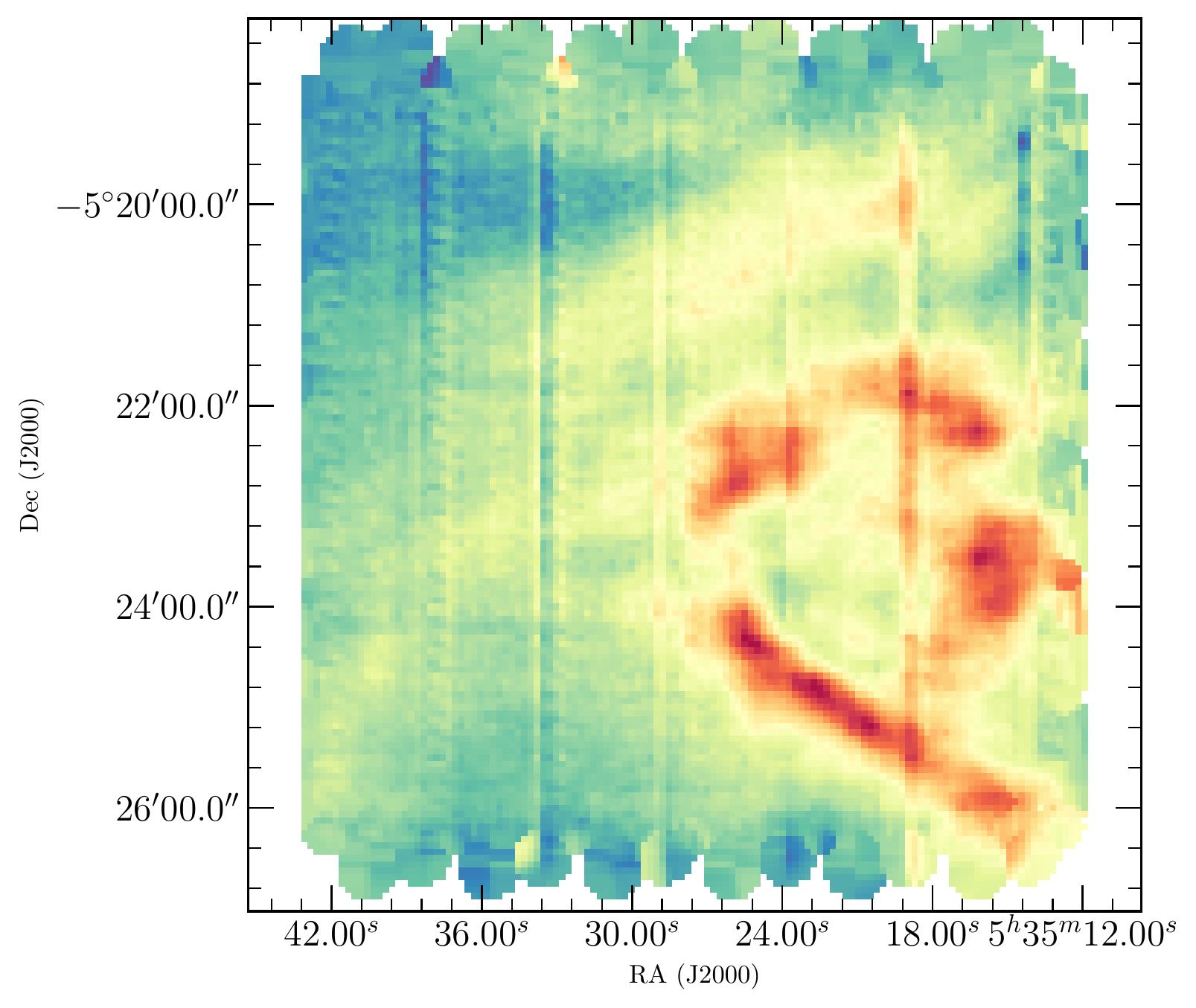}
\includegraphics[width=0.32\linewidth]{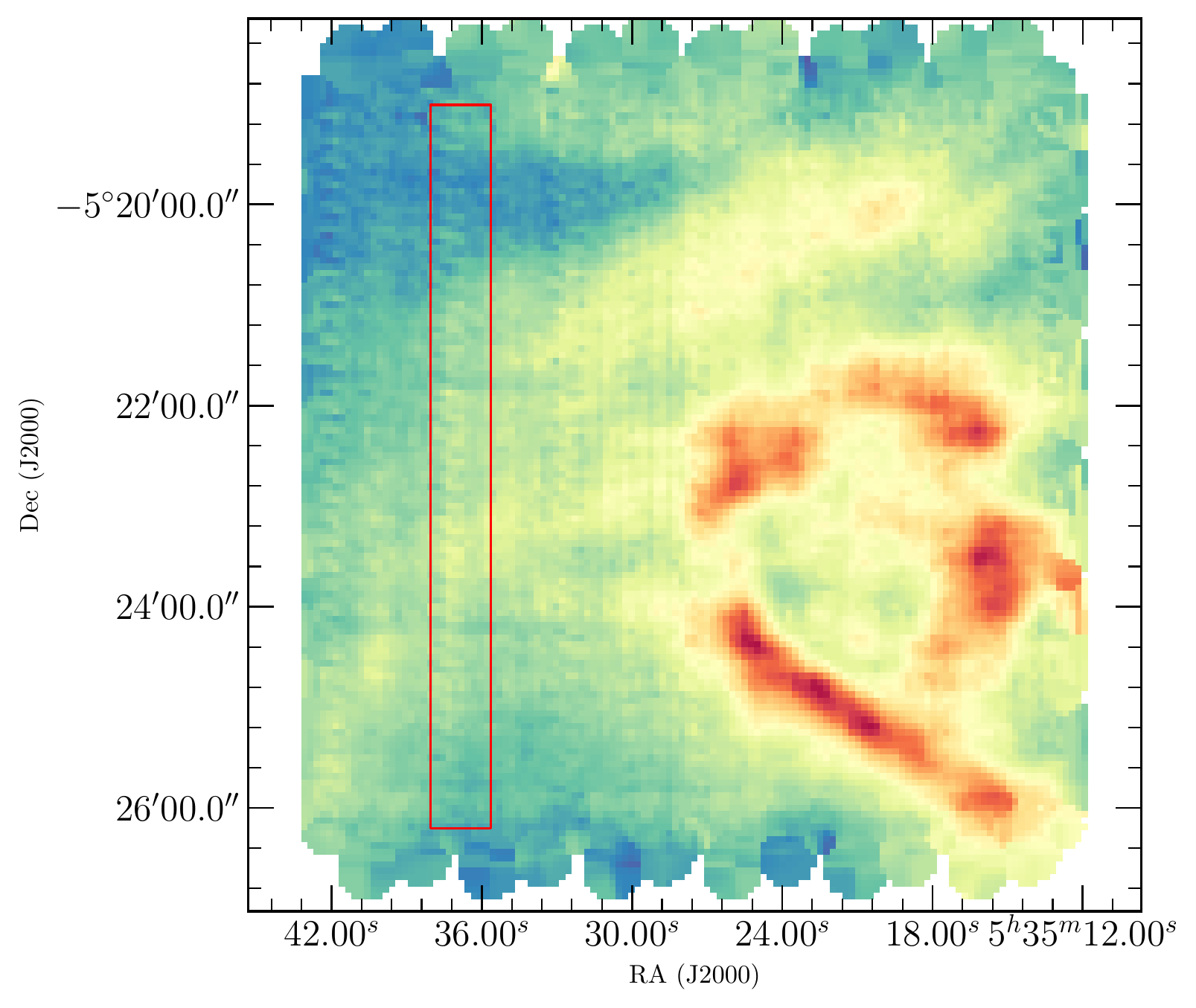}
\includegraphics[width=0.32\linewidth]{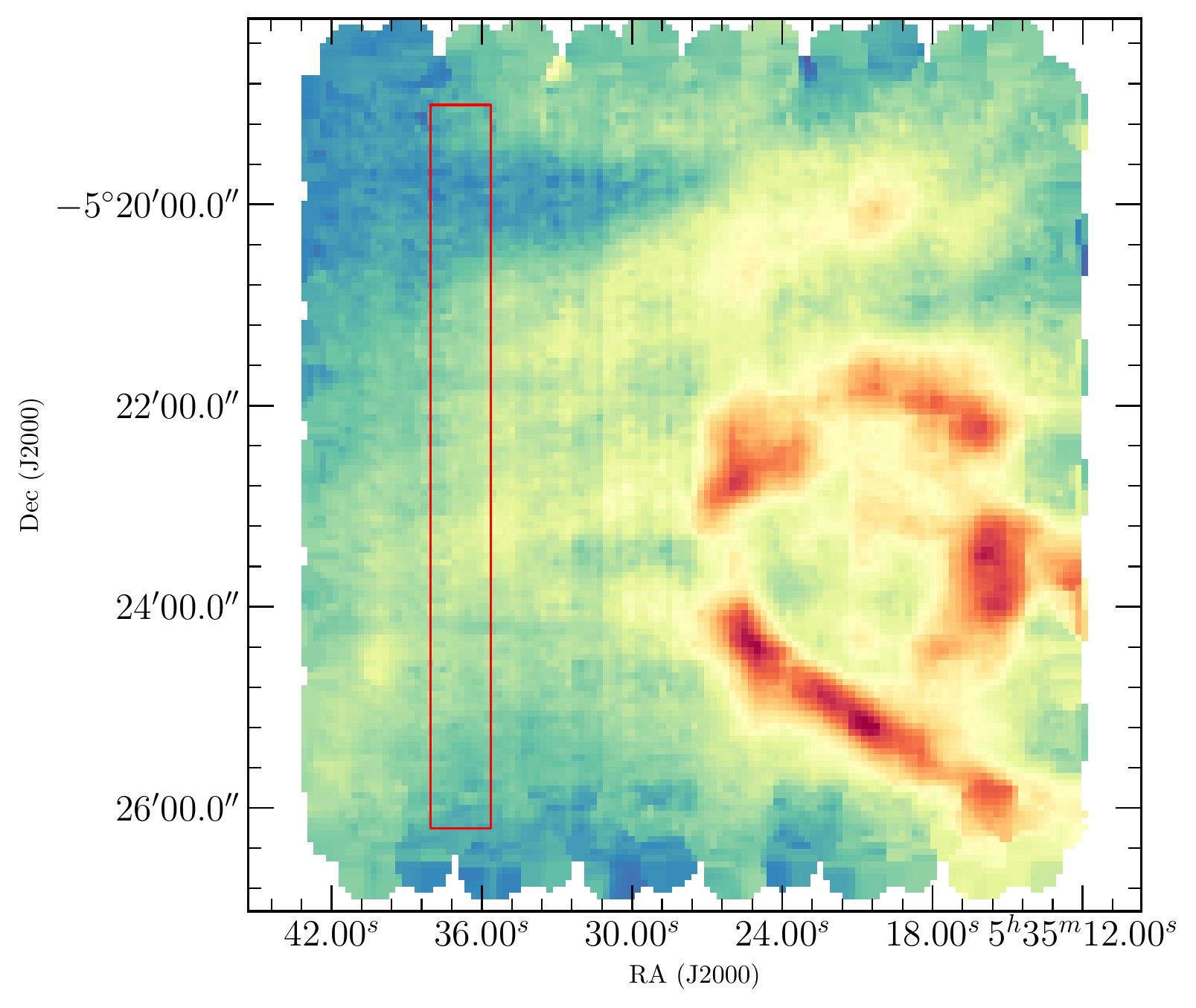}
\caption{\textit{Left panel:} Integrated intensity between -5 and 15 \kms\         for the center tile using a polynomial order 3 correction. \textit{Center panel: }Same tile, using a spline correction approach. \textit{Right panel:} Using a spline approach and filtering pixels with gain instabilities. The red box in the center and right panels highlights a region of gain instability that is detected as a zipper-like artifact in the map. This is only apparent after the spline baseline correction (for more details, see section \ref{sec:gain_instability}).}
\label{fig:spline_map_before_after}
\end{figure*}

The average difference between ON- and OFF-source scans in an ideal system should be zero in channels that do not contain an ON-source signal. The signals common to both ON- and OFF-source positions, such as thermal noise from the sky and receiver-intrinsic noise contributions, are identical. In practice, the unavoidable drifts of system components together with the longer or shorter time lag between the ON- and OFF-source integration leave residual baseline features in the calibrated spectrum.
\par 
The standard and commonly accepted procedure to correct for these instrument deficiencies is to fit 
a polynomial or sinusoidal model to the data outside of the line region. Ideally, the instrument effect is a slowly changing variation, and interpolation below the line region is sufficient to correct the spectrum. 
However, it cannot quantitatively determined whether this approach is correct. 
In particular, in sources where the line emission is a significant portion of the spectrum (Galactic center, extragalactic sources), this classic approach is limited and can potentially corrupt the underlying astronomical data.
\par
Baseline features come in different families. Careful selection of a model that can correct for the baseline feature best is crucial to correctly remove instrument effects. A common 
baseline effect is due to 
the standing-wave pattern 
that 
is associated with back-scattering from the secondary mirror. This causes a sinusoidal modulation of the 
continuum level in the spectrum and thus 
scales with the continuum strength of the observed source. 
The pattern shape itself is relatively constant and similar for all data of a map; in particular, it is weak for  the areas of the source that do not have a strong continuum. 
In the case of upGREAT, experience shows that the 
higher-frequency channels 
are less affected by these secondary standing 
waves; a scattering cone is installed in the central part of the SOFIA secondary mirror (covering the area corresponding to the part of the primary mirror beam that is obscured by the tertiary mirror tower and mirror). 
The dominant baseline structure affecting upGREAT data comes after the signal detection in the intermediate frequency amplification chain.
\par
The back-scattering in this case occurs between electrical components in the amplification chain, in this case, between the mixer and the first amplifier. The period of the associated pattern is a function of the coaxial cable length connecting the two components. Unlike the classic secondary mirror detector cavity, the reflection and transmission properties of the electrical component vary significantly over the IF bandwidth of the mixer, leading to a complex residual baseline structure  (see \cite{2009SPIE.7215E..0LH} for more details). 
With  the hot electron bolometer (HEB) used for upGREAT, the impedance and hence the mixer reflection properties are dependent on the pump level of the mixer. If there is a small change in receiver local oscillator (LO) power during the course of an observation, the mixer impedance state changes during the acquisition of the ON and OFF phase, which 
then leads 
to nonideal baseline structure residuals. The shape of this baseline structure is complex and varies from one upGREAT pixel to the next. Furthermore, some pixels are more sensitive to LO power fluctuations than others. 
To remove these artifacts from the calibrated data, 
a simple polynomial fit is not sufficient. This is especially a problem for sources in which the line emission is a significant portion of the total spectrum 
(extragalactic or Galactic center observations), and it is difficult to disentangle instrument residuals and astronomical emission. These baseline structures also affect the narrow-line emission (5-20 \kms \ wide) Orion data, in particular with regard to proper identification of the weak [$^{13}$CII] satellites (see section \ref{sect:13cii}). 
\par 
One characteristic of a standing wave in the electronics is that it is independent of LO frequency (different LO frequencies return the same baseline structure). The shape of the baseline structure is primarily a function of the impedance state of 
the 
reflecting elements and the cable length between them (see \cite{nuimeprn2584} for a detailed discussion). 
In the case of upGREAT, the two reflecting elements are the mixer and the first low-noise amplifier. This causes spectra that are observed minutes apart to have a similar baseline structure. This property can be used to generate a catalog of baseline structures from the residuals between OFF measurements, which can be used to correct the ON-OFF data, with the advantage that the residual OFF spectra are devoid of astronomical emission.
\par 
To facilitate the matching of baseline catalog spectra, each spectrum in the OFF residual catalogue is fitted with a spline profile, and the fit parameters are stored in a table. Each ON spectra is then compared to each spectra in the OFF residual spline catalog. Each spline is scaled to the ON data, and the best match is determined by the residual with the minimum chi-squared. This is similar to the approach developed by \cite{kester2014correction}, who used a Bayesian approach to determine the best fit. 
\par
Figure \ref{fig:spline_correction_process} shows an example of a individual baseline correction process. The data are first smoothed so that the underlying baseline structure is more apparent. This smoothed spectrum is then compared with scaled spectra regenerated from the spline catalog. The plot shows an example of an OFF residual spectrum fitted with a spline and multiplied by a factor of -1.23, which best matches the ON-calibrated spectra. This scaled OFF residual spectrum is then subtracted from the ON spectrum. Figure \ref{fig:spline_otf_scan} shows an example of the process applied to a single OTF scan (84 spectra). The waterfall plots on the right side best illustrate the nature of the baseline feature. The profile has troughs and rises at the same velocity. The only variation in the profile seen over the course of the OTF scan is the scaling. The left panels show the individual spectra for each OTF, and the average spectra before and after baseline correction are overplotted.
\par 
This process has no effect on the underlying intensity calibration. Figure \ref{fig:spline_H_V_comparions} shows the same data as in Figure \ref{fig:spline_otf_scan}, but with a comparison to the nearby LFAH2 pixel. The LFAH2 pixel is not affected by data-quality issues and provides a good comparison to the spline-corrected LFAV2 pixel. It should be noted, however, that the LFAV2 and LFAH2 pixels are not coaligned and are offset by 2 arcseconds on the sky. This causes some difference in line emission: In this case, there is a $\sim$ 5 kelvin peak difference. The difference is maintained before and after baseline correction, which gives us confidence that our correction method does not affect the data calibration. To underline this point, another example is shown in Figure \ref{fig:cmz_average_otf_scan} for a Galactic center observation. We show the average spectra before and after spline correction. The line emission is not discernible for the V0 pixel before correction, but it shows a consistent line intensity with the H0 pixel afterward.
\begin{figure*}[ht]
\includegraphics[width=1.0\linewidth]{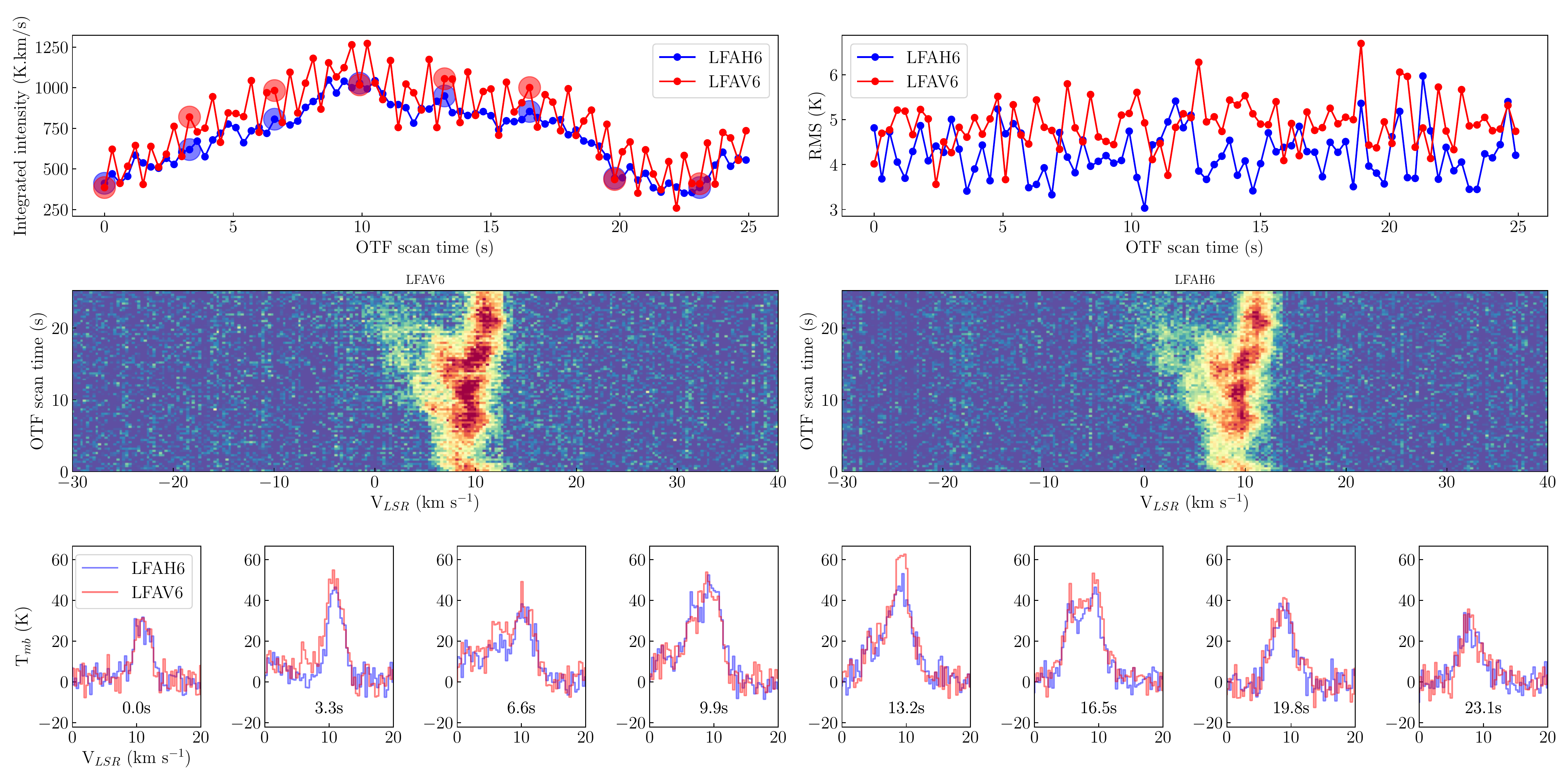}

\caption{Example of gain instability seen in pixel LFAH6 for a single OTF scan. \textit{Top left panel:} Integrated line intensity between 0 and 15 \kms \ vs. OTF dump time. \textit{Top right panel:} RMS over a velocity range from 30-40 \kms. \textit{Center panel:} Map of 84 spectra associated with a single OTF scan. LFAV6 is shown on the left, and LFAH6 is shown on the right. The noisy behavior between adjacent dumps is associated with the V pixel that is not seen in the H pixel. \textit{Bottom panel:} Eight example spectra associated with the solid circles shown in the top left panel.}.
\label{fig:gain_instability}
\end{figure*}
\par 
The generation of the OFF catalog does not require a special observation mode, and the OFFs taken as part of the regular observation process are sufficient. Ideally, the OFF catalog should be generated from OFF spectra nearby in time, up to 30 minutes. This depends on the atmospheric stability, but it is possible that atmospheric emission varies between OFF measurements. This can add another dimension of instability to the baseline that is not related to the receiver. For example, note the strong emission between -300 and -200 \kms \ in figure \ref{fig:spline_correction_process}, which is associated with atmospheric line emission (this region is therefore duly ignored for baseline fitting). These non-receiver baseline shapes are discarded during the fitting procedure because regions with variable atmospheric emission lead to large chi-squared residuals in that region of the spectra. Figure \ref{fig:cmz_sky_diff} shows an example of an OFF catalog for the LFAV0 and LFAH0 mixer from a Galactic
 center project. The variation in baseline shape for pixel V0 is strong, and pixel H0 shows relatively flat baselines. Figure \ref{fig:cmz_correction} shows the spline correction process for this dataset. The spline fit to the bold blue spectra shown in Figure \ref{fig:cmz_sky_diff} is scaled by a factor -1.56, which provides an accurate fit to the baseline shape and recovery of the weak-line emission.
\par 
Figure \ref{fig:tiles_rms_before_spline} shows the RMS distribution for each map tile of the large Orion map before and after spline correction for two different pixels. The RMS is taken over a 
spectral range 
of -75 to 80 \kms \ and the central line region 
is blanked out. 
Baseline problems are typically identified 
for a particular pixel and over a particular time period 
by a long 
non-Gaussian 
tail toward higher RMS values. After the spline correction, the RMS distribution approaches a 
distribution that is closer to 
Gaussian. Pixels LFAV2 and LFAH0 are shown for comparison. The baselines of pixel LFAV2 were particularly poor and 
significantly 
benefited from the spline correction. Pixel LFAH0 pixel 
performed better. The RMS distribution before and after spline correction shows no discernible difference.
\par
Figure \ref{fig:spline_map_before_after} shows the central tiles of the map with spline 
baseline correction 
method and with a typical polynomial third-order correction. Only part of the 
map is affected by poor baseline line structure. This causes the striped structure in the map that is not corrected for spline.
The third map shows a second processing step that we discuss in the next section.
\par
The spline fitting and minimization was undertaken using the python/SciPy library\citep{2020SciPy-NMeth}. The SciPy library was used from within the GILDAS environment using the pyclass library \citep{2013pyclass}. The code to fit a spline to data and generate a catalog of baseline shapes is available at \href{https://github.com/KOSMAsubmm/kosma_gildas_dlc}{https://github.com/KOSMAsubmm/kosma\_gildas\_dlc}. The Galactic center data shown in figure \ref{fig:cmz_average_otf_scan} are available as a demonstration dataset. The repository also contains functions integrating the pandas table \citep{mckinney-proc-scipy-2010} and matplotlib plotting \citep{Hunter:2007} libraries into the GILDAS ecosystem via the \textit{pyclass} interface. 

\subsection{Gain instability}
\label{sec:gain_instability}

When the baseline issues are resolved, the second-order effects become apparent. Figure \ref{fig:spline_map_before_after} shows an example of one such effect highlighted with a red box. It looks like a zipper in the map. The effect is not visible in a single spectrum and only becomes apparent in the integrated intensity over an OTF scan. Figure \ref{fig:gain_instability} shows the corresponding OTF scan in the red box of figure \ref{fig:spline_map_before_after}. This region of the map was covered by a Y scan with pixels H6 and V6. The integrated intensity for the H pixel shows a smooth profile as the telescope moves through a 435 arcsecond OTF scan. The V pixel shows a similar pattern, but with a larger scatter than for the H pixel. The source of this instability can be traced back to a vibration caused by the cryogenic cooler. The cold head, which keeps the mixers at their 4 Kelvin operating temperature, oscillates at a frequency of 1.3 Hz. This vibration modulates the LO power pumping the mixer, which causes the mixer pump state to oscillate with the period of the cold head. This behavior is particularly noticeable in the V array. The baseline features discussed in the section \ref{sec:spline_correction} are one manifestation of this vibration. Another more subtle effect is the modulation of the mixer gain. This is shown in figure \ref{fig:gain_instability}. The integrated line intensity is seen to oscillate during the OTF scan. This effect is most prominent in pixel V6 and to a lesser degree in pixel V2. A similar variation is expected in 
all pixels associated with a given LO, but this is not the case. Further investigations are needed to understand the origins of this effect.
\par 
Unlike the baseline effects seen in the previous section, the gain variations are difficult to detect and mitigate. For this dataset we developed a filter to detect this gain instability. Gain instability is not seen for all flights or pixels. The filter looks at the variability in integrated intensity for a given OTF scan. By running a rolling standard deviation over bins of 5 OTF dumps, a clear distinction between the stable and unstable regime is detected. Each OTF scan is processed using 
these 
criteria, and if an OTF scan  exceeds a given threshold, the data are flagged and are not included in the map reduction. Figure \ref{fig:gain_instability_detection} shows a rolling standard deviation of the integrated intensity time line shown in figure \ref{fig:gain_instability}. The V6 pixel shows a higher variability than the H6 pixel. 
\par
Gain instability might be exptected to be detected  in the spectrum noise, but this is not the case. The top right panel of figure \ref{fig:gain_instability} shows the RMS over a 30-40 \kms 
window. A gain variation might be expected to affect the mixer sensitivity and hence the measured RMS of the spectra. There is no apparent correlation between the variations in integrated intensity and spectrum RMS.

\subsection{Radio interference during flight}
\label{sec:rfi}

\begin{figure}
\centering
\includegraphics[width=1.0\linewidth]{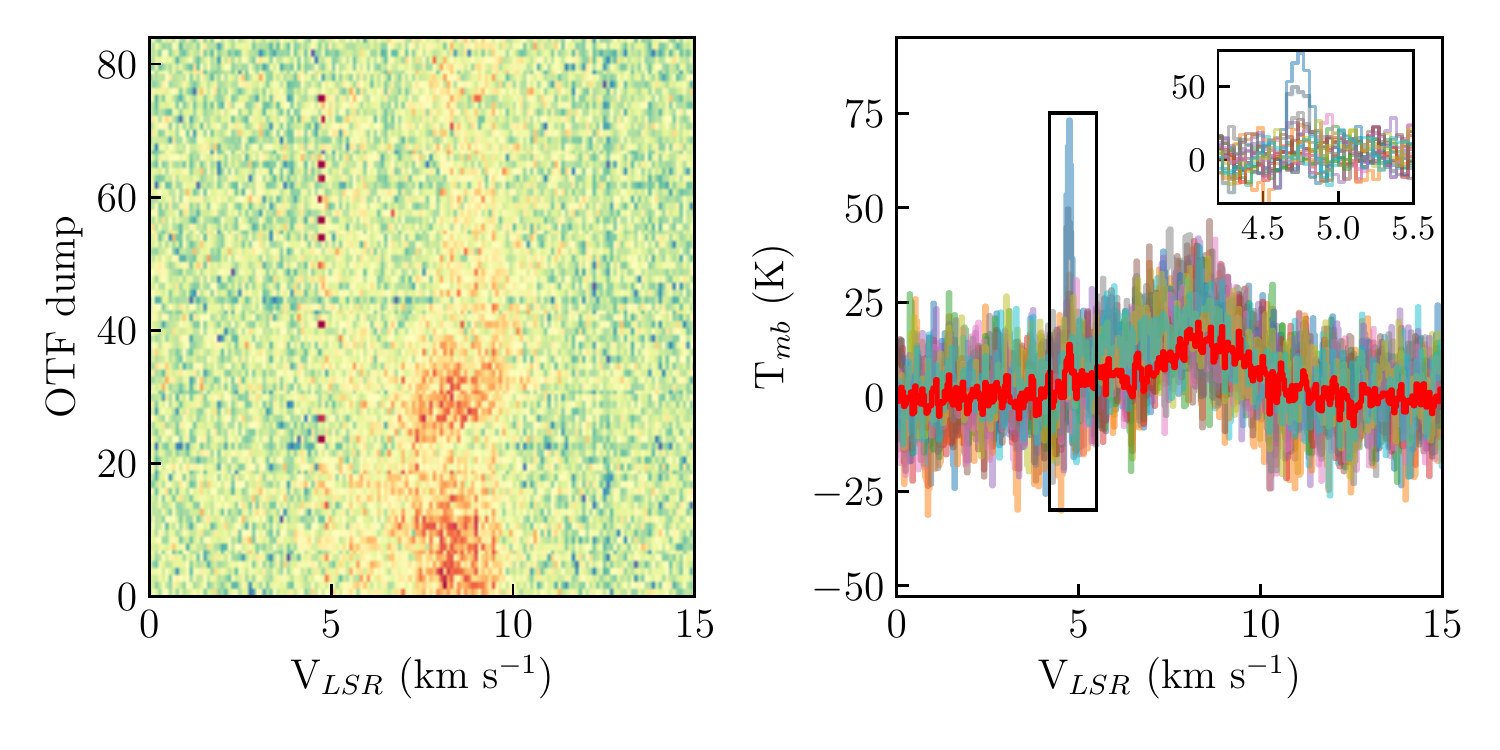}
\caption{Example interference from a cellphone during the flight of February 9th 2017. 
\textit{Left panel:} Example of 84 dump OTF scans showing RFI (red dots) at 4.75 \kms. Line emission is seen centered around 9 \kms. \textit{Right panel:} Example spectra showing RFI effects at 4.75 \kms. \textit{Right inset:} Zoom on the RFI emission. The emission extends over several channels and has a maximum peak of 75 Kelvin.}
\label{fig:rfi_example_in_spectra}
\end{figure}

Radio interference is parasitic man-made radio 
emission from terrestrial sources. It can corrupt astronomical observations \citep{2001A&A...378..327F}. RFI is a major problem for low-frequency ($<$ 5 GHz) observatories such as LOFAR \citep{2013A&A...549A..11O}, the SKA \citep{2004ExA....17..261E}, and the Effelsberg 100m telescope \citep{2010rfim.workE..42F}. It is also becoming an issue for higher frequencies with the onset of radars in self-driving cars (up to 300GHz, see \cite{kohler2013feasibility}) and new satellite internet systems (Starlink from SpaceX, see \cite {McDowell_2020}). SOFIA is also prone to a plethora of aviation RFI sources such as distance-measuring equipment (DMEs), instrument landing systems (ILS), and secondary surveillance radar (SSR) observed between 900 to 1200 MHz. Ideally, this should not be an issue for a high-frequency radio receiver ($>$ 1THz) such as upGREAT, but because it down-converts into frequencies between 0 and 4GHz, it is also susceptible to RFI effects. 
\\
Figure \ref{fig:rfi_example_in_spectra} shows an example of RFI effects seen in pixel H3. An intermittent spike, multiple channels wide, is observed over the course of an 84 dump OTF scan. In the final 
map, the RFI 
is then detected as a scratch-like artifact 
at a given velocity channel. RFI effects are fortunately seen for only 2 of the 14 upGREAT pixels, namely pixels H3 and H6. 
Although the cause is not understood, we speculate that this may be due to a leaky IF connector in the IF chain of these pixels. 
The RFI shown in figure \ref{fig:rfi_example_in_spectra} corresponds to a frequency of 1.9 GHz, which corresponds to the frequency of a cellphone. The example shown was observed during flight number 372 (on 9 February 2017). 
\par
Fortunately, this signal was only detected on this flight and therefore affected only 10\% of the total data. Other RFI signals are also detected, such as bluetooth at 2.4 GHz and a number of aircraft-related signals, but they do not fall close to the downconverted [CII] (or [$^{13}$CII]) line emission. 
As mentioned in the overview section, RFI can corrupt baseline fitting methods and can also contaminate other data  when the corrupted channels are convolved with healthy channels from pixels. One option would be to ignore the affected spectrum completely, but this discards data that are useful, except for 
the ten affected channels. 
Another option would be to replace the affected channels with an equivalent noise, but because the RFI effects shown in figure \ref{fig:rfi_example_in_spectra} are within the line emission velocity range, this could contaminate the line emission in a more subtle way. Fortunately, the GILDAS software team have implemented a method for weighting the significance of a given spectral channel in the map-making process using an associated array. 
These methods enable flagging an RFI affected channel and then weighting these channels down to zero when the final map is generated. With this approach, it is possible to ignore the RFI affected channels but retain the other healthy channels from the affected spectra. The crucial component in this process is the associated array support in GILDAS. This allows the user to store an additional array in parallel to the intensity array. For this reduction, a dedicated 
``RFI associated array'' 
was used to track outlier channels in a spectrum. RFI-affected channels were flagged using a rolling standard deviation and a threshold generated from the local RMS. The code to perform this operation is available in the \emph{despike} method in the same github repository as the spline correction code. 
\par 
Great efforts are made by the SOFIA observatory (preflight briefing info, onboard request to switch off RFI sources) to mitigate RFI, but given the prevalence of radio-emitting devices today, it is difficult to police completely. To mitigate these effects, an RFI detector will be deployed on future flights to detect bluetooth, WiFi, and cellphone signals. This system will alert observatory staff who can locate the RFI source.

\subsection{OFF correction}
\label{sec:off_correction}

\begin{figure}[!bh]
\centering
\includegraphics[width=1.0\linewidth]{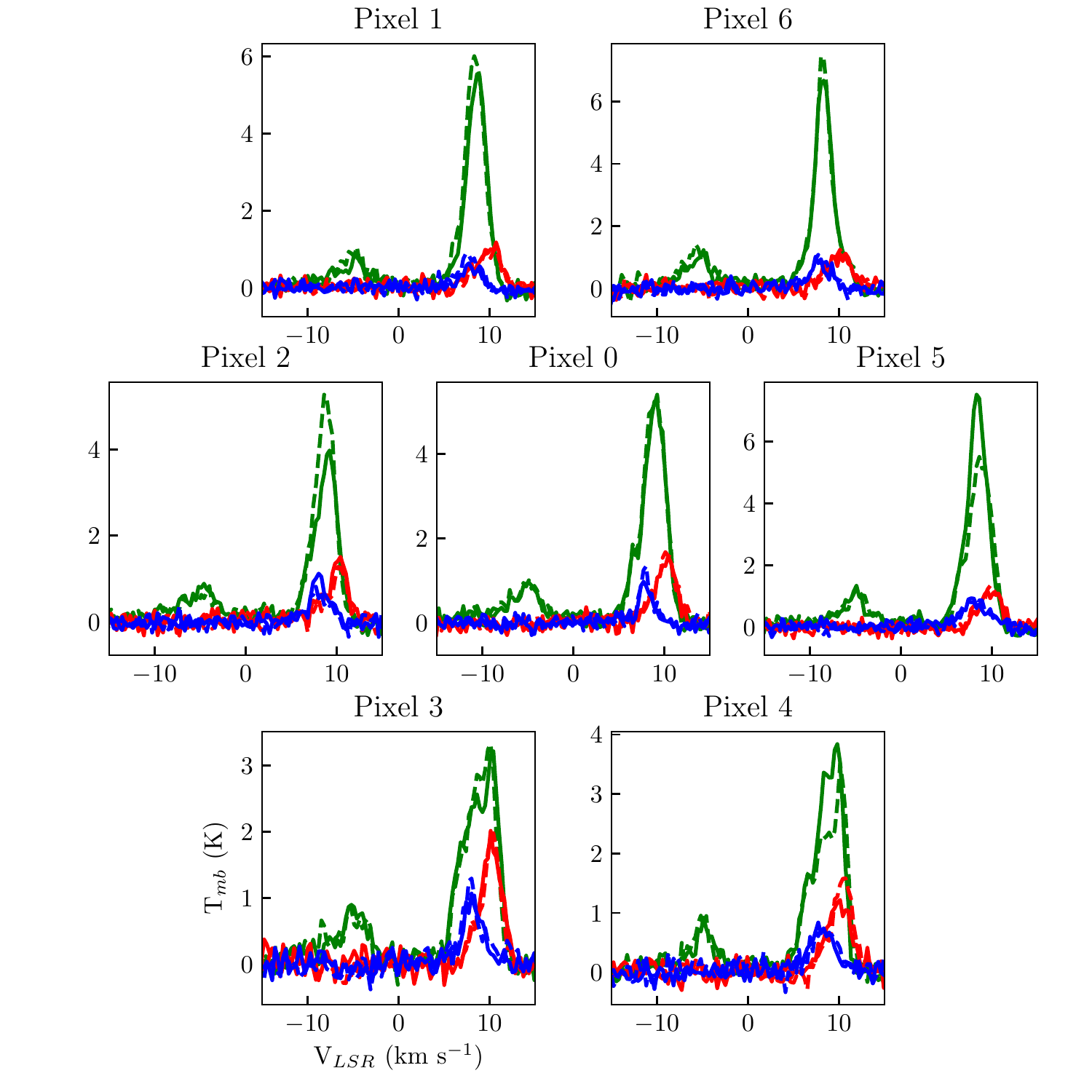}
\caption{Overview of emission from the three OFF positions for H-array pixels. Each position is observed at the X and Y  scan direction orientation (Y is shown as the dashed line in the plot). The colors correspond to the color shown in the tile overview in figure \ref{fig:tile_overview} (blue shows COFF-SE1, green shows COFF-C, and red shows COFF-OFF1).}
\label{fig:off_emission}

\end{figure}

\begin{figure*}
\centering
   \includegraphics[width=1.0\linewidth]{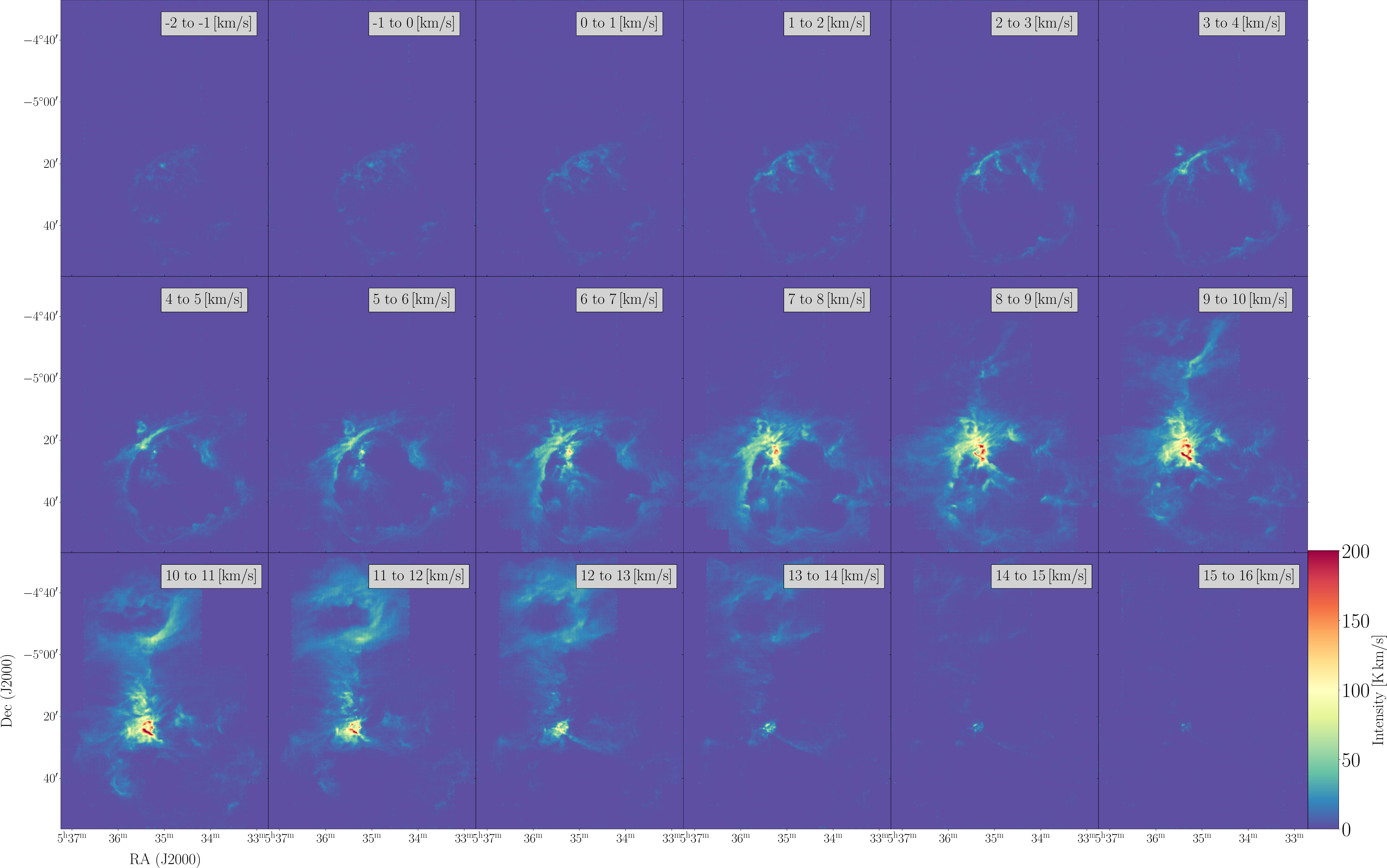}
     \caption{Final velocity cube generated with a Gaussian kernel of 18 arcseconds projected onto a pixel grid of 3.5 arcseconds square.}
     \label{fig:final_map}
\end{figure*}

\begin{figure*}[!h]
\centering
\includegraphics[width=0.45\linewidth]{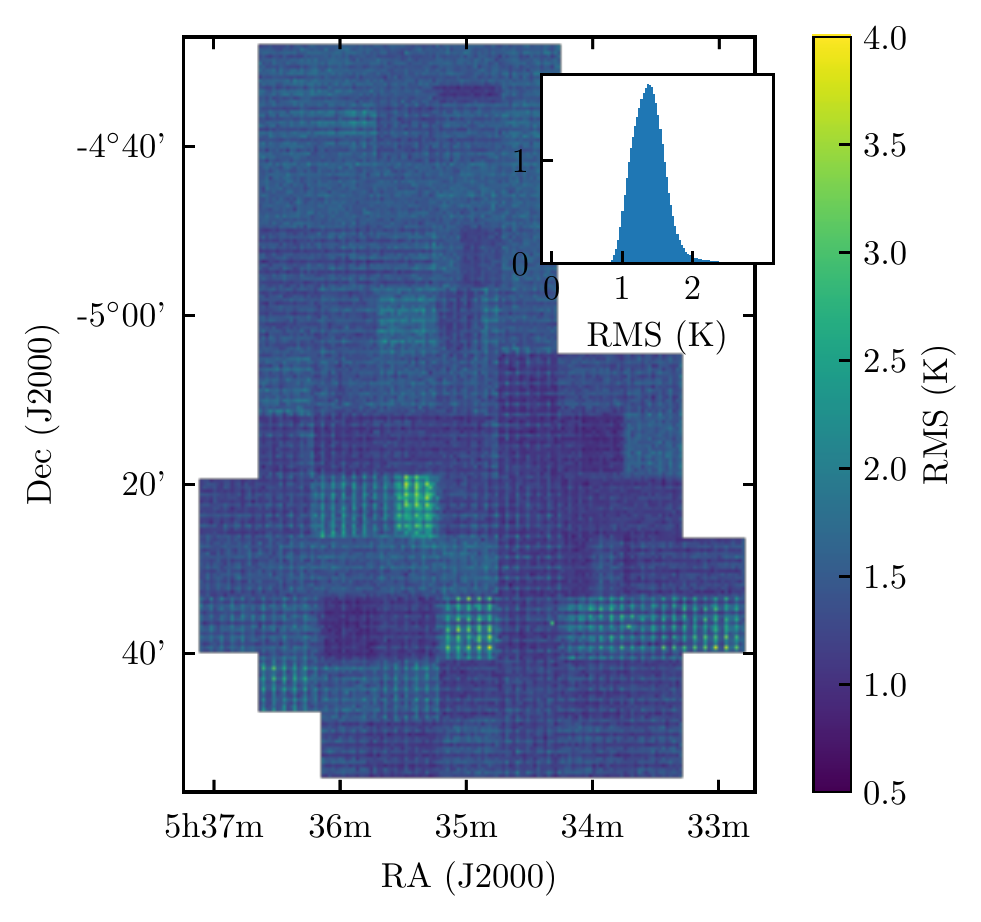}
\includegraphics[width=0.45\linewidth]{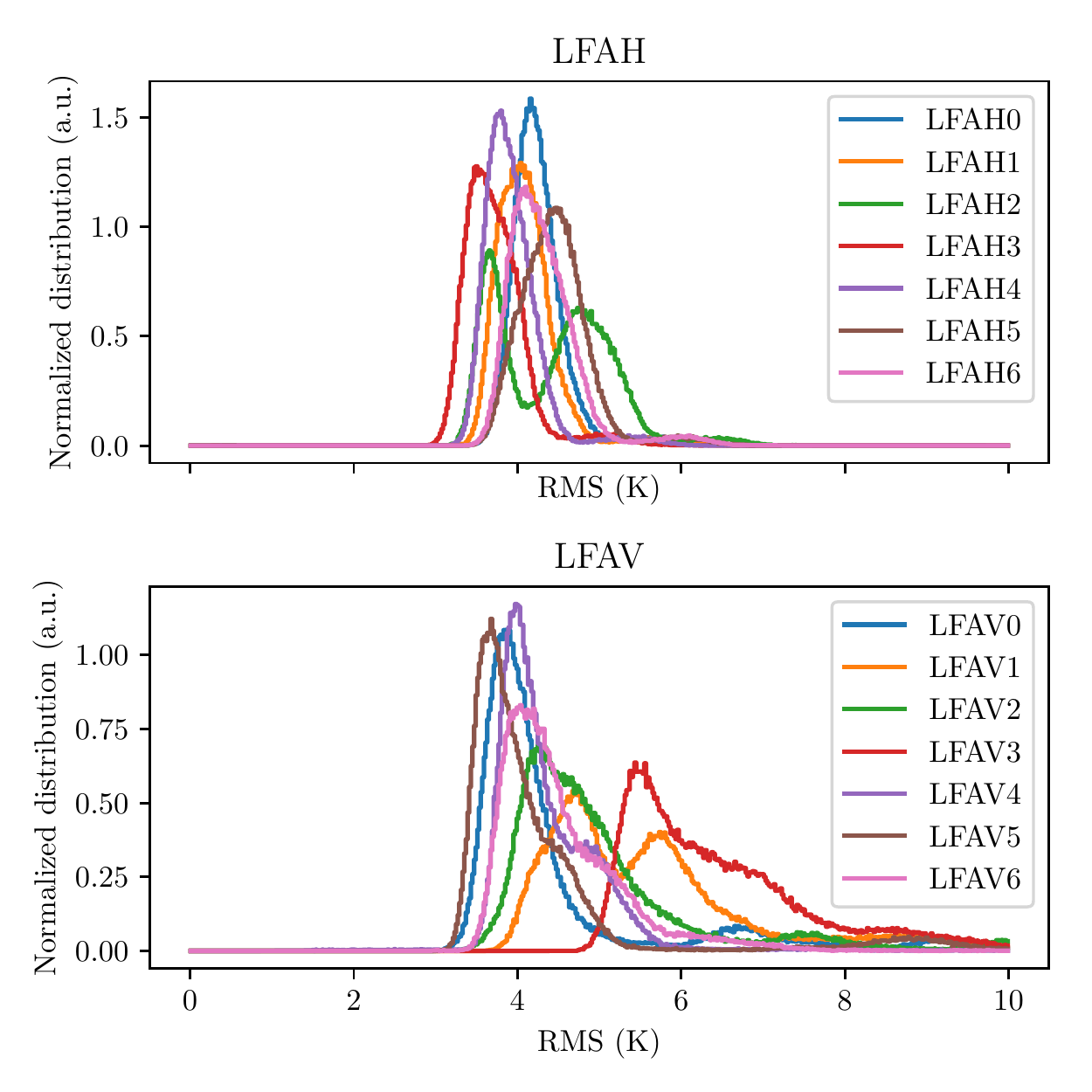}
\caption{\textit{Left panel:}Distribution of RMS over final map. RMS taken over a line free range from -50 to -20 \kms \ with a 0.3 \kms \ spectral resolution. \textit{Left inset panel:} RMS map shows a histogram of the distribution of RMS over the entire map with a peak value of 1.63K. \textit{Right panel:} Distribution of spectrum RMS before map generation with a 0.3 \kms spectral resolution. Some pixels have two peaks (V1 and H2) or a trailing tail (V3) in their distribution, indicating a change in RMS performance over the two flight series.
}
\label{fig:rms_before_after_deconvolution}
\end{figure*}

As mentioned in section \ref{sect:obs_strat}, one of the concerns in undertaking such a large map of a large-scale extended source, requiring far off OFF positions, was the effect of slew time to the OFF position on overall data quality. A further issue was the change in atmospheric transmission between observation phases when the angular distance between the ON and OFF phase is large. It was decided to minimize the effect of slew times (and therefore distances) by using three OFF positions around and inside the map region. 
The drawback to this approach is that all chosen OFF positions are contaminated with [CII] emission. Dedicated observations were undertaken of the OFF position coupled to a known far-OFF position free of emission (see table \ref{tab:position_summary} for a summary of the positions used). The resulting spectra of the OFF emission are shown in figure \ref{fig:off_emission}. At each pixel/position combination, two spectra 
correspond to the OFF positions of each pixel resulting from the array angle, first for the X (solid line) and then for the Y (dashed line) scan direction. 
Rotating the array on sky takes a few seconds of time. The array angle is kept constant between the OTF scan and OFF position acquisition to improve observing efficiency. For some positions, there are significant differences between the OFF spectra taken at the array angles for the X and Y scan (see the position of pixel 5, COFF-C, for an example). 
\par
OFF contamination is a common effect when extended sources are mapped. This manifests itself as 
an apparent 
absorption 
dip 
in a spectrum and over the entire map. 
The standard 
data reduction procedure is to add the contaminated emission back to the On sky emission. This procedure is 
significantly more complex 
with an array receiver because the OFF contamination is unique to each pixel. This is further complicated as the X and Y scans also has a unique OFF emission. For the Orion [CII] map, there are 84 unique OFF spectra for 14 pixels at three OFF positions and in two scan directions. Figure \ref{fig:off_emission_correction} shows an example of the average spectrum over a tile before and after OFF emission correction for each pixel.



\subsection{Summary}
Channel maps of the final data product 
are 
shown in figure \ref{fig:final_map}. The final map is generated using the \textit{xy$\_$map} function in GILDAS convolved with a kernel of 18 arcseconds and a pixel size of 3.5 arcseconds. Figure \ref{fig:rms_before_after_deconvolution} shows an overview of the spectrum RMS before and after the map generation. The RMS distribution of the individual 0.3-second OTF dumps is shown on the right side of figure \ref{fig:rms_before_after_deconvolution}. The performance  of the H and V arrays is different. The broadness of the distribution for the V array is related to a changing performance of pixels in later flights in the series. This is best illustrated in figure \ref{fig:trec_summary}, which shows an increasing system temperature toward later flights. Figure \ref{fig:tiles_rms_before_spline} shows the variation in RMS for the different tiles. The RMS increases for pixel V2 toward the northern part of the map.
\par 
The final map RMS is shown in the left panel of figure \ref{fig:rms_before_after_deconvolution}. The first feature of the map is the stripe feature that moves through the map. This indicates some of the nonuniform performance of the pixel over the course of the flight series, which is also seen in the receiver temperature summary plot (figure \ref{fig:trec_summary}) and also in the RMS distribution before deconvolution. These variations in performance are difficult to avoid and are still a topic of investigation within the upGREAT team. This can be a combination of changing LO performance and illumination of the array because the array response does not degrade uniformly over the flight series. Stripes also show regions in which spectra were dropped due to other performance issues, such as RFI and the gain variations discussed in previous sections.

\section{Data integrity}

\subsection{Consistency between flights}
\label{sec:consistency}
\begin{figure}
\centering
\includegraphics[width=1.0\linewidth]{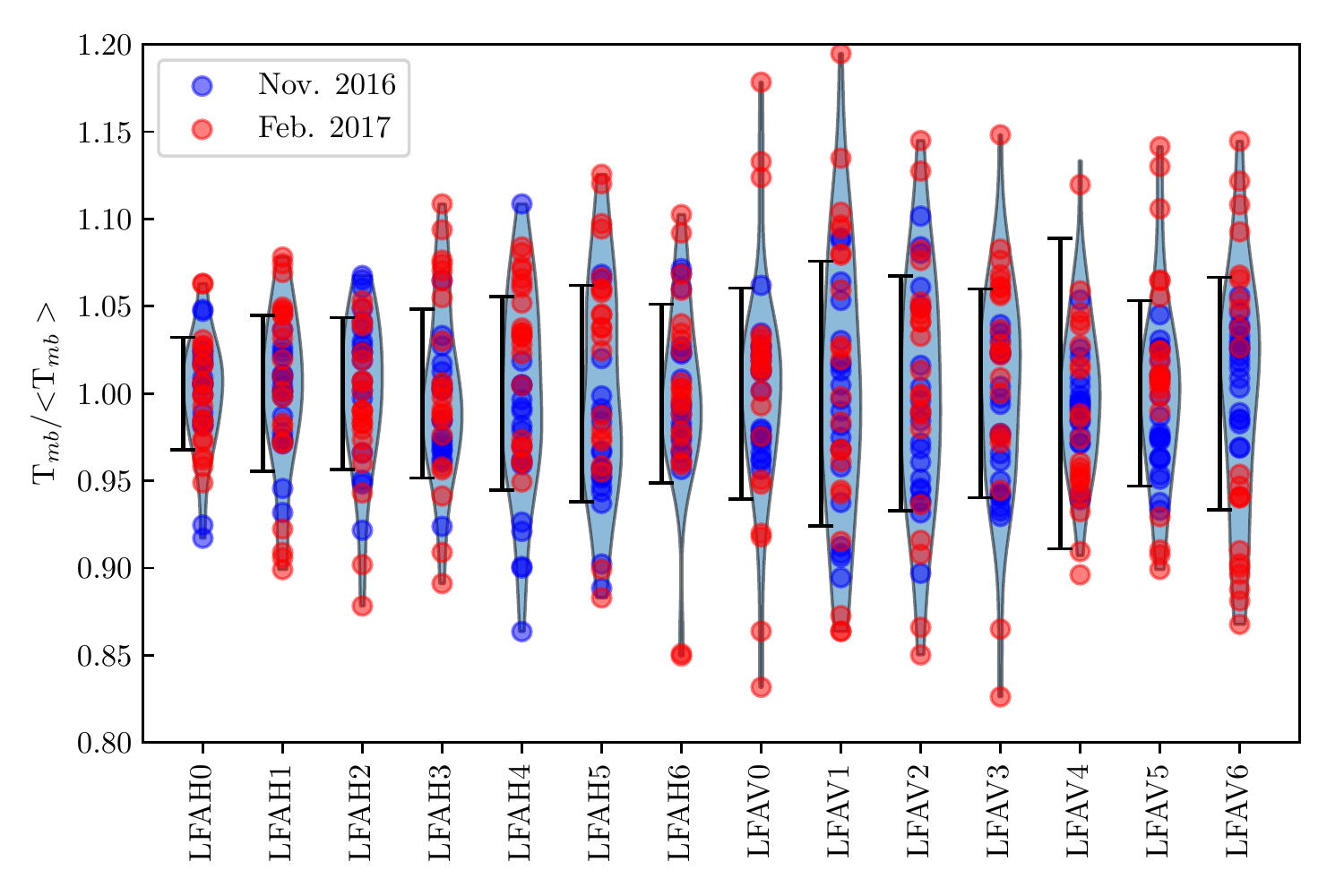}
\caption{Mean normalized integrated intensity of Orion bar reference observations per pixel. Red dots represent individual observations, and black bars show the standard deviation. The blue shaded region is a violin plot of the Kernel density estimation.}
\label{fig:data_integrity_violin}
\end{figure}

\begin{figure}
\centering
\includegraphics[width=1.0\linewidth]{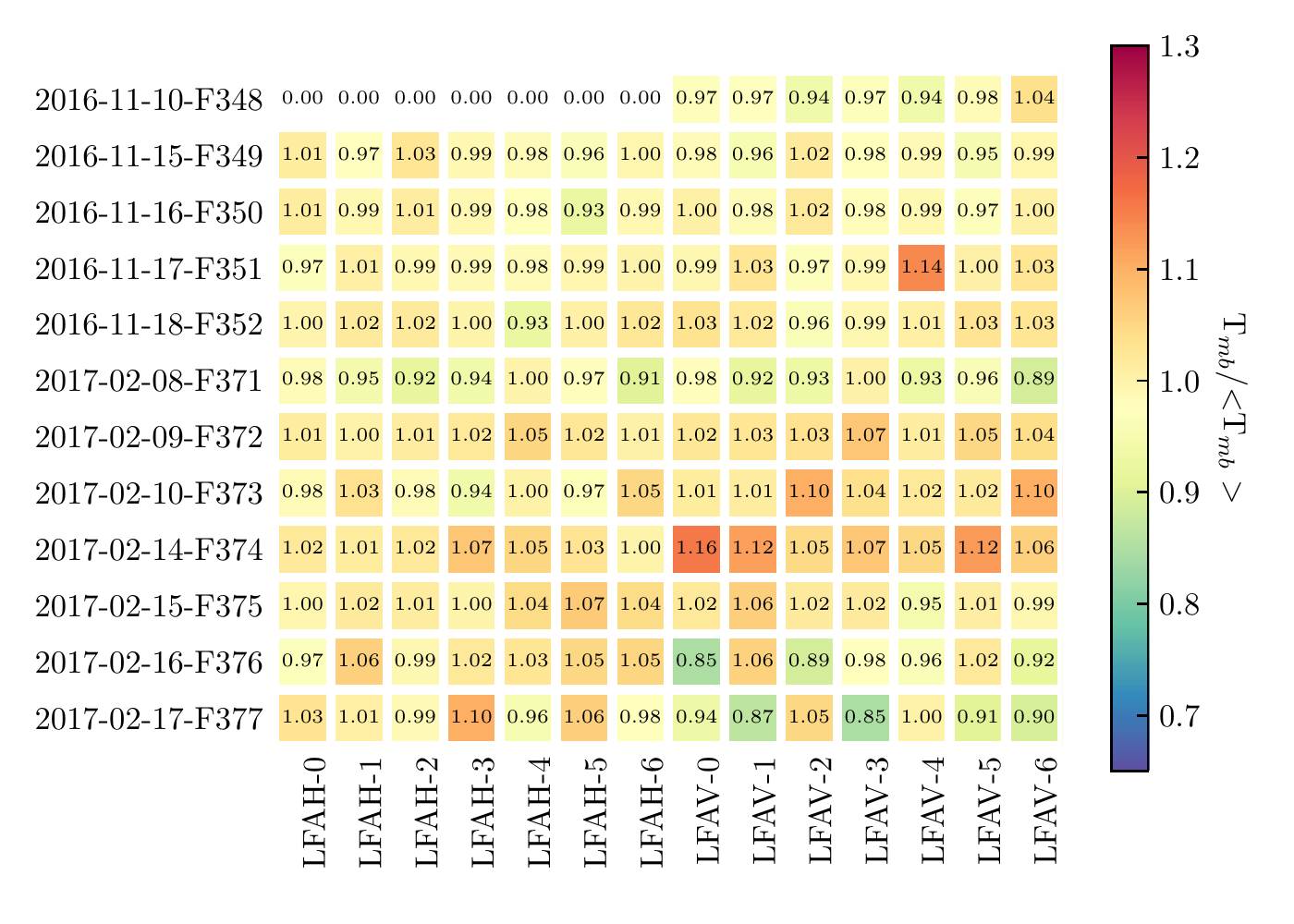}
\caption{Heatmap of the mean normalized integrated intensity of Orion bar observations over all flights and pixels.}
\label{fig:data_integrity_heatmap}
\end{figure}

The large-scale OMC [CII] map 
was the first time that a project of this scale was undertaken with the upGREAT instrument. 
The lessons learned from the careful data reduction as described here are of particular importance as they pave the path for the currently ongoing SOFIA [CII] mapping legacy programs \citep{Schneider2020,Pineda_2020}. 
The observations were spread over 13 flights and two flight series (November 2016 and February 2017), providing a unique opportunity to test the repeatability of upGREAT observations. A number of factors might affect the repeatability of upGREAT observations. They include the atmospheric calibration, pointing repeatability, and receiver stability. Furthermore, additional factors affect the repeatability of observations between flight series, such as the repeatability of the receiver mounting on the telescope flange.  The main concern from the scientific goals of the project was the repeatability of observations from one flight to the next. 
The concern was that the overlap between tiles could be an issue based on previous experiences from HIFI \citep{2015ApJ...812...75G} and IRAM 30m 
\citep{2014ApJ...795...13B}. This 
would require additional data reduction to correct the instrument response from one tile to the next. 
\par
This 
necessity and importance 
of monitoring 
the 
upGREAT performance was identified in advance, and a reference observation was scheduled for each flight. A region in the Orion bar (array centered on 5h35m20.90  $-5^\circ25{}^\prime04.8{}^{\prime\prime}$) was chosen, and a single-point total power observation of 3 minutes total duration was observed at least once on each flight. Twenty observations in total were observed. The position is shown in Figure \ref{fig:hifi_upgreat_noise_comparison}. Figure \ref{fig:calibration_orion_bar_spectra} shows an overview of each of these reference observations for each pixel. Figures \ref{fig:data_integrity_violin} and \ref{fig:data_integrity_heatmap} provide a summary of the normalized integrated line intensity over a -5 to 15 \kms \ window for all pixels in these reference observations. Normalization was achieved by dividing each spectrum by the average line emission over all reference observations for that pixel. Ideally, all points should be close to 1. 
\par
Taking the standard deviation of each pixel and averaging over all pixels returns a repeatability of +/- 6\% between flights. When we break this up further, the H array has a repeatability of close to 5\%, while the V array is closer to a 7\% repeatability. This is consistent with the other trends seen in the data quality: the H array was seen to perform better than the V array. The varying performance between the two arrays could be linked to different LO performance between the two arrays. We observe an increase in receiver temperature toward later flights; see figure \ref{fig:trec_summary}. These effects are not uniform across the array, however, which suggests that some pixels are more sensitive to LO performance than others. This could be due to the method used to split the LO signal between the seven mixers. The single-beam LO output signal is divided into seven beams through phase grating. Small changes in alignment of the phase grating and mixers can affect the illumination of each pixel differently. The degradation of receiver temperature over the course of a flight series points to an LO performance issue, however. The LFAV LO was therefore shipped back to the manufacturer for maintenance. After this maintenance, a more uniform performance of the two arrays was seen.


\subsection{Consistency of the H and V array}
\begin{figure}
\centering
\includegraphics[width=1.0\linewidth]{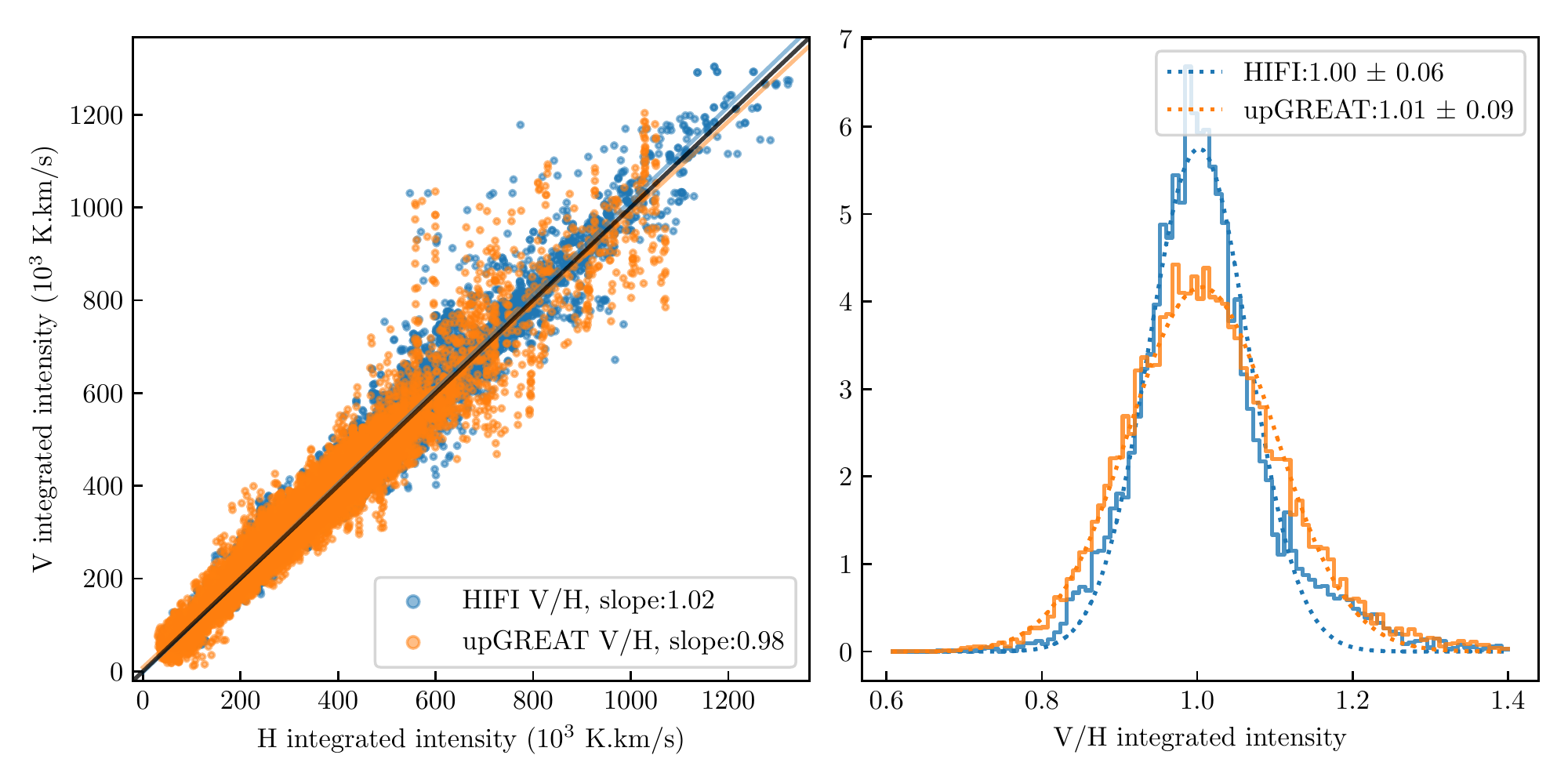}
\caption{Comparison of the line intensity between the H and V polarization over a velocity range of -5 to 15 \kms. Data are shown for upGREAT and HIFI. \textit{Left panel:} V polarization plotted against H polarization. The legend shows the slope of a linear fit through the origin. \textit{Right panel:} Histogram of V/H integrated intensity. The legend shows the fit offset and width of a Gaussian profile.}
\label{fig:h_v_consistency}
\end{figure}

An additional check of the instrument performance is the consistency of the H and V integrated intensity. Figure \ref{fig:h_v_consistency} 
shows the H-integrated intensity plotted against the V-integrated intensity for the central region of the map shown in figure \ref{fig:hifi_upgreat_side_by_side}. Plotting  V against H data provides an overview of the linearity between the two  polarizations. This linearity is maintained to the highest intensities observed in the map. The histogram of V divided by H polarization provides a  statistical measure of the consistency of the two polarizations. The upGREAT histogram is centered at 1.01. The linear fit to the V versus H plot shows a slope of 1.02 of the V and H intensity, indicating that the upGREAT V array is on average 2\% brighter than the H array. For consistency, the HIFI V and H performance (the comparison of HIFI and SOFIA/upGREAT data is discussed in detail in the next section) is also plotted. This shows a slightly better performance, but this is just a two-pixel comparison. The line fit to the HIFI H and V data shows a similar 2\% difference of the polarizations. The upGREAT data are a comparison between two 7-pixel arrays, and it is expected to have more scatter given the different main-beam efficiencies and associated uncertainties (see table \ref{tab:main_beam_effeciency}).

\subsection{Intensity comparison with a HIFI map}

\begin{figure}

\includegraphics[width=1.0\linewidth]{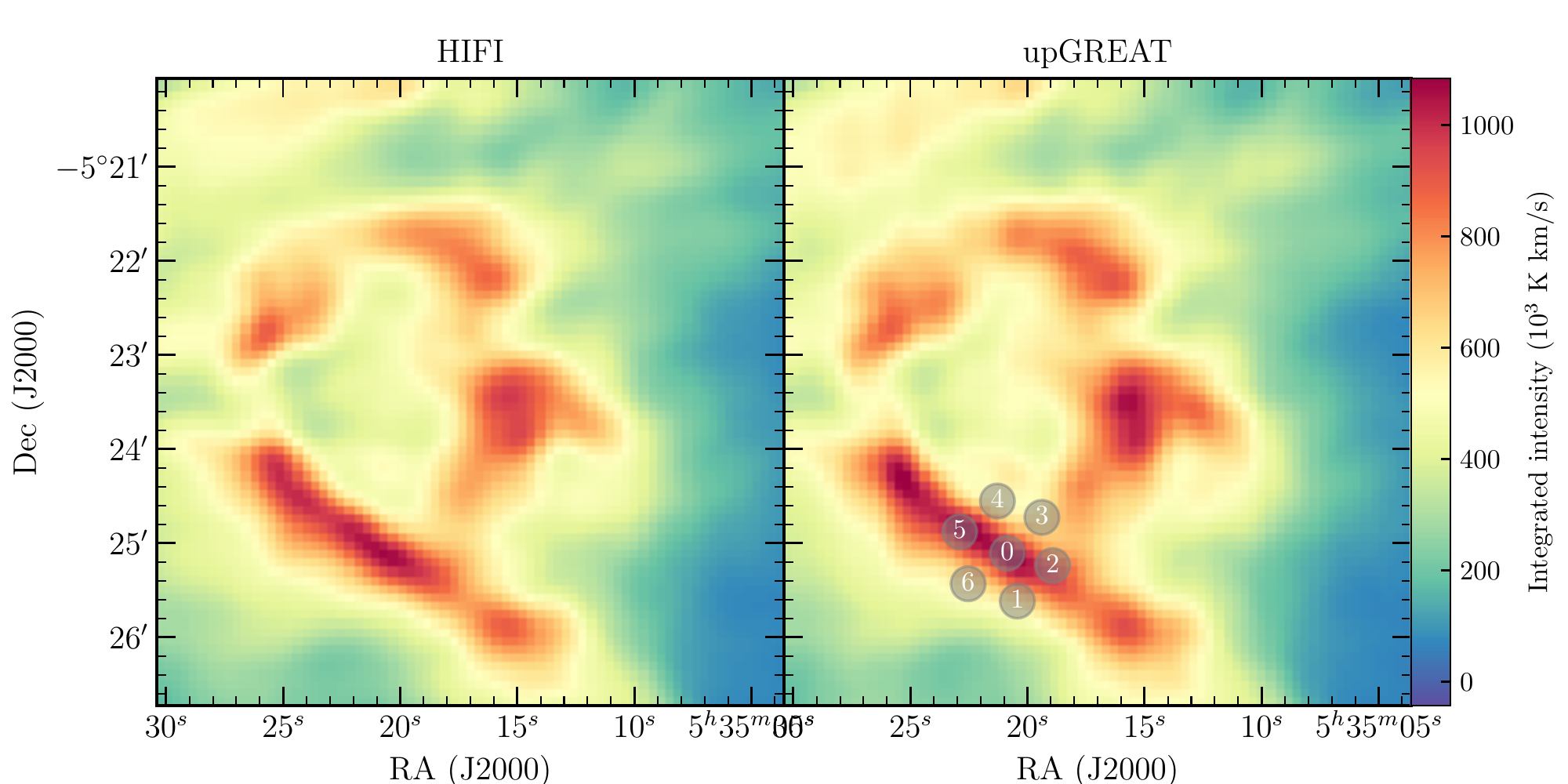}
\caption{Side-by-side plot of HIFI and upGREAT data. Both maps are generated with a 22 arcsecond mapping kernel, and both datasets are in main-beam antenna temperature. The integrated intensity between -5 and 15 \kms is shown. The array positions highlighted in gray show the location of the Orion bar consistency observation discussed in Section \ref{sec:consistency}.}
\label{fig:hifi_upgreat_side_by_side}
\end{figure}

\begin{figure}

\includegraphics[width=1.0\linewidth]{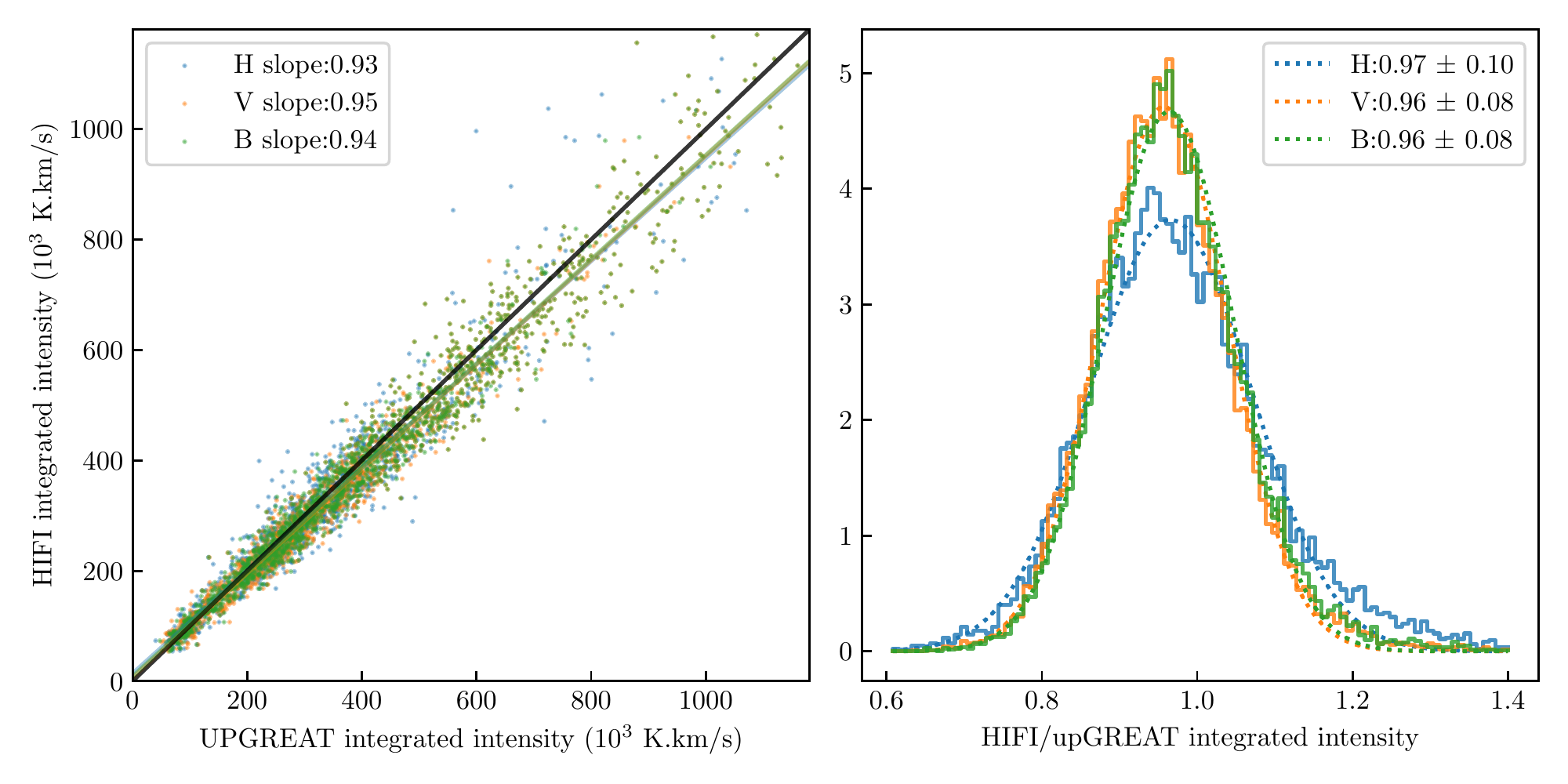}
\caption{\textit{Left panel:} Integrated intensity from HIFI and upGREAT maps plotted against each other. Three upGREAT maps are shown: separate maps for the H and V polarizations, and a map with both polarizations together. The HIFI map is generated from both polarizations. \textit{Right panel:} Distribution of the HIFI map divided by upGREAT maps with a Gaussian fit. The plot legend shows the fit Gaussian width and offset.}
\label{fig:hifi_upgreat_overlay}
\end{figure}


Another factor that affects the repeatability of observations between flights is the atmosphere. The 
atmospheric transmission calibration process 
makes a number of assumptions, for instance, on 
the layering of the atmosphere above the 
aircraft, on the stability of the atmospheric transmission during the ON-OFF cycle, and on the validity of the atmospheric model over time. 
This can lead to an over- or underestimation of the atmospheric transmission, leading to inconsistency between flights. This section addresses this issue by comparing the upGREAT [CII] Orion map with one taken by the Herschel/HIFI spectrometer \citep{2015ApJ...812...75G,2010A&A...518L...6D}. As a satellite mission, HIFI was free of atmospheric contribution and so provided a unique dataset to compare with the upGREAT map. 
\par 
Figure \ref{fig:hifi_upgreat_side_by_side} shows a side-by-side comparison of the HIFI and upGREAT maps. Both maps used a Gaussian kernel of 22.1 arcseconds on a grid of 6 arcseconds. The larger mapping kernel was used to account for the undersampled nature of the original HIFI map. Figure \ref{fig:hifi_upgreat_side_by_side_under_sampled} shows a similar comparison, but with a mapping kernel of 18.1 arcseconds. The blocky nature of the HIFI map arises because the OTF map was not Nyquist sampled and has a grouping of spectra every 10 arcseconds on average due to the load chop-observing mode. See Figure \ref{fig:HIFI_upGREAT_map_sampling} for a comparison of the spectra distribution between upGREAT and HIFI. From a visual inspection of figure \ref{fig:hifi_upgreat_side_by_side}, both maps show a good spatial correlation. 
\par
The bottom panels give a more quantitative view of the two datasets. The integrated intensity of each HIFI map pixel is plotted against the same upGREAT map pixel. For completeness, three upGREAT maps are generated, one with the H array and V array each, and then both arrays 
combined (labeled B). 
The resulting cloud of points is fit with a linear curve going through zero. In an ideal situation, when the two instruments measure the same intensity, the fit slope would be 1 (see the black line in the plot). 
In this case, the upGREAT intensity is 4\% brighter on average than the average HIFI map. The histogram in the right panel of figure \ref{fig:hifi_upgreat_overlay} shows the distribution of the HIFI intensities divided by the upGREAT distribution. This provides an additional view of the two datasets and shows that each HIFI and upGREAT map pixel is within a factor of 0.96 of each other on average. 
\par
HIFI had a calibration accuracy of 10$\%$ based on extensive preflight and in-orbit testing (see HIFI Handbook). UpGREAT states a calibration accuracy of $\pm$20\% \citep{2016A&A...595A..34R}. The good correlation of the two dataset provides confidence in the atmospheric calibration code and overall calibration of the upGREAT receiver. 
\subsection{Performance comparison with HIFI map} 

\begin{figure}

\includegraphics[width=1.0\linewidth]{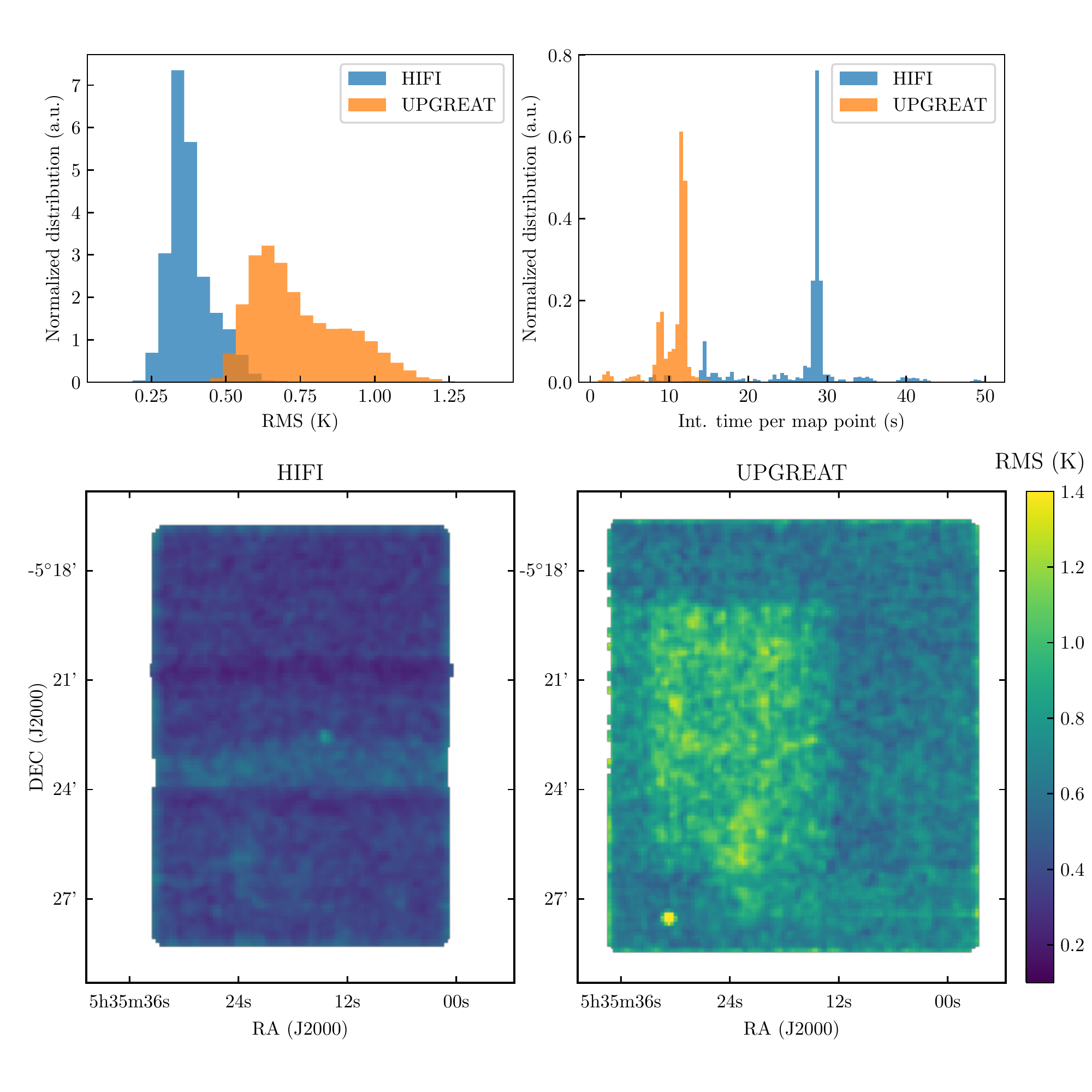}
\caption{RMS comparison of HIFI and upGREAT between -50 and -20 \kms. \textit{Top left panel:} Histogram of the RMS map shown in the bottom panels. \textit{Top right panel:} Histogram of the integration time per map point. \textit{Bottom panels:} Spatial distribution of the RMS for HIFI and upGREAT. The region of higher RMS in the upGREAT map is from the first flight when the LFAH LO was not available. Dark horizontal bands in the HIFI map show lower RMS regions that correspond to the overlap between the map tiles.}
\label{fig:hifi_upgreat_noise_comparison}
\end{figure}

Figure \ref{fig:hifi_upgreat_noise_comparison} shows a comparison of the instrument noise of the two instruments. The final map RMS from HIFI and upGREAT is shown in the lower panels. The HIFI map has a lower RMS than the upGREAT map because HIFI has a longer integration time per map point. The top right panel in figure \ref{fig:hifi_upgreat_noise_comparison} shows the distribution of integration time per map point. upGREAT has a better overall system temperature than HIFI, but the longer integration time for the HIFI map leads to a lower RMS. 
\par 
Figure \ref{fig:tsys_summary_hifi_upgreat} shows an overview of system temperatures from HIFI and upGREAT. upGREAT has an average single-sideband receiver temperature of 2600K versus 3600K for HIFI. Even with the additional noise from the atmosphere, the system temperature of upGREAT is a factor 1.4 lower than that of HIFI. From the radiometer equation, this difference means that an upGREAT pixel is twice as efficient as a HIFI pixel, or alternatively, that a single upGREAT pixel can achieve the RMS of a HIFI pixel in half the time. With the difference of a factor 7 in pixel count, 2 for HIFI compared to 14 for upGREAT, upGREAT can map a region to the same RMS 14 times quicker than HIFI would take, which is consistent with a similar analysis of the CII Horsehead map by \cite{2016A&A...595A..34R}.
\par
Another dimension in the comparison of HIFI and upGREAT is the observing efficiency. This is nominally defined as the fraction of on-source time for the total observation time. A direct comparison of HIFI and upGREAT Orion maps is not possible because the observing modes are significantly different. The observing efficiency of an airborne observatory is overall expected to be higher than that of a satellite because of the higher accelerations and slew speed that is possible on SOFIA. Herschel has a maximum slew speed of 0.1 degrees per second compared to 1 degree per second for SOFIA. However, other factors also play a role, such as the coordination of the instrument and telescope. A detailed discussion of the observing efficiency is beyond the scope of this paper. The efficiencies quoted in the Herschel Explanatory Supplement Volume I (section 4.3.2.2) of between 30 to 50\%  are largely consistent with SOFIA/upGREAT observations.
\par 
 
\par 
The large leap in progress from HIFI to upGREAT 
was only achievable 
using an airborne telescope, for which the instrument team can upgrade the receiver after each flight series and continuously advance the development. This rate of progress is not possible with satellite-based instruments: the receiver is not accessible 
in orbit, 
and the design is frozen early in the development process. During this project, upGREAT had 
an 
LO 
performance 
issue during the first flight that was quickly resolved for the next flight. Furthermore, the degrading performance of the LO during the last two flights of the project was resolved after the flight series, in time for the next upGREAT campaign. In addition to the hardware flexibility, the software system is also flexible and continuously updated to improve observing efficiencies and track data-quality issues (e.g. RFI effects on board).

\section{Scientific outlook}
The ISM and its interaction with massive stars are central to the evolution of star-forming galaxies. Mechanical and radiative energy input by massive stars stirs up the interstellar medium and heats interstellar gas. This controls the characteristics (density and temperature) of the phases (diffuse clouds, warm intercloud, and hot intercloud) of the interstellar medium, sets the thermal pressure of the gas, and the distribution of the gas over these phases \citep{1969ApJ...155L.149F,1995ApJ...443..152W,1979ARA&A..17..213M, 1977ApJ...218..377W}. Stellar feedback also governs star formation because ionized evaporative flows, stellar winds, and supernova explosions will disrupt molecular clouds, thereby stopping star formation (negative feedback), while at the same time, gas is compressed in shells, promoting gravitational instabilities (positive feedback) \citep{2011EAS....51...59E,1977ApJ...214..725E,2014MNRAS.445..581H}. The relative importance of these feedback processes and their dependence on the characteristics of nearby massive stars and the surrounding ISM is not well understood. 
\\
Radiative energy input by massive stars also controls the emission spectrum of star-forming galaxies. Extreme-UV (h$\nu> 13.6$ eV) photons ionize the gas and heat it to $\sim$ 7500K. This photoionized gas cools through optical and UV collisionally excited lines and through (H) recombination lines. These processes are well understood, and HII regions and their properties have been extensively studied on Galactic and extragalactic scales \citep{2006agna.book.....O}. H gas is transparent to far-UV photons ($6<h\nu <13.6$ eV), but trace species with low-ionization potentials can still be ionized. The neutral HI gas is heated through photoelectrons from polycyclic aromatic hydrocarbon molecules and small dust grains. These photons therefore travel largely unimpeded through HII regions, creating a layer of warm gas in so-called photodissociation regions (PDRs) that separate the ionized gas from the surrounding molecular cloud. On the scale of a galaxy, PDRs also create the cold neutral medium (diffuse clouds) \citep{1999RvMP...71..173H}. Of the relevant trace species, carbon is the most abundant and [CII] is the main ion in far-UV illuminated gas. This neutral HI gas cools through far-infrared fine-structure lines of [CII] at 1.9 THz, for example. As the dominant cooling lines can only be observed with airborne or space-based platforms, much less is known about the properties of PDRs.
\\
Recent observations reveal that much of the molecular gas mass is in a form that is not probed by CO (so-called CO-dark molecular gas), its standard tracer \citep{2005Sci...307.1292G,2011A&A...536A..19P}. For diffuse
clouds, hydrogen is atomic and carbon is ionized ([CII]). In CO-dark molecular gas, hydrogen is in H2, but [CII] rather than CO is the dominant form of carbon  \citep{2009A&A...503..323V,2010ApJ...716.1191W}. The HI and CO-dark
phases have so far eluded detailed characterization because the main tracer of the diffuse ISM, the HI 21 cm line, does not constrain the physical conditions of the gas (e.g., temperature, density) well, and optical
and UV absorption line studies, which are good probes of the physical conditions, are limited to pinhole experiments toward bright, nearby stars with only low column densities of gas. As a result, the physical conditions of CO-dark molecular gas and its relation to other phases in the ISM are not well known. This CO-dark molecular gas is heated by far-UV photons from massive stars through photoelectrons from large molecules and very small grains \citep{1994ApJ...427..822B} and cools through the [CII] 1.9 THz line \citep{2010ApJ...716.1191W}. 
\\
Observations of regions of massive star formation in the [CII] 1.9 THz line carry the promise of providing much insight into all of these aspects of the interaction of massive stars with their surroundings and the structure and characteristics of the ISM. We briefly summarize the anticipated results of the [CII] square-degree survey of Orion. Section \ref{mech_effects} discusses mechanical feedback aspects, including the Veil bubble created by the stellar wind of $\theta^1$ Ori C and the Spitzer-type expansion of M43 and NGC 1977. Section \ref{fuv_effects} highlights the radiative heating of PDR gas by far-UV photons in the Orion region. Section \ref{veil_co} discusses the relation of [CII] and CO emission in the Veil bubble. In section \ref{filaments}, we consider the interaction of radiative and mechanical energy input on the filamentary structure of molecular clouds and the effects of star formation. In section \ref{sect:13cii}, the synergy of [CII] 1.9 THz observations with radio recombination lines of carbon is illustrated. Finally, in section 5.6, we describe the effects of optical depth on the [CII] 1.9 THz emission through observations of the [$^{13}$CII] isotopolog.


\subsection{[CII] observations and mechanical feedback from massive stars}
\label{mech_effects}
The large [CII] dataset at high spectral resolution enables a systematic study of the large-scale kinematics of the region \citep{2019Natur.565..618P,2020A&A...639A...2P}. In the [CII] data several expanding structures can be identified. Three of them are parsec-scale bubbles affected by stellar feedback by the respective central stars. Specifically, the data reveal that the shell structure surrounding the hot gas within the extended Orion nebula, which dominates the morphology of photometric images, is indeed part of a coherently and rapidly expanding shell, whose dynamics is controlled by the stellar wind of the most massive Trapezium star, $\theta^1$ Ori C. The results show that the mechanical energy from the stellar wind is converted very efficiently into kinetic energy of the swept-up shell and, rather than photoionization and evaporation, dominates the disruption of the Orion molecular core 1 \citep{2019Natur.565..618P}. The HII regions, M43 and NGC 1977, are also associated with expanding shells. These HII regions are powered by the less massive stars NU Ori and 42 Orionis, respectively, with feeble stellar winds. In contrast to the Veil surrounding M42, the thermal expansion of the warm ($\sim$7500K) ionized gas in these regions drives the expansion, the shell dynamics, and the disruption of these regions \citep{2020A&A...639A...2P}.
\par
The [CII] line is one of the brightest far-infrared cooling lines of the ISM and has been proposed as a valuable tracer of star formation across cosmic timescales. The Orion [CII] map provides a valuable dataset that can be used to better understand the origin of [CII] emission and its relation to other gas and dust tracers. Specific correlations that can be studied include the relation of [CII] emission, IRAC 8 micron emission, PACS 70 micron emission, total far-IR emission, and CO line emission (Pabst et al 2021). Using large-scale datasets allows us to study these correlations across several orders of magnitude and to relate their dependence to local physical conditions.
\par 
Analysis shows that the [CII] emission correlates well, but nonlinearly, with the 8 micron PAH and 70 micron dust emission. This results in a so-called [CII] deficit in regions characterized by warm dust. For Orion, this deficit is caused by a decreased heating efficiency in UV-bright regions, an increased importance of other cooling lines, and the contribution of deeply embedded star formation to the far-IR dust emission (Pabst et al 2021; Higgins et al 2021, both in prep.). 

\subsection{[CII] observations and radiative feedback from massive stars}
\label{fuv_effects}
The [CII] line is a key diagnostic for analyzing the heating and physical
conditions in the neutral gas (e.g., Kaufman et al. 2006, Pabst et al. 2017). 
For moderate-density photodissociation regions and for the bulk of 
the molecular cloud surface, the [CII] line is the dominant gas coolant. 
In thermal equilibrium,  the [CII] line measures the gas heating 
regardless of the exact heating process. Whether by radiative heating 
via photoelectric ejection of electrons from grains as expected for 
photodissociation regions or by mechanical heating from shocks or turbulence,
the heating rate is measured by the [CII] line. The ratio of [CII] to the 
infrared continuum, [CII]/$L_{IR}$ is a measure of the fraction of 
radiative energy that is converted into gas heating and is expected to be
$0.1-1$\%\ for PDRs. The theoretical heating rate is a function of 
$G_0 T^{1/2}/n_{e}$ , where $G_0$ is the incident far-ultraviolet radiation field and $n_e$ is the electron density.
We can use the true heating rate as measured by [CII]/$L_{IR}$
to calibrate the theoretical prediction. 
\\
Large-scale mapping allows
for a continuously varying distribution of heating rates from the
dense gas and intense radiation fields found near the Trapezium 
cluster to the lower density and lower radiation fields  
found in the outer edges of the molecular cloud, while the fainter outer edges of molecular clouds are generally overlooked as a source of [CII] emission when in fact they might dominate the [CII] emission on large scales \citep{2017ApJ...842....4A,2020A&A...639A.110A}. [CII] observations of extragalactic objects typically sample regions of massive star formation on a scale size of $\sim$400 pc.

Likely, the [CII] flux from these regions is also dominated by the emission from PDR surfaces on molecular clouds illuminated by moderate radiation fields.   When combined  with large-scale maps of the infrared 
continuum and the CO (2-1) from the IRAM 30m (PI: Goicoechea) along 
with theoretical PDR modeling tools \citep{2008ASPC..394..654P},
the [CII] maps provide a measure of the 
gas temperature, density, and incident far-ultraviolet field strength 
across the cloud. Analysis shows that the thermal gas pressure scales with the incident radiation field to the power 3/4, as expected for pressure equilibrium between photoionized gas and the surrounding PDR (Pabst et al 2021, in prep).

\subsection{Relation of [CII] emission and the large-scale molecular cloud}
\label{veil_co}
In order to obtain the molecular emission counterpart to [CII], we have started to expand previous $^{12}$CO and \mbox{$^{13}$CO (2-1)} maps of the central region of Orion~A taken with the IRAM\,30\,m telescope \citep[see][]{2014ApJ...795...13B}. The new  maps cover more extended and diffuse CO-emitting regions and are part of the $\sim$165\,h  Large Program ``Dynamic and Radiative Feedback of Massive Stars'' \mbox{(PI: J. R. Goicoechea)}. These maps have angular and spectral resolutions of 11$''$ ($\simeq$4,500\,AU) and  0.25 \kms , respectively (comparable to those of our [\CII]\,158\,$\mu$m map). The wide  bandwidth of the EMIR receivers  enable  mapping lines from  $^{12}$CO,  $^{13}$CO,  C$^{18}$O, and other species simultaneously 
\citep[for details, see][]{Goico20}.

A combined analysis of velocity-resolved [\CII]\,158\,$\mu$m and CO maps at this high spatial resolution provides  a powerful diagnostic tool for studying the radiative and mechanical feedback of massive stars.
While this type of  analysis has previously been pursued at arcmin$^2$ scales
to probe the interaction between UV radiation from massive stars and their  natal molecular cores  (see e.g.,~\citet{2015ApJ...812...75G} for OMC-1 or \citet{Pabst17} for the \mbox{Horsehead} nebula), much less was known about their square-degree spatial  distribution. This large-scale emission samples entire star-forming complexes, their vast (perhaps \mbox{CO-dark}) surrounding halos, and the swept-up material blown away  in the form of expanding shells.
This coherent [\CII] and CO data base, used in tandem with archival photometric images of the mid-IR PAH and far-IR dust emission, allows us to quantify the role of UV radiation and  stellar winds in great detail. While photometric images do not reveal the gas kinematics of the expanding shells, photoevaporative flows, and hydrodynamical instabilities at molecular cloud interfaces, this is a unique aspect we investigate with these  maps. In particular, by studying the [\CII] and CO emission at velocities different to those of OMC, we can search for any faint molecular emission  outside the star-forming  cloud: in the swept-up material of the shells, at the edges of \HII~regions, or in the form of small globules that may lead to the formation of very low mass stars \citep[][]{Goico20}.

Models of PDRs show that the
\mbox{[\CII]/CO~(2-1)} line intensity ratio depends on  physical conditions ($G_0$ and $n_{\rm H}$). The \mbox{[\CII]/FIR versus CO/FIR} relation is used to infer 
these parameters not only in local star-forming regions, but also toward
distant star-forming galaxies \citep[e.g.,][]{Stacey10}. With  more than two million
 [\CII]\,158\,$\mu$m and  \mbox{CO} spectra (and measurements of
 the far-IR dust emission) toward regions of Orion in which $G_0$, $n_{\rm H}$, and $A_V$ vary by orders of magnitude, we  can accurately calibrate the diagnostic power of the \mbox{[\CII]/CO} intensity ratio. More broadly, we  use these [\CII]\,158\,$\mu$m and CO~(2-1) maps to observationally constrain the physical conditions of the gas, not only $G_0$, $n_{\rm H}$, and $T_k$, but also to characterize the gas turbulence and the associated nonthermal contribution to the total pressure balance.

\subsection{[CII] emission and the filamentary structure of molecular clouds}
\label{filaments}
The unique spatial and kinematic information that the [CII] map encompasses allow a comprehensive look into the nature of the filamentary structure in a cloud that is engulfed by stellar feedback. In an ongoing effort, we are working on a detailed comparison of large-scale $^{13}$CO and C$^{18}$O emission \citep{2018ApJS..236...25K} and [CII] emission maps toward a number of filaments (previously identified by \citep{1969ApJ...155L.149F,2019A&A...623A.142S}). This study reveals for the first time the PDR nature of the filaments, their fate in an irradiated cloud, and their relation to ongoing star formation (Suri et al. in prep).

\subsection{Probing PDRs with [CII] 1.9 THz line and carbon radio recombination lines}

The physical properties of the atomic and CO-dark molecular gas are largely unknown because their probes are faint, for instance, carbon radio recombination lines (CRRLs). Using multiple CRRLs that are well spread in frequency, the gas temperature and density can be precisely determined (e.g., \citep{2017MNRAS.465.1066O}). The analysis of CRRLs can be greatly aided through the combined use of [CII] and CRRLs (e.g., \citep{1994ApJ...428..209N, 2014AstL...40..615T,2017MNRAS.467.2274S}). The large area covered by this [CII] survey enables the joint analysis of these lines, from subparsec to parsec scales, through a synergy with radio telescopes such as ALMA and LOFAR \citep{2019A&A...626A..70S}. This analysis demonstrates that the thermal pressure in the [CII] layers of the PDR on the surface of OMC-1 exceeds that of the ionized gas in the HII region. The strong pressure gradient toward us from the self-gravitating core OMC1 to the PDR surface, to the ionized gas, to the hot plasma, and to the foreground Veil nebula is at the basis of the evaporative ``champagne'' flow of the ionized gas \citep{2019Natur.565..618P}.

\subsection{[$^{13}$CII] emission and [CII] optical depth effects}
\label{sect:13cii}
\begin{figure}
\includegraphics[width=1.0\linewidth]{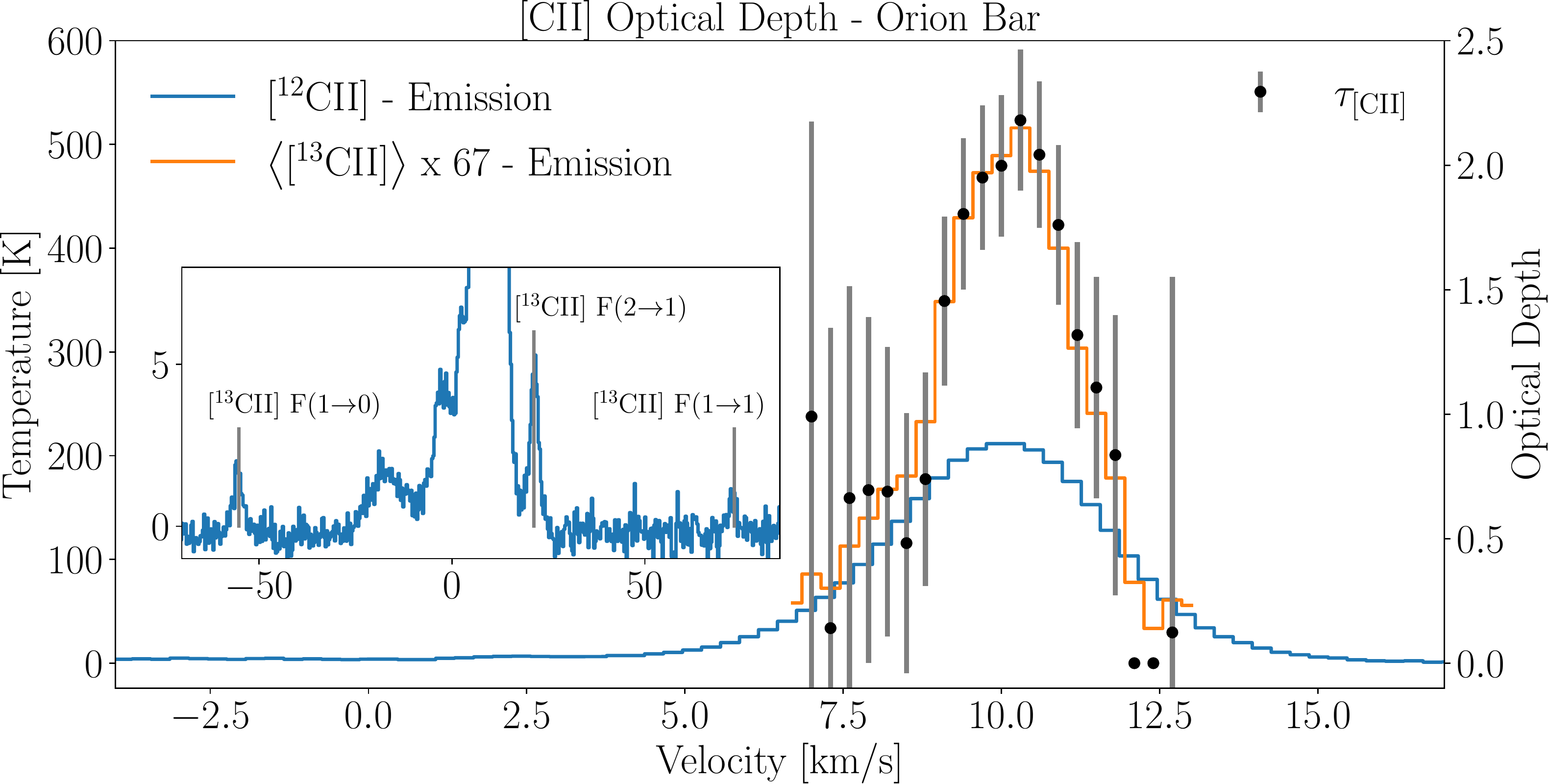} 
\caption{Averaged [$^{12}$CII] and [$^{13}$CII] emission originating from the ridge of the Orion bar.  The blue spectrum shows the  [$^{12}$CII] line, and the orange spectrum shows the velocity-corrected  $[^{13}\mathrm{CII}]$ emission averaged over the three hyperfine components and multiplied by 67, the carbon isotopic ratio in Orion \citep{Langer1990}. The black data points with error bars show the determined optical depth for each velocity channel, which is above unity around the peak of the line  [$^{12}$CII]. The inner panel shows the three  [$^{13}$CII] hyperfine transitions lines detected alongside the [$^{12}$CII] line. }

\label{fig:cii_spectrum}
\end{figure}

In contrast to the single [$^{12}$CII] $^{2}P_{3/2} \to\, ^{2}P_{1/2} $ transition, the [$^{13}$CII] emission is split into three hyperfine transitions by the unbalanced spin of the additional neutron. The location and relative intensity of the  [$^{13}$CII] hyperfine components are shown in Table \ref{tab:transition_parameter} \citep{Ossenkopf2013,Guevara2020}. 

\begin{table} [htb!]
        \caption{[$^{12}$CII] fine structure and [$^{13}$CII] hyperfine transition parameter.} 
        \label{tab:transition_parameter}
        \begin{tabular}{lcccrr}
                \hline
                \hline
                Transition line &  \multicolumn{2}{c}{Weight}   & Freq. & Velocity & Relative \\
                                &  &                            &       & offset & intensity   \\         
                & $g_u$ & $g_l$ &$\nu$& $\Delta v_{\mathrm{F\to F'}}$ & $s_{\mathrm{F\to F'}}$ \\
                & & & [GHz] & [km/s] \\
                \hline
                [$^{12}$CII] $\mathrm{^2P_{3/2}\to ^2P_{1/2}}$ & 4 & 2 & 1900.5369 & 0 & 1 \\
                
                [$^{13}$CII] $\mathrm{F = 2\to 1}$ & 5 & 3 & 1900.4661 & + 11.2 & 0.625 \\
                
                [$^{13}$CII] $\mathrm{F = 1\to 0}$ & 3 & 1 & 1900.9500 & - 65.2 & 0.250 \\
                
                [$^{13}$CII] $\mathrm{F = 1\to 1}$ & 3 & 3 & 1900.1360 & + 63.2 & 0.125 \\
                \hline
        \end{tabular}
\end{table}  

The weak  [$^{13}$CII] emission from the three transition lines require a low spectral rms and a flat baseline structure over a large velocity range $\pm 100\,\mathrm{km/s}$ to distinguish the emission from the noise floor. This means that recovering bad spectra using the novel spline approach by removing baseline features such as drifts and standing waves and therefore reducing the overall spectral noise are crucial for the  [$^{13}$CII] analysis.
\\
However, the detection of the three [$^{13}$CII] satellites requires a high signal-to-noise ratio, much higher that the signal-to-noise ratio for a given map pixel. The square-degree [CII] map allows averaging over arcminute-size regions. This results in a greatly improved signal to noise
 in the average spectra, facilitating the detection of the isotope emission. The velocity-resolved [CII] spectrum averaged over the Orion bar is shown in Fig. \ref{fig:cii_spectrum}. We observe the [$^{12}$CII] at $10\,\mathrm{km/s}$ and the three hyperfine transitions. The strongest [$^{13}$CII] F$(2\to1)$ line is shifted by $11.2\,\mathrm{km/s}$ from the [$^{12}$CII] line. This close line is often affected by the wing of the main isotope and is therefore not useful for an analysis. The second strongest component is sufficiently far away from the main component, $-65.2\,\mathrm{km/s,}$ but has only $25\,\%$ of the total intensity; see Table \ref{tab:transition_parameter}. 

The simulation of [$^{12}$CII] and its isotope  [$^{13}$CII] enables us to determine possible optical depth effects in [CII] cloud emission. Previous studies (e.g. \cite{Graf2012, Ossenkopf2013, Guevara2020, 2019A&A...631L..12O}) have shown that the [$^{12}$CII] line can be heavily affected by optical depth effects. Thus, the observed [$^{12}$CII] intensity or line shape do not necessarily reflect the extent of [CII] emission at this position, which further highlights the necessity of velocity-resolved [CII] observations.

A comparison of the line intensities and shape in [$^{12}$CII] and [$^{13}$CII] can reveal whether the main component is affected by optical depth effects. The optical depth of the ionized carbon atom can be determined by \citep{Guevara2020} 

\begin{equation}
\frac{\mathrm{T_{mb, [^{12}CII]}}(v)}{\mathrm{T_{mb, [^{13}CII]}}(v+\Delta v_{\mathrm{F\to F'}})} s_{\mathrm{F\to F'}}= \frac{1-\exp(-\tau(v))}{\tau(v)} \alpha 
,\end{equation}

\noindent
with the carbon abundance ratio in Orion $\alpha = 67 \pm 3$ determined by \cite{Langer1990} and the relative intensity $s_{\mathrm{F\to F'}}$ of the hyperfine transition of the ionized carbon isotope. 

The [$^{12}$CII] line and the [$^{13}$CII] line scaled by the carbon isotopic ratio are plotted in Fig. \ref{fig:cii_spectrum}. We observe that the scaled [$^{13}$CII] overshoots the [$^{12}$CII] emission, indicating optical depth effects as observed by \cite{Ossenkopf2013} in the Orion bar. The velocity-resolved optical depth is shown by the black data points with error bars. The optical depth increases toward the peak of the spectrum,  reaching a maximum value of $\tau_{\mathrm{[CII]}}=2.2\pm0.3$. Integrating both spectra between $7-13\,\mathrm{km/s}$ gives us an averaged optical depth for the Orion bar of $\langle\tau_{\mathrm{[CII]}}\rangle=1.3\pm0.1,$ similar to the value previously determined by \cite{Ossenkopf2013} using [CII] data obtained with Herschel/HIFI.

As the analysis of the ionized carbon spectrum in the Orion bar has shown, the observed [CII] spectra might be  affected by optical depth effects, which could affect their observed intensity and shape. An unbiased in-depth study of the [CII] optical depth effects in the [CII] map is underway, using tools to group similar regions together to increase the signal-to-noise ratio \citep{Kabanovic2020}.
\section{Summary}
We have introduced the largest velocity-resolved map of [CII] observed so far. We described the observing strategy and data reduction. The usage of a catalog of splines to remove baseline artifacts was demonstrated. The observed data were compared to a similar smaller map taken with the Herschel/HIFI spectrometer. The HIFI and upGREAT data agree to within 4\% of each other.  The techniques discussed here will benefit upcoming large-scale mapping projects such as the SOFIA Cycle 8 legacy project Feedback \citep{Schneider2020} and potentially the GUSTO balloon mission \citep{2019AGUFM.A33Q2940B}. 
\par
The final section of the paper discusses the scientific possibility of this dataset. The scientific topics discussed will be expanded into stand-alone articles. The final map shown in figure \ref{fig:final_map} is available from the NASA infrared archive (IRSA).\begin{acknowledgements} 
This work is based on observations made with the NASA/DLR Stratospheric Observatory for Infrared Astronomy (SOFIA). SOFIA is jointly operated by the Universities Space Research Association, Inc.(USRA), under NASA contract NAS2-97001, and the Deutsches SOFIA Institut (DSI) under DLR contract 50 OK 0901 to the University of Stuttgart. This work is carried out within the Collaborative Research Centre 956, sub-project [A4], funded by the Deutsche Forschungsgemeinschaft (DFG) – project ID 184018867. We thank the Spanish MICIU for funding support under grant AYA2017-85111-P.
\end{acknowledgements}

\bibliographystyle{aa}       
\bibliography{bibliography}

\begin{appendix}

\section{Supplementary material}
\begin{table*}
\centering
\begin{tabular}{llll}
\toprule
Flight ID            & Tiles &                                         Tile IDs &  Spectra count\\
\midrule
2016-11-10\_GR\_F348 &     2 &                                       0505, 0605             & 34272 \\
2016-11-15\_GR\_F349 &     9 &               0105, 0205, 0304, 0305, 0306, 0307,0405, 0705, 0805 & 272160 \\
2016-11-16\_GR\_F350 &     8 &   0206, 0301, 0302, 0303, 0406, 0506, 0606, 0706             & 226800 \\
2016-11-17\_GR\_F351 &     8 &   0106, 0107, 0206, 0207, 0407, 0507, 0607, 0608             & 226800 \\
2016-11-18\_GR\_F352 &     8 &   0204, 0404, 0408, 0504, 0508, 0604, 0704, 0804             & 244440 \\
2017-02-08\_GR\_F371 &     8 &   0004, 0104, 0204, 0206, 0503, 0505, 0603, 0605             & 181440 \\
2017-02-09\_GR\_F372 &    10 &   0003, 0103, 0203, 0403, 0502, 0602, 0603, 0702, 0703, 0803 & 277200 \\
2017-02-10\_GR\_F373 &     5 &                     0102, 0202, 0402, 0501, 0601             & 138600 \\
2017-02-14\_GR\_F374 &     8 &   0101, 0201, 0401, 0509, 0601, 0609, 0708, 0709             & 209160 \\
2017-02-15\_GR\_F375 &     8 &   0309, 0310, 0409, 0410, 0509, 0510, 0610, 0710             & 236880 \\
2017-02-16\_GR\_F376 &     7 &         0311, 0411, 0511, 0611, 0612, 0711, 0712             & 185220 \\
2017-02-17\_GR\_F377 &     9 &   0308, 0312, 0403, 0412, 0512, 0612, 0703, 0707, 0803       & 228060 \\
\midrule
\multicolumn{3}{r}{Total Spectra }                                                           &         2461032\\  
\bottomrule
\end{tabular}
\caption{Summary of flights and observed tiles. The project was observed over 13 flights. 12 flights were used for map observing. A separate flight was dedicated to calibration and OFF check observations.}
\label{table:flight_summary}
\end{table*}

\begin{figure*}
\centering
\includegraphics[width=1.0\linewidth]{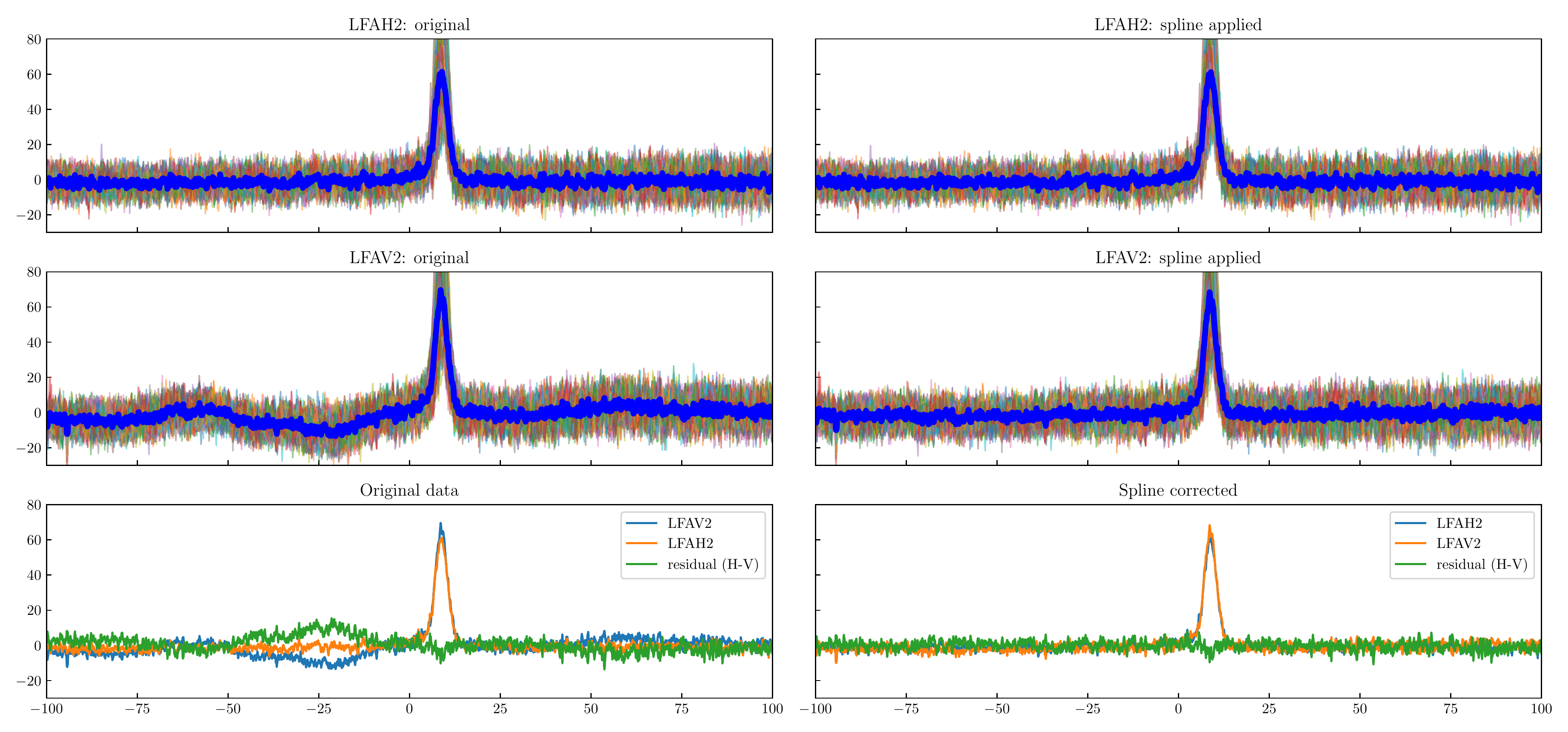}
\caption{Comparison of the LFAH2 (left panels) and LFAV2 (right panels) pixels with and without spline baseline correction. Dark blue lines show the average spectrum over 84 spectra. The lower panels show the residual between the H and V data with and without spline correction. There is a 5 kelvin difference between the peak intensity before and after spline correction, indicating that spline correction has not affected the line intensity. LFAH2 and LFAV2 are not coaligned and are offset on the sky by 2 arcseconds, which might account for the different line intensity.}
\label{fig:spline_H_V_comparions}

\end{figure*}

\begin{figure*}
\includegraphics[width=1.0\linewidth]{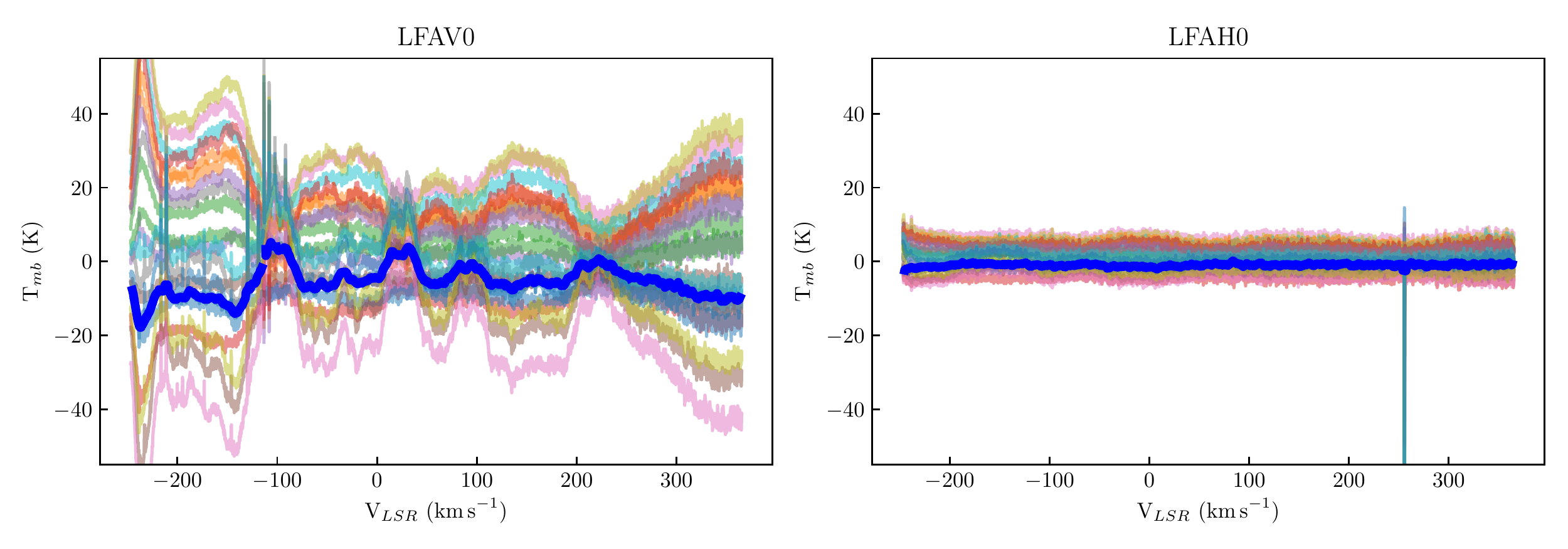}
\caption{Calibrated residual between different OFF measurements taken over a flight leg. This is used to generate a catalog of baseline shapes. Note the large residuals for the LFAV0 pixel (left panel) compared to the more stable LFAH0 pixel (right panel).}
\label{fig:cmz_sky_diff}
\end{figure*}

\begin{figure*}
\includegraphics[width=1.0\linewidth]{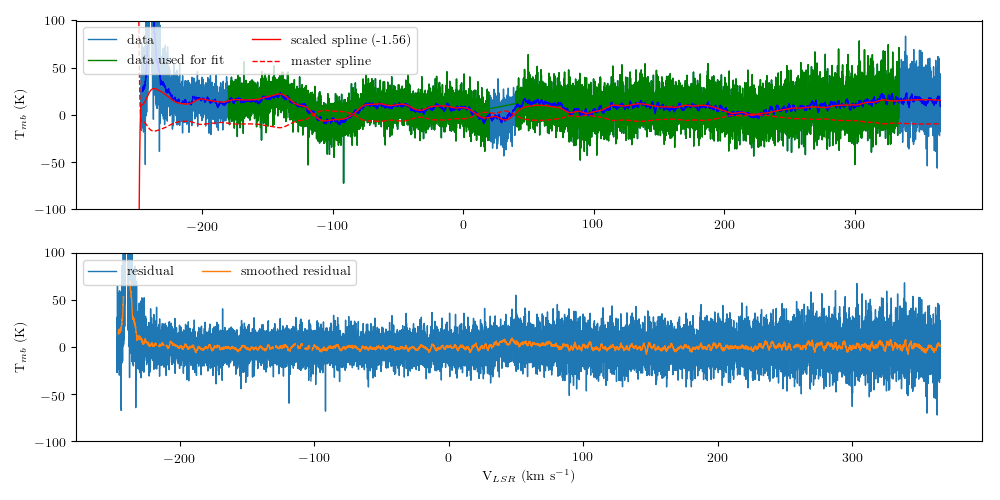}
\caption{Spline correction of a single OTF dump for the Galactic center project. \textit{Upper Panel:} Master spline (shown as a dashed red line) used in the correction is also shown in dark blue in figure \ref{fig:cmz_sky_diff}. This master spline is scaled by a factor of -1.5, which provides a good match to the original data. \textit{Lower panel:} Residual after scaled spline subtraction.  The weak broad emission seen between 20 and 70 \kms \ in the spline-corrected spectrum is not visible in the original data.}
\label{fig:cmz_correction}
\end{figure*}

\begin{figure*}
\includegraphics[width=1.0\linewidth]{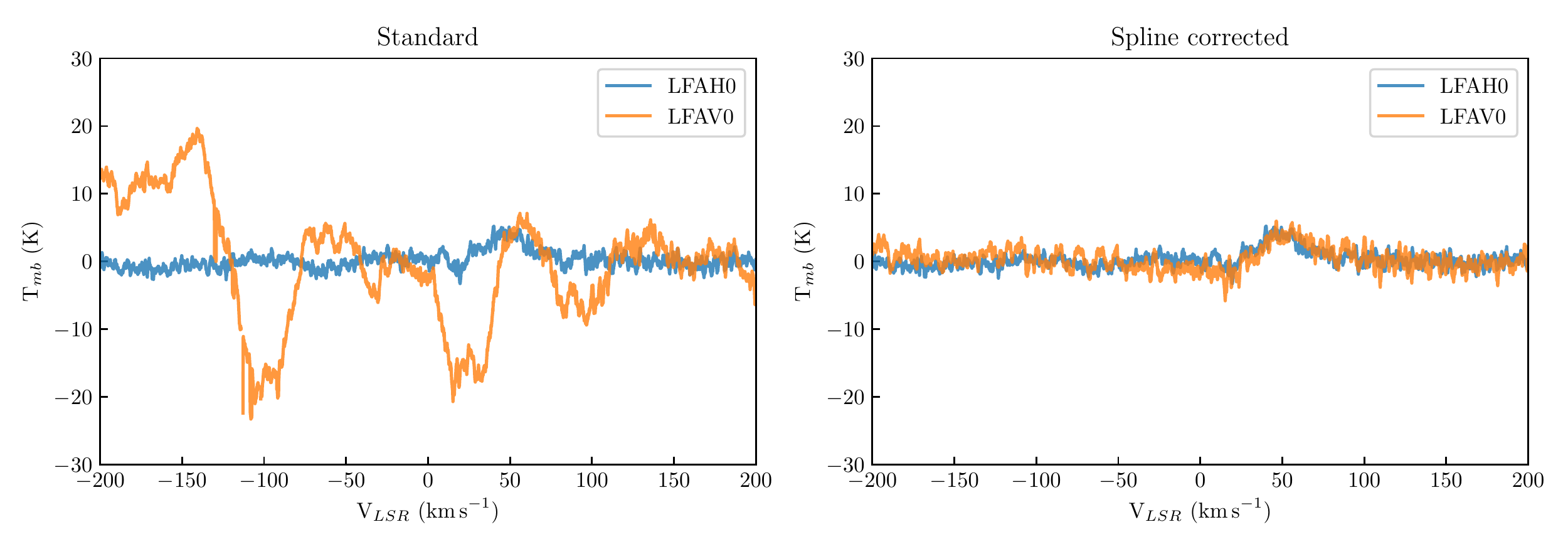}
\caption{Before and after spline correction. Average of 80 spectra from an OTF scan taken as part of the [CII] Galactic center mapping project (data shown in figure \ref{fig:cmz_correction} are part of this average). \textit{Left panel:} Standard reduction with a third-order polynomial baseline correction. \textit{Right panel:} Reduction using the spline catalog approach. We note [CII] emission from 20 to 70 \kms \ and consistent line emission seen between coaligned ($\sim$ 2 arcseconds) H and V pixels after correction.}
\label{fig:cmz_average_otf_scan}
\end{figure*}

\begin{figure*}
\centering
\includegraphics[width=0.47\linewidth]{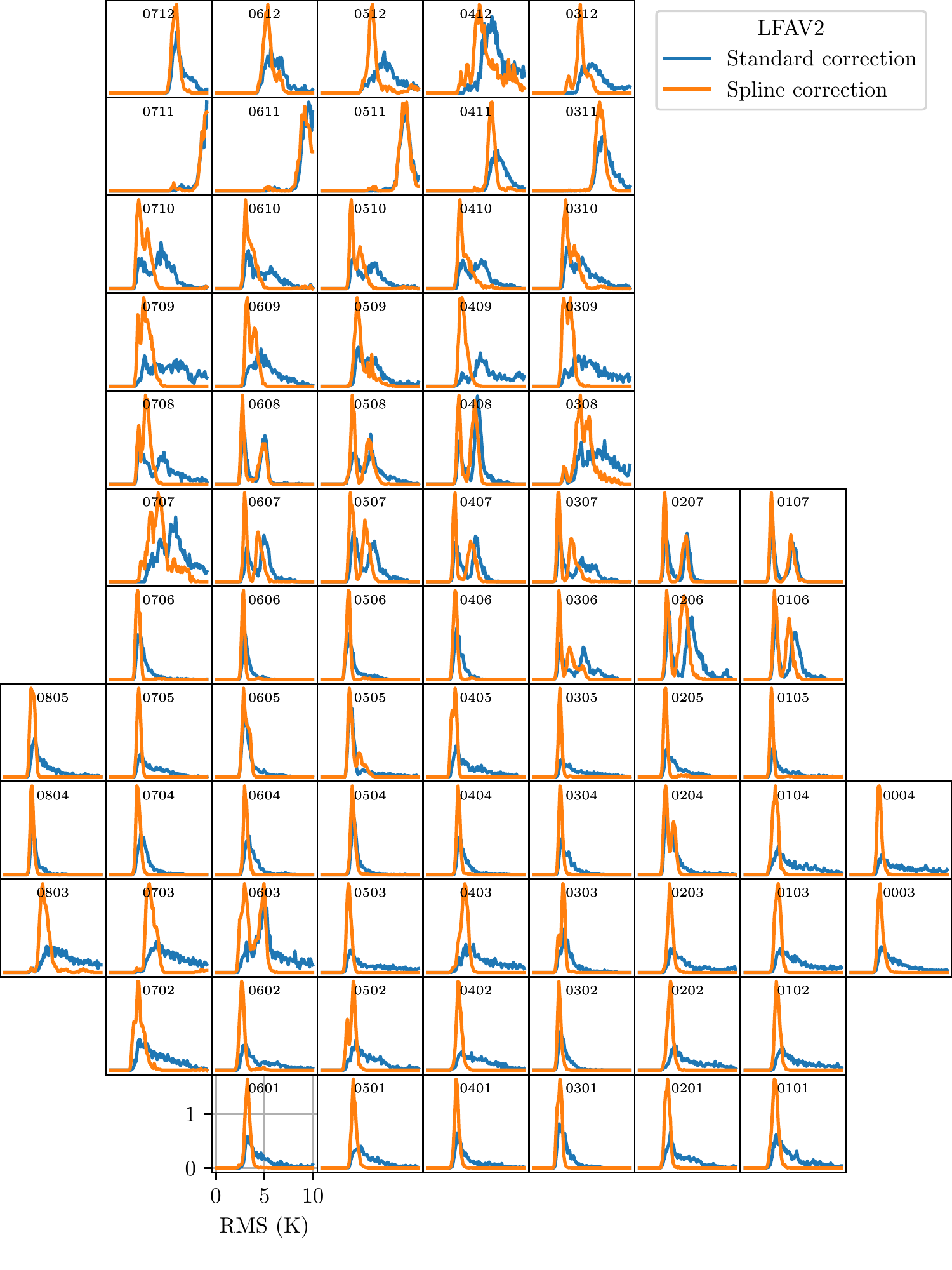}
\includegraphics[width=0.47\linewidth]{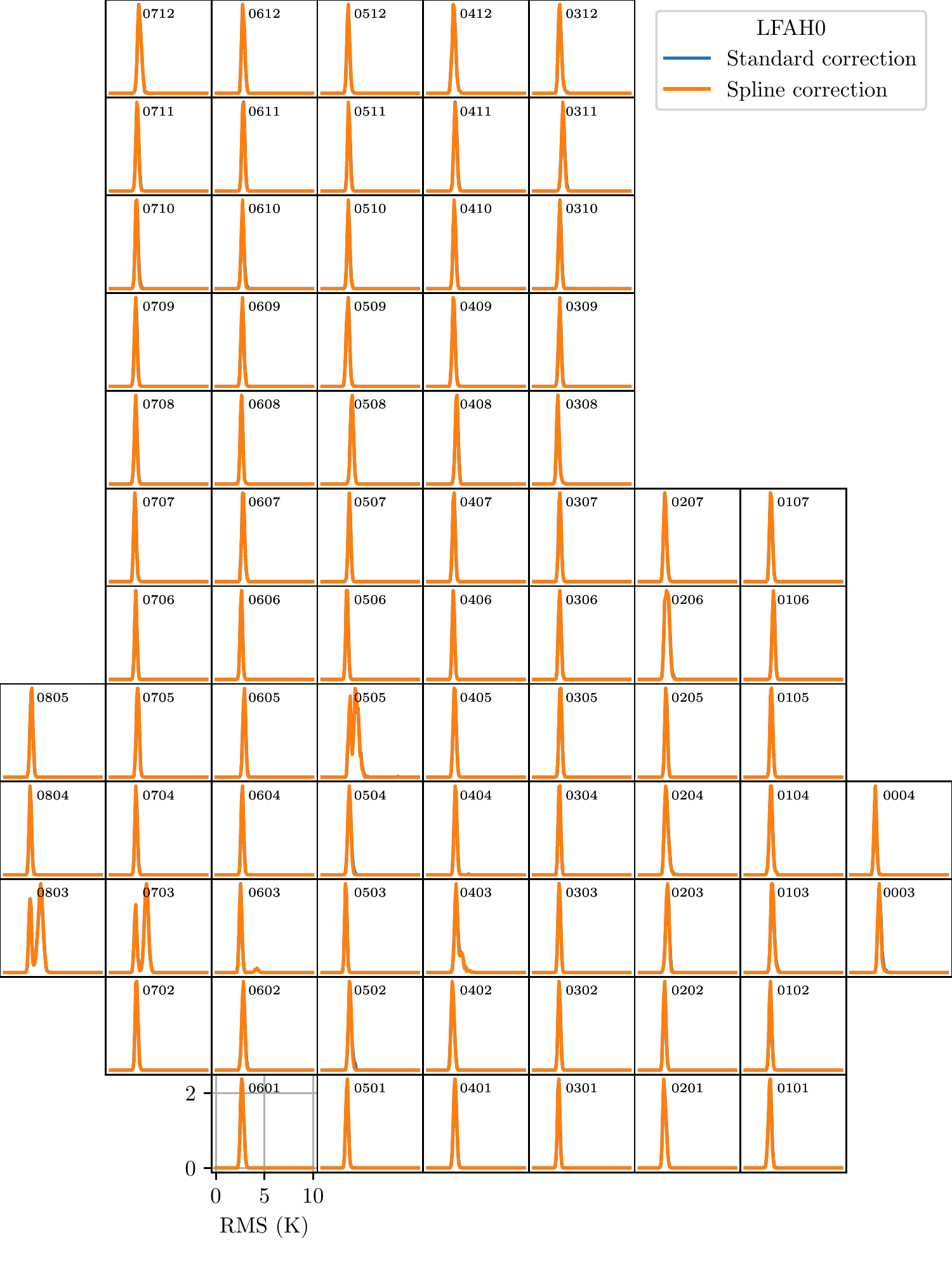}
\caption{RMS distribution for all spectra for pixel LFAV2 (left panel) and LFAH0 (right panel) per map tile. The RMS distribution is shown before (blue) and after baseline spline correction (orange). Baseline issues in the RMS distribution are shown by a long tail toward higher RMS values. Pixel LFAV2 shows a noticeable improvement in RMS distribution compared to the standard polynomial baseline correction. Conversely, LFAH0 has no baseline performance issue, and so the application of the spline correction made no significant difference to the RMS distribution. The RMS is taken over a range of -100 to 100 km/s in which the central region from -20 to 30 km/s was ignored for the RMS determination. The spectra have a 0.2 km/s resolution.}
\label{fig:tiles_rms_before_spline}

\end{figure*}

\begin{figure}
\includegraphics[width=1.1\linewidth]{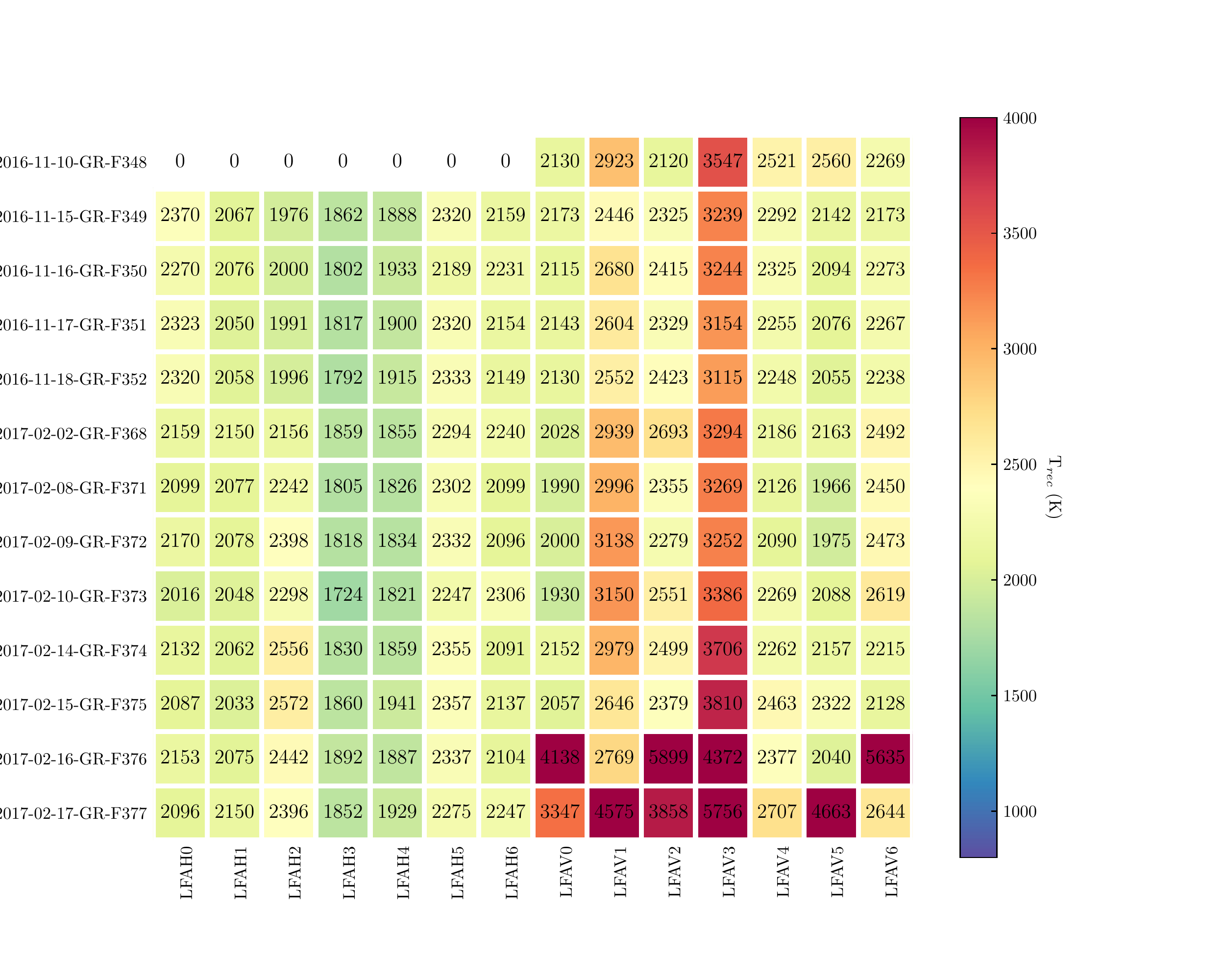}
\caption{Summary of upGREAT LFA single-sideband receiver temperature for the 13 project flights. The receiver temperature for the V array increases toward later flights. This was linked to a degradation in LFAV LO performance. The LO was shipped back to the manufacturer for maintenance after the flight series. The H LO was not available during the 2016-11-10 flight and is recorded here with 0 receiver temperature.}
\label{fig:trec_summary}
\end{figure}

\begin{figure}
\centering
\includegraphics[width=1.0\linewidth]{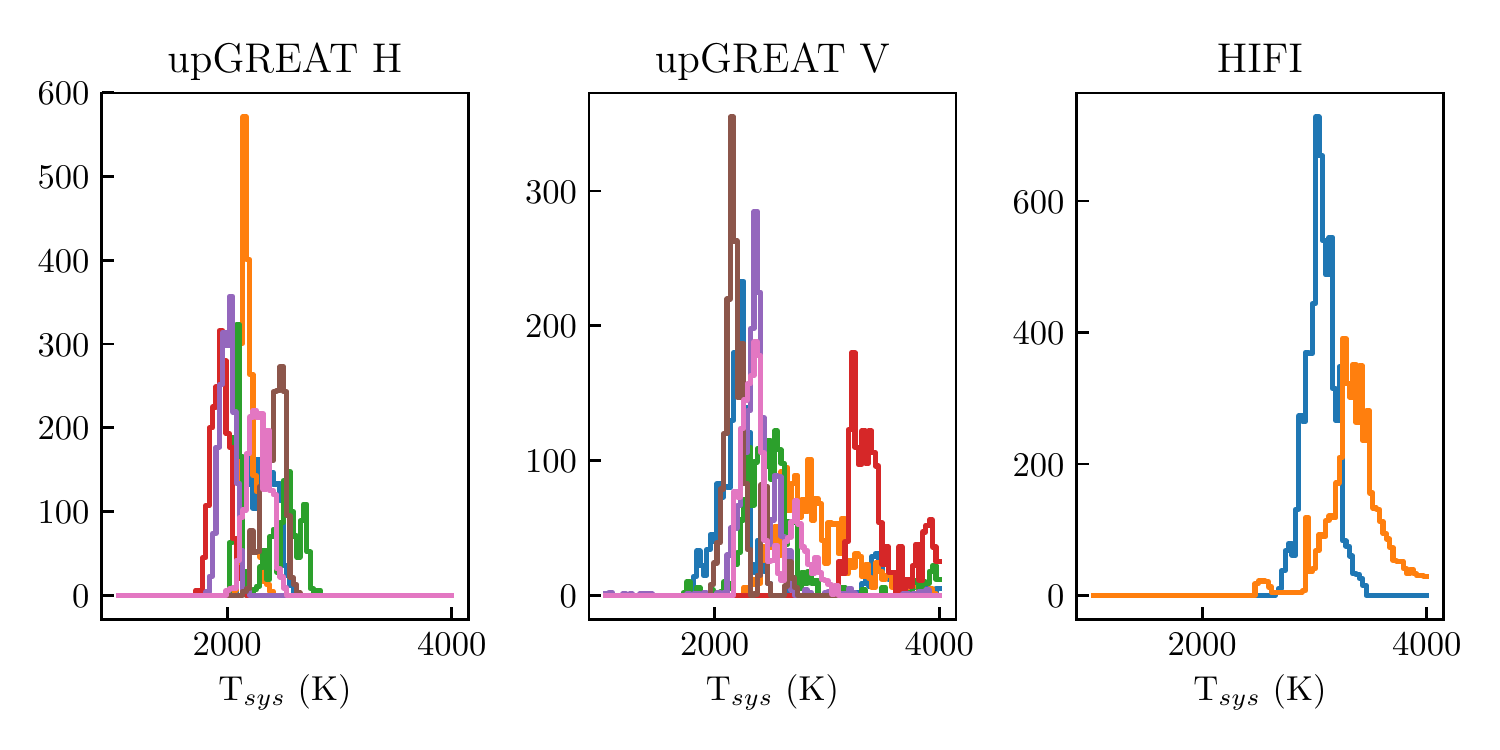}
\caption{Distribution of single-sideband system temperature for each pixel in all 13 flights with a comparison to HIFI system temperatures for the Orion map. The receiver temperatures are measured in the IF band where the astronomical signal falls	. In the case of upGREAT, this is at an IF of 1.9 GHz (see figure \ref{fig:trec_example_upgreat}).}
\label{fig:tsys_summary_hifi_upgreat}
\end{figure}

\begin{figure}
\centering
\includegraphics[width=1.0\linewidth]{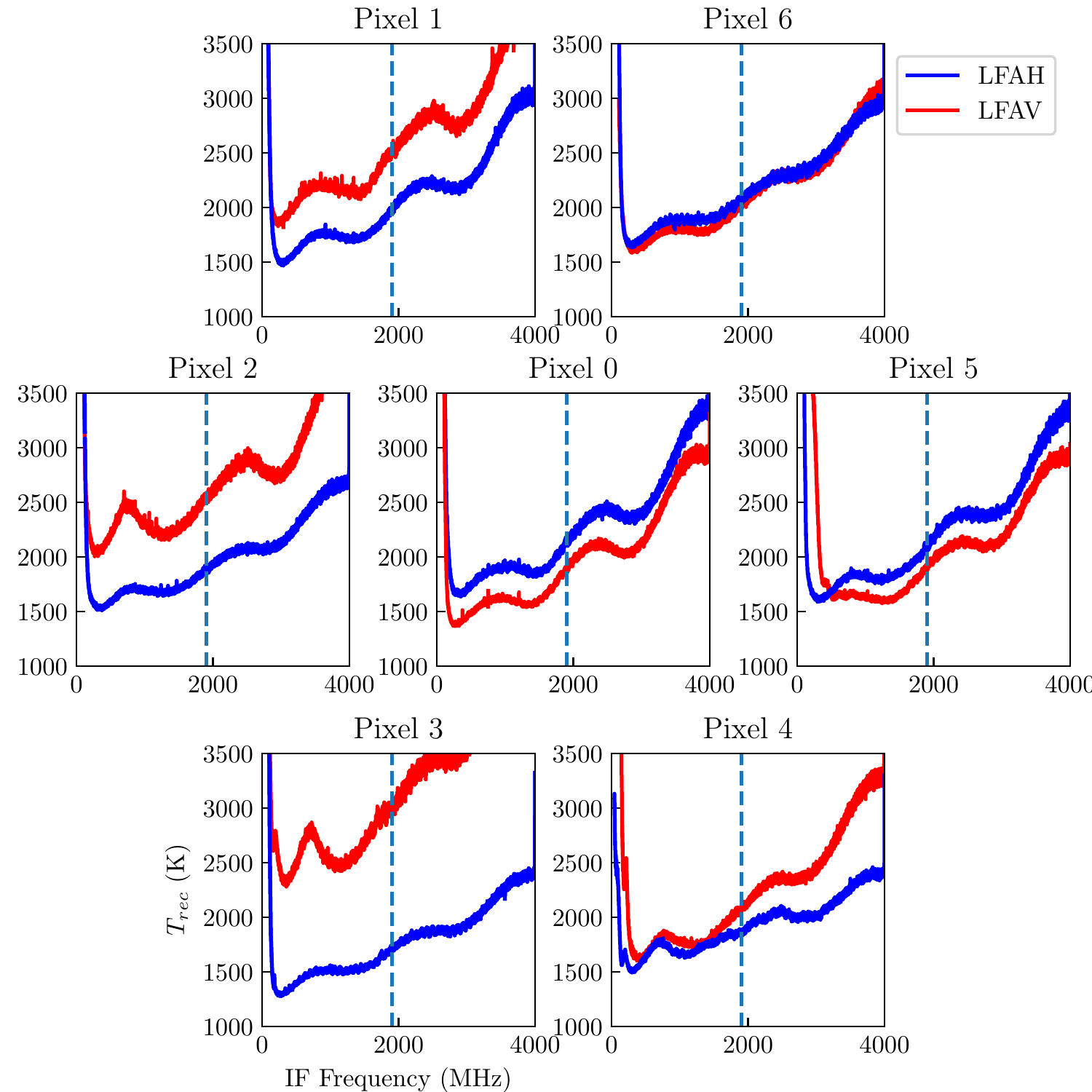}
\caption{Example of upGREAT single-sideband receiver temperatures. The IF frequency used to set the LO frequency is marked at 1.9 GHz, which corresponds to a $V_{LSR}$ of 10 km/s. The slope in receiver temperature increases toward higher IF frequencies. Lower receiver temperatures are available at lower IF frequencies, but they can be susceptible to LO instabilities or spurious output.}
\label{fig:trec_example_upgreat}
\end{figure}

\begin{figure}
\centering
\includegraphics[width=1.0\linewidth]{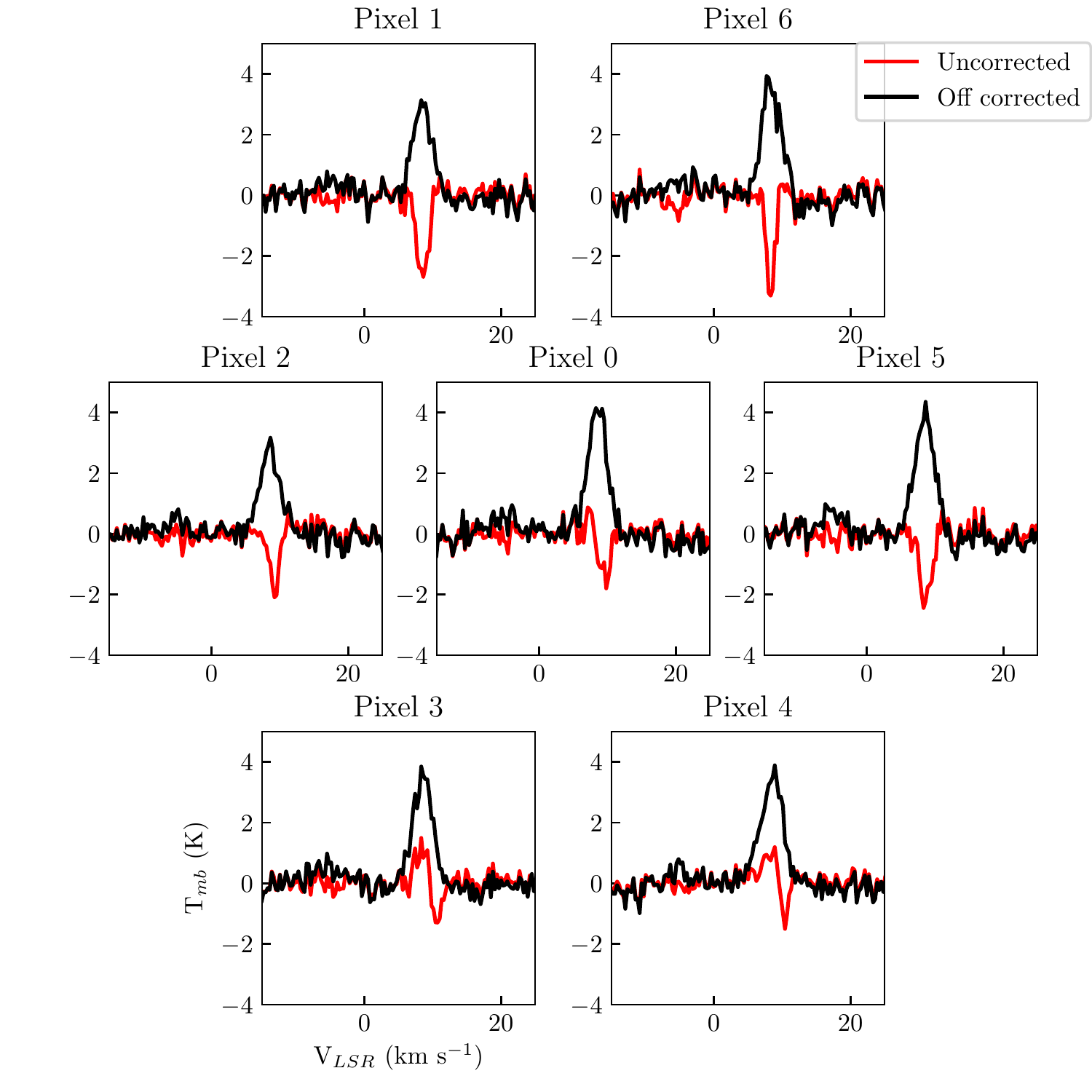}
\caption{Example of OFF correction process for H array showing average emission for tile 0303 before and after correction for OFF contamination from the COFF-C position.}
\label{fig:off_emission_correction}
\end{figure}

\begin{figure*}

\includegraphics[width=0.45\linewidth]{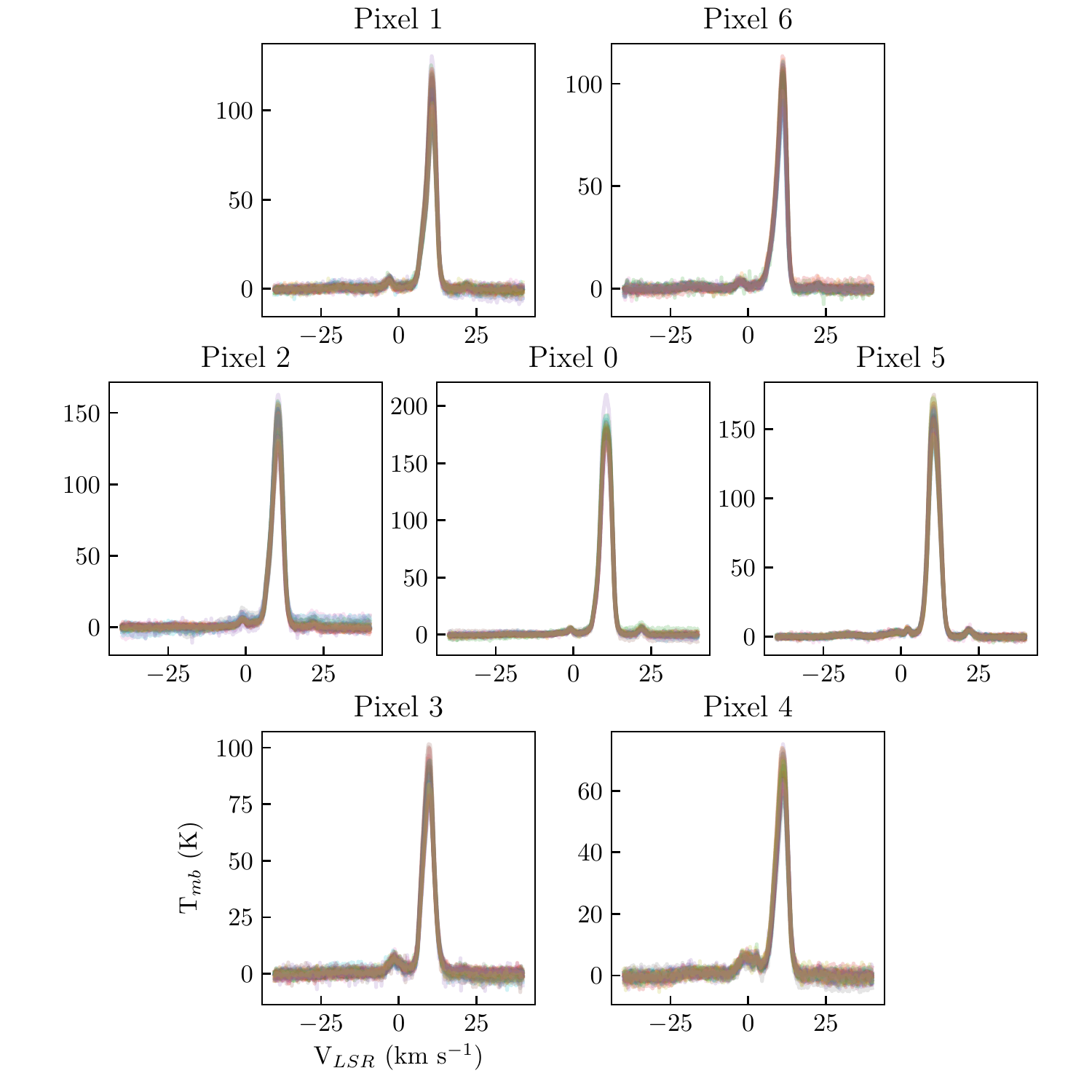}
\includegraphics[width=0.45\linewidth]{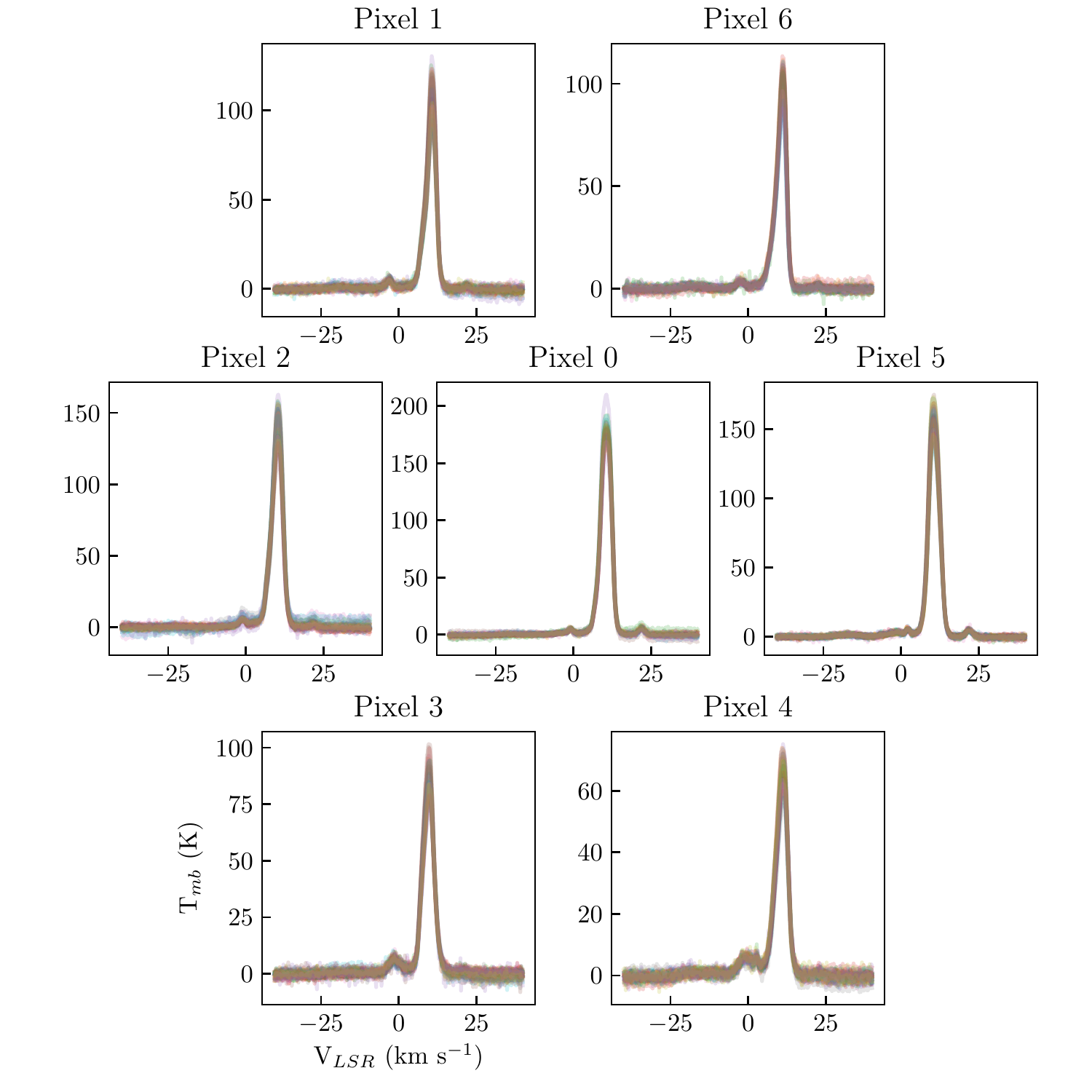}

\caption{Overview plot showing all calibration observations of the Orion bar observed in 13 flights, 20 single-point total power observations in total. See section \ref{sec:consistency} for more details. The LFAH array is shown in the left panel, and LFAV is presented in the right panel.}
\label{fig:calibration_orion_bar_spectra}
\end{figure*}

\begin{figure*}
\includegraphics[width=1.0\linewidth]{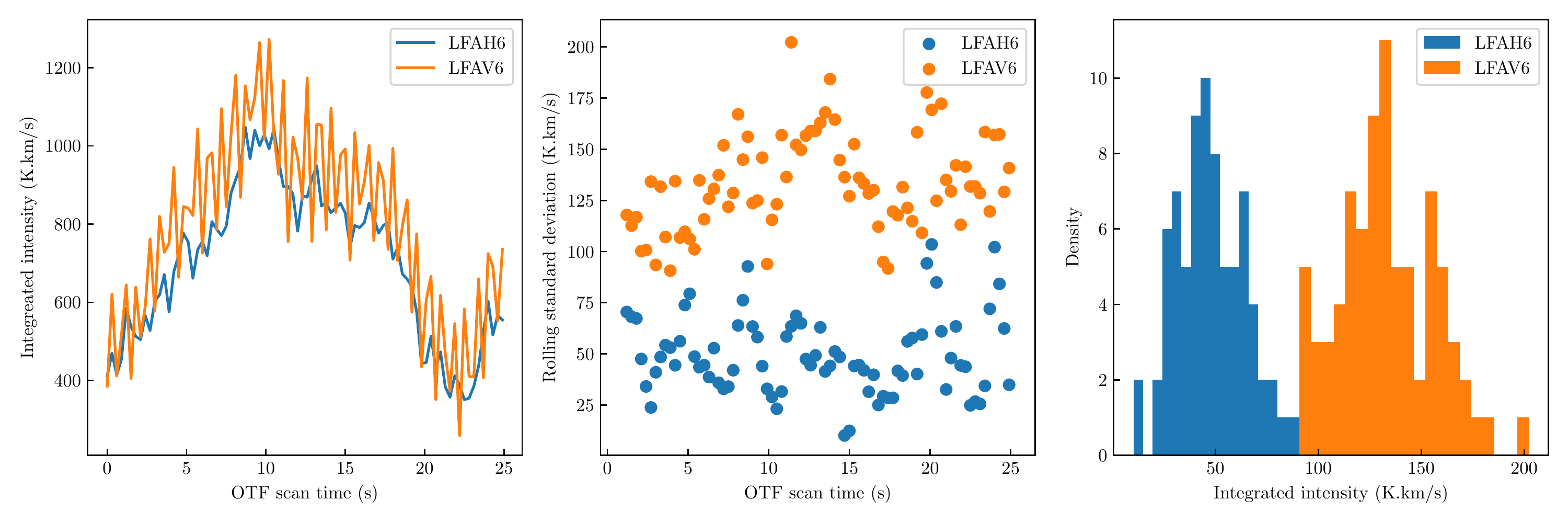}
\caption{Detection of gain instability in an OTF scan using a rolling standard deviation over the integrated intensity of an OTF scan. \textit{Left panel:} Integrated intensity variation over a 25-second OTF scan with 84 spectra. The variation for LFAV6 is stronger than that in LFAH6. \textit{Center panel:} Rolling standing deviation over five samples. A rolling standard deviation removes the sky signal variation during the OTF scan and leaves just the OTF sample to sample variations. \textit{Right panel:} Distribution of rolling standard deviations. The unstable pixel, LFAV6, shows a larger mean standard deviation than pixel LFAH6. This property can be used to detect gain instability in a given pixel.}
\label{fig:gain_instability_detection}
\end{figure*}

\begin{figure*}

\includegraphics[width=1.0\linewidth]{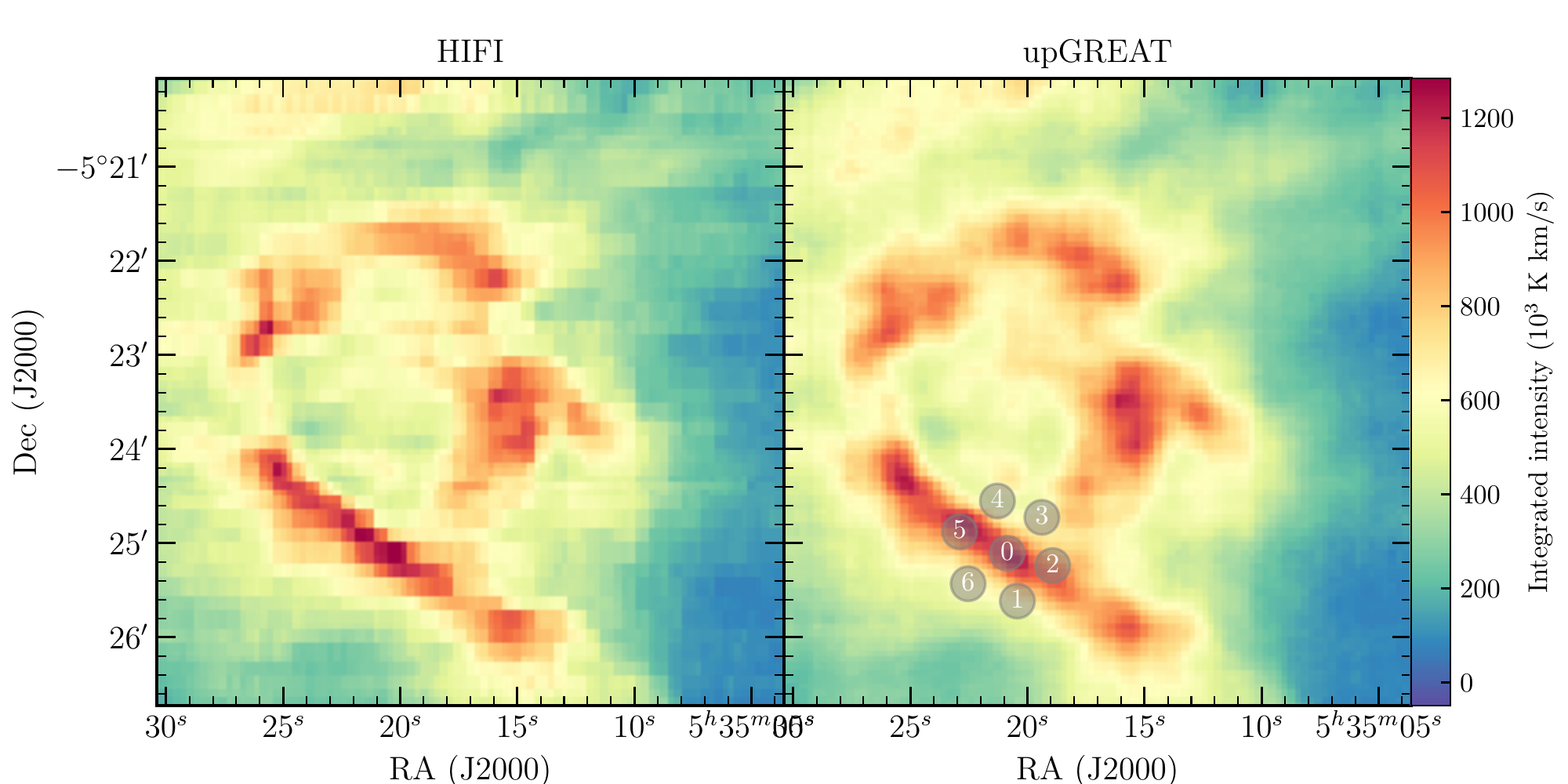}
\caption{Side-by-side plot of HIFI and upGREAT data gridded with an 18 arcsecond beam. The variation in integrated intensity between -5 and 15 km/s is shown. The array positions highlighted in gray show the location of the Orion bar consistency observation. The map is generated with a 18.1 arcsecond kernel. The smoother map shown for upGREAT shows the difference between a fully sampled (5.2-arcsecond sample) and the HIFI partially sampled grid (10-arcsecond sample). See figure \ref{fig:HIFI_upGREAT_map_sampling} for an overview of the OTF dump positions.}
\label{fig:hifi_upgreat_side_by_side_under_sampled}
\end{figure*}

\begin{figure*}
\includegraphics[width=1.0\linewidth]{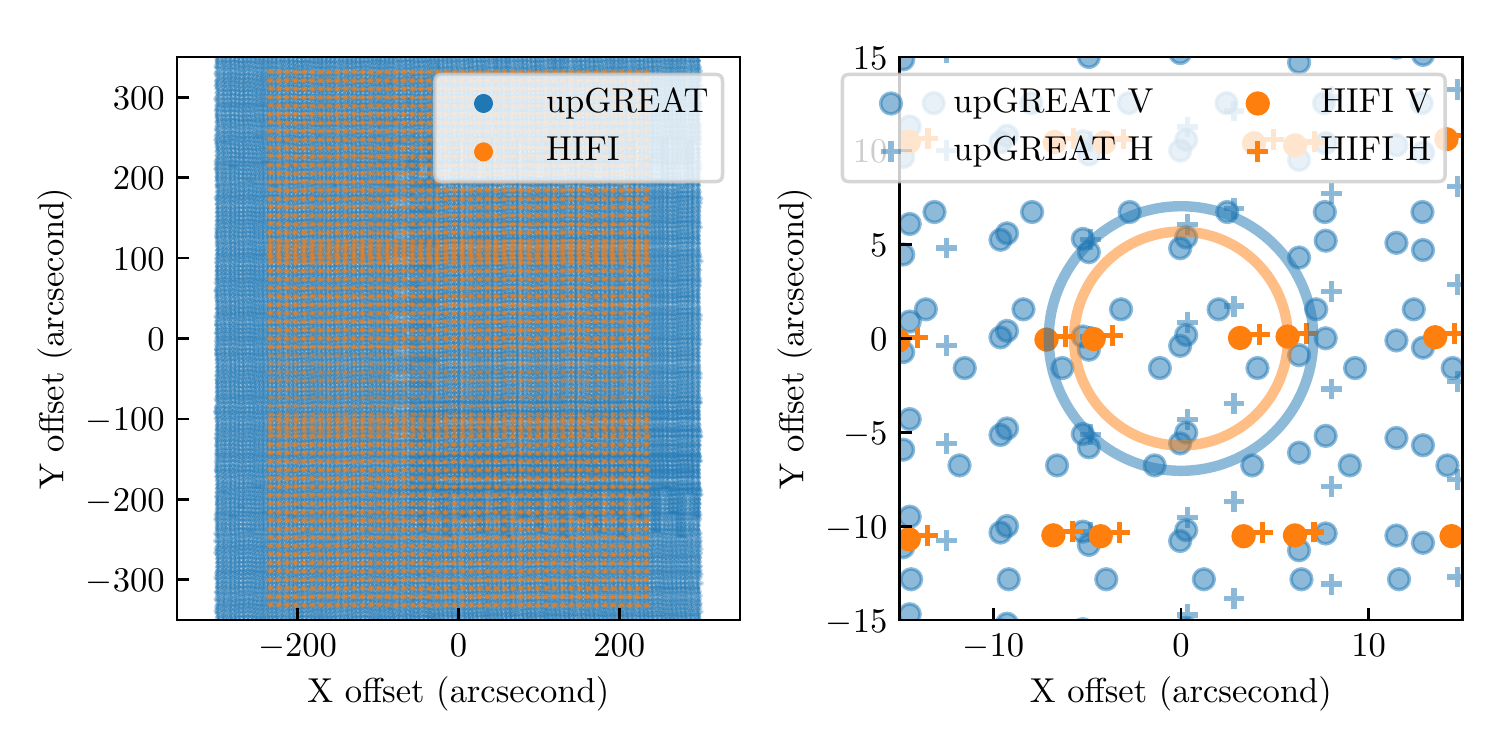}

\caption{\textit{Left panel:} Overview of spectrum positions for HIFI and upGREAT maps. \textit{Right panel:} Zoom into the central map region. The sparse sampling of the HIFI map with 11-arcsecond spacing between OTF scans arises because half the OTF scan is spent observing the internal cold load  andthere are gaps in coverage in the scan direction. HIFI observed in a ABBA format, so that we see two ON spectra taken in succession, followed by two load observations. HIFI and upGREAT beam size are overplotted for comparison at 11.4 and 14.1 arcseconds, respectively.}
\label{fig:HIFI_upGREAT_map_sampling}
\end{figure*}

\end{appendix}
\end{document}